\documentclass[twocolumn]{aastex63}

\renewcommand{\deg}{\mbox{$^{\circ}$}}
\newcommand{\kms}{\mbox{km\,s$^{-1}$}}

\newcommand{\x}{\mbox{$\times$}}
\newcommand{\Msun}{\mbox{M$_{\odot}$}}

\newcommand{\E}[1]{\hbox{$10^{ #1 }$}}

\def\HII        {\hbox{H \small{II}}}

\usepackage{graphicx}
\usepackage{txfonts}
\usepackage{natbib}

\defcitealias  {Cortes2016}{Paper I}

\begin{document}
\title{THE SEVEN MOST MASSIVE CLUMPS IN W43-MAIN AS SEEN BY ALMA: DYNAMICAL EQUILIBRIUM AND MAGNETIC FIELDS}

\author{Paulo C.\,Cortes}
\affiliation{Joint ALMA Office, Alonso de Cordova 3107, Vitacura, Santiago, Chile}
\affiliation{National Radio Astronomy Observatory, Charlottesville, VA 22903, USA}

\author{Charles L. H. Hull}
\affiliation{National Astronomical Observatory of Japan, NAOJ Chile Observatory, Alonso de C\'ordova 3788, Office 61B, 7630422, Vitacura, Santiago, Chile}
\affiliation{NAOJ Fellow}
\affiliation{Joint ALMA Office, Alonso de Cordova 3107, Vitacura, Santiago, Chile}

\author{Josep M. Girart}
\affiliation{Institut de Ci\`encies de l’Espai (ICE, CSIC), Campus UAB, C/ Can Magrans S/N, 08193 Cerdanyola del Vall\`es, Catalonia, Spain}
\affiliation{Institut d’Estudis Espacials de Catalunya (IEEC), 08034 Barcelona, Catalonia}

\author{Carlos Orquera-Rojas}
\affiliation{Departamento de F\'isica, Universidad Cat\'olica del Norte, Av. Angamos 0610, Antofagasta, Chile}
\affiliation{Instituto de Astrof\'isica, Facultad de F\'isica, Pontificia Universidad Cat\'olica de Chile, Santiago, Chile}
\affiliation{Millennium Institute of Astrophysics (MAS), Nuncio Monsenor Sótero Sanz 100, Providencia, Santiago, Chile}

\author{Tirupati K. Sridharan}
\affiliation{Harvard-Smithsonian Center for Astrophysics, 60 Garden St., Cambridge, MA 02138, USA}

\author{Zhi-Yun Li}
\affiliation{Astronomy Department, University of Virginia, Charlottesville, VA 22904, USA}

\author{Fabien Louvet}
\affiliation{Departamento de Astronomia - Universidad de Chile}

\author{Juan R. Cortes}
\affiliation{Joint ALMA Office, Alonso de Cordova 3107, Vitacura, Santiago, Chile}
\affiliation{National Radio Astronomy Observatory, Charlottesville, VA 22903, USA}

\author{Valentin J. M. Le Gouellec}
\affiliation{European Southern Observatory, Alonso de Córdova 3107, Vitacura, Santiago, Chile}
\affiliation{AIM, CEA, CNRS, Université Paris-Saclay, Université Paris Diderot, Sorbonne Paris Cité, F-91191 Gif-sur-Yvette, France}

\author{Richard M. Crutcher}
\affiliation{Astronomy Department, University of Illinois at Urbana-Champaign, IL 61801, USA}

\author{Shih-Ping Lai}
\affiliation{Institute of Astronomy and Department of Physics, National Tsing Hua University, Hsinchu 30013, Taiwan}

\begin{abstract}
{
Here we present new ALMA observations of polarized dust emission from six of the most
massive clumps in W43-Main.
The clumps MM2, MM3, MM4, MM6, MM7, and MM8, have been resolved into
two populations of fragmented filaments. From these two populations
we extracted 81 cores (96 with the MM1 cores) with  masses between 0.9 \Msun\ to 425 \Msun\
and a mass sensitivity of 0.08 M$_{\odot}$.
The MM6, MM7, and MM8 clumps show significant fragmentation, but the polarized intensity appears to be  sparse and compact.
The MM2, MM3, and MM4 population shows less fragmentation, but with a single proto-stellar core dominating the emission 
at each clump. Also, the polarized intensity is more extended and significantly stronger in this population.
From the polarized emission, we derived detailed magnetic field patterns
throughout the filaments which we used to estimate field strengths for 4 out of the 6 clumps.
The average field strengths estimations were found between  500 $\mu$G to 1.8 mG.
Additionally, we detected and modeled infalling motions towards MM2 and MM3 from single dish
HCO$^{+}(J=4 \rightarrow 3)$ and HCN$(J=4 \rightarrow 3)$ data 
resulting in mass infall rates of $\dot{\mathrm{M}}_{\mathrm{MM2}} = 1.2 \times 10^{-2}$ \Msun\ yr$^{-1}$  and
$\dot{\mathrm{M}}_{\mathrm{MM3}} = 6.3 \times 10^{-3}$ \Msun\ yr$^{-1}$.
By using our estimations,
we evaluated the  dynamical equilibrium of our cores by computing the total virial parameter $\alpha_{\mathrm{total}}$.
For the cores with reliable field estimations, we found that 71\% of them appear to be
gravitationally bound while the remaining 29\% are not. We
concluded that these unbound cores, also less massive, are still accreting and have not yet reached a critical mass.
This also implies different evolutionary time-scales, which essentially
suggests that star-formation in high mass filaments is not uniform.
}
\end{abstract}

\keywords{ISM: Magnetic Fields, ISM: clouds, ISM: Kinematics and dynamics}


\section{INTRODUCTION}\label{se:INTRO}

The formation of high mass stars is still eluding a comprehensive and detailed theoretical model.
As these high mass stars are born inside giant complexes of molecular gas and dust (GMCs),
which are, mostly, more than 1 kpc away from the Sun, resolving them into detail images have been
an historical problem. This has radically changed by the rise of a new generation of 
powerful millimeter and sub-millimeter facilities, such as ALMA, which have
provided us with unprecedented data about the GMCs and the high mass star formation process. 
Given the large amounts of gas and dust needed to assemble them, the high mass star formation process
is much more dynamic and complex than what we see in the nearby low mass star 
forming regions. 

High mass stars are usually encountered in dense clusters composed up to 
many millions of stars \citep{Weidner2010}. The formation of such clusters requires
large reservoirs of gas, which are likely the result of 
the fragmentation of massive clumps of, weakly ionized and magnetized, gas and dust inside GMCs.
Thus, the formation of high mass
stars cannot be decoupled from the study of associations where the cluster environment
plays a significant role. 

 There are currently two competing theoretical views for the high mass star formation 
process: (1) The core accretion model, which starts from a fragmented  clump of self-gravitating, 
centrally condensed, core of gas and dust which will collapse into one or more stars. The final mass is
determined by the accretion of  material from
their surroundings. This scenario predicts that high mass stars will form disks
around them as accretion proceeds. (2) The competitive accretion model  assumes instead that 
the formation process is more chaotic and controlled by turbulence. In this model there is no
gravitationally bound pre-stellar phase and the final mass is determined by competitive
accretion from the swarm of cores produced by turbulence. If the density is sufficiently
high, star collisions might also contribute to the final mass of the star. For a detailed
review  see \citet{Tan2014}.

 The W43-Main is a large molecular complex located within the W43 region at
5.5 kpc from the sun \citep{Motte2003,Zhang2014a}. The cloud is
near $l=31^{\circ}, b=0^{\circ}$ and at an interface with an extended \HII\ region powered by a cluster
of OB type and Wolf-Rayet stars \citep{Cesaroni1988,Liszt1995,Mooney1995,Blum1999}.
Although the exact number is uncertain, it is believed that the W43 cluster harbors around 50 OB star
with a total luminosity of about $3.5 \times 10^{6}$ $L_{\odot}$. Some of the stars
in the cluster are suggested to be WN7 class W-R stars, where W43-1 has an estimated 
bolometric magnitude  of $M_{bol} \sim 11.5$ mag compared with $M_{bol} \sim 11.7$ mag for the Pistol
star in the Galactic center Quintuplet cluster \citep{Blum1999,Rahman2013}. These stars 
belong to the most massive end of the distribution and therefore, their influence in the
surrounding gas should not be under-estimated.

\begin{figure*}
\includegraphics[width=0.98\hsize]{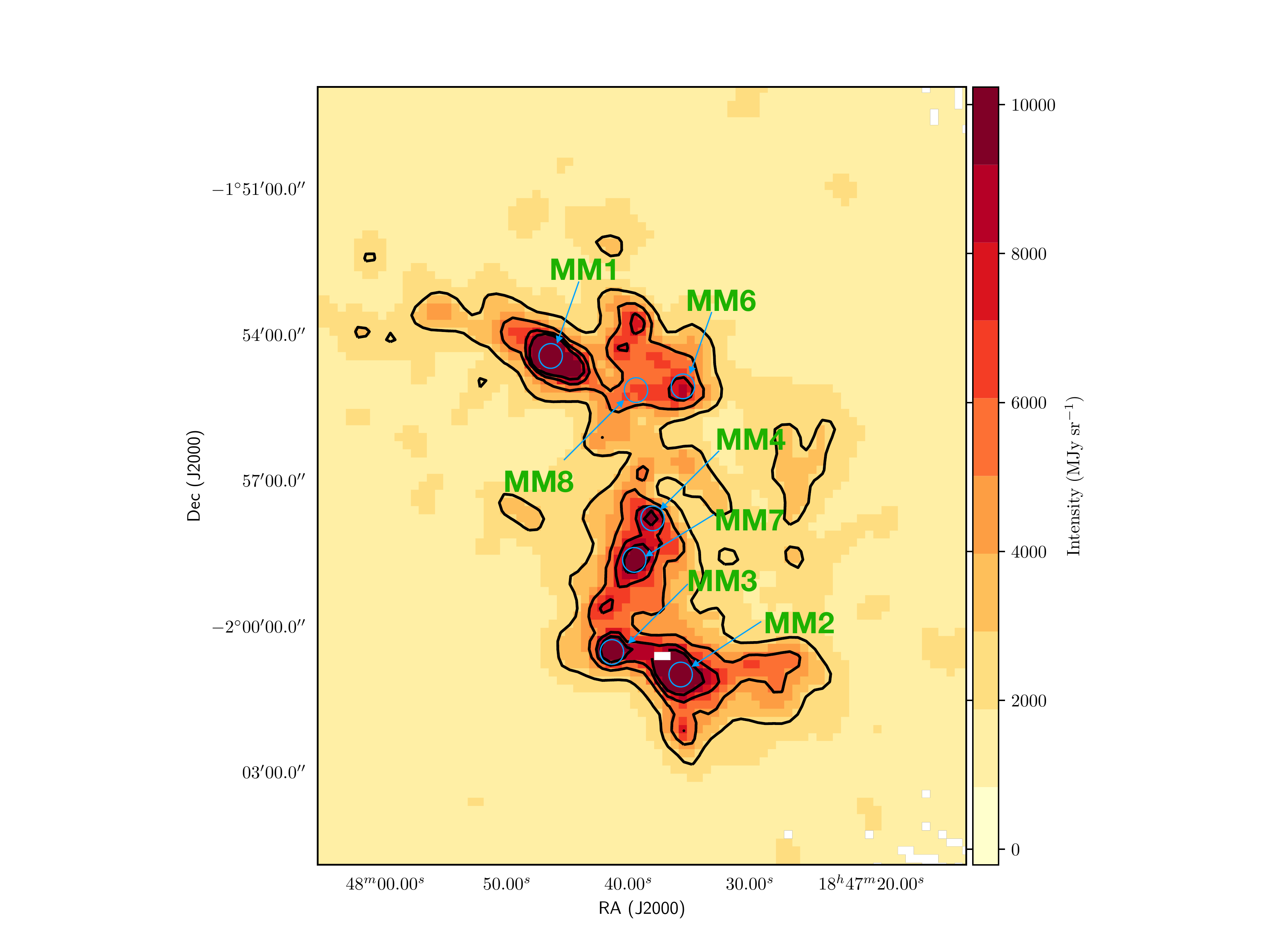}
\caption{The Figure shows an overview of the W43-Main molecular complex with the location of its most massive clumps. Here we show Shark 350 $\mu$m
data from W43-Main in black contours superposed to the color scale \citep[data taken from Figure 1b from ][]{Motte2003}. 
In blue circles are the approximate
locations of the clumps observed with ALMA and their respective names. The circles do not represent size and are used for indicative 
purposes only.
\label{w43main}
}
\end{figure*}

 The W43-Main complex (see Figure \ref{w43main} for an overview) has been well studied in 
continuum from 1.3 mm to 70 $\mu$m \citep{Motte2003,Bally2010} as
well with large surveys of CO \citep{Carlhoff2013}, HCO$^{+}$(3-2) \citep{Motte2003}, and 
HCO$^{+}$(1-0) and H42$\alpha$ \citep{Nguyen-Luong2017}.
These studies identified a large sample of clumps in the W43-Main molecular complex.
The W43-MM1 clump, has already been revealed as the most massive one
with an estimated mass of about 3000 \Msun\ and a deconvolved size of $\sim$ 0.1 pc \citep{Cortes2006a,Sridharan2014,Louvet2014}.
The clump appears to be a fully fragmented and magnetized filament, with large-scale
infalling motions \citep{Cortes2010}, and with  a massive condensation dominating most the flux from the clump 
\citep[core $A$ in][]{Sridharan2014,Cortes2016}. Recently, \citet{Motte2018a} presented a detailed, high resolution
(0.01 pc scales) ALMA data from W43-MM1. From their data, they detected 131 cores with masses 
ranging from 1 to 102 \Msun, where the core $A$ was resolved into
a binary system of massive cores. The derived core mass function (CMF) has a slope
which is consistently shallower than the currently assumed IMF \citep{Bastian2010}. This challenges
the origin of the IMF as it cannot be simply inherited from the shape of the CMF.

 Given the magnetized nature of these structures, mapping the magnetic field is paramount.
Initial studies of the magnetic field, using polarized dust emission with single dish telescopes,
focused on the large scales while later individual cores were mapped with interferometers.
From these studies, there is mounting evidence about the dynamical importance of the magnetic 
field from the ``hour-glass'' morphologies seen at different length-scales \citep{Schleuning1998,Girart2009,Qiu2014}.
For a detailed review see \citet{Hull2019}.
In W43-Main, our group started by mapping the brigthest clump in the millimeter, W43-MM1
\citep{Cortes2006a,Sridharan2014}, where \citet{Cortes2016} (hereafter \citetalias{Cortes2016}) presented the first, 
high spatial resolution, ALMA observations of polarized dust emission. From this study, we obtained magnetic field
strengths on the order milli of Gauss for densities of a few 10$^{7}$ cm$^{-3}$. The field 
morphology appears to be ordered over most of the filament, which is consistent with a strong field. 
From estimations of the strength of the field, we determined that the magnetic field appears
to be weaker than gravity and thus the cores in W43-MM1 would be collapsing, unless additional
sources of energy are considered.
Estimations of the mass-to-magnetic flux ratio suggest 
that the cores are mostly super-critical and the field lines are being dragged by gravity around the
dominant binary system. 

 To proceed with the investigation about star formation in W43-Main, we have selected
the following 6 most massive clumps from the survey done by  \citet{Motte2003}. Thus,
in this paper we report the ALMA observations of polarized dust emission towards W43-MM2, MM3, MM4, MM6, MM7, and MM8
clump done in band 6 at 1.3 mm.  Here,
section \ref{se:OBS} reports the observations and calibration, section \ref{se:CE} the continuum emission results,
source extraction, the magnetic field results, the line emission results and
infall modeling for W43-MM2 and W43-MM3, section \ref{se:discussion} is the discussion, and 
section \ref{se:sum} is the summary and discussion.

\section{OBSERVATIONS}\label{se:OBS}

\subsection{ALMA OBSERVATIONS}

The data presented here corresponds to ALMA observations at 1.3 mm (band 6), project 2013.1.00725.S, done on May 30th, 2015 
over the seven brightest clumps from W43-Main \citepalias[where W43-MM1 results were presented in ][]{Cortes2016}.
As part of a multi-field observation, the phase center coordinates for each clump
were taken from Table 1 in \citet{Motte2003}.
An array of 35 antennas was used reaching an angular resolution of $\sim$ 0.5$^{\prime \prime}$ 
($\sim$ 0.01 pc scales at a distance of 5.5 kpc).  The spectral configuration was set to 
continuum mode (or TDM), with 64 channels per spectral window, giving a spectral resolution of 31.25 MHz 
in full polarization mode ( where $XX, YY, YX$, and $XY$ are the visibilities produced by the ALMA baseline correlator).
Each spectral window was centered at the standard ALMA
band 6 polarization frequencies (224, 226, 240, and 242 GHz).  
At the reference frequency, $\nu_{\mathrm{ref}} = 233$ GHz, the 12 meter antenna
primary beam is $\sim 25^{\prime \prime}$.
Two successful executions were done using the session scheme, which is 
described in \citet{ALMATechnical}.
Calibration and imaging was done using the Common Astronomical Software Applications (CASA) 
\citep[version 4.7 ][]{McMullin2007}.
In total, we obtained about 12 minutes per field (clump), which gave
an average rms for Stokes I of $\left< \sigma_{\mathrm{I}} \right> \sim 200 \mu$Jy beam$^{-1}$
while the average rms for the polarized intensity is $\left< \sigma_{\mathrm{poli}} \right> \sim 54 \mu$Jy beam$^{-1}$. 
This is equivalent
to a mass sensitivity of 0.08 \Msun\ under the assumptions made in section \ref{2se:DUST}. 

\subsection{Calibration and imaging in full polarization mode}\label{se:calib}
 The ALMA antennas are equipped  with linearly polarized receivers.
After reception of the incoming radiation, the wave is
divided into two orthogonal components ($X$ and $Y$) by a wave splitting device \citep{ALMATechnical}.
This operation is not perfect and there is always a residual, or projection, from
one polarization onto the other which is known as the instrumental polarization, or D-terms \citep{Sault1996}.
Given that the antenna frame uses azimuth and elevation coordinates, the frame of the sky rotates
with respect to the antenna introducing an angular dependence which is parameterized by the parallactic angle. 
Additionally to the D-terms, the $X$ and $Y$ polarizations have sligthly different analog signal paths 
inside the receivers before digitization, which 
introduces a relative delay between both polarizations.
Also, the interferometric calibration scheme for amplitude and phase requires the
use of a reference. This reference breaks the degeneracy intrinsic to the array and thus 
we do not measure absolute phase values but relative ones with respect to the reference (where
phases are set to zero in both polarizations). By doing this, we introduce an additional
phase bandpass between the {\em XY} and {\em YX} cross correlations.

To calibrate all these quantities, an ALMA polarization observation samples a strong, un-resolved, 
polarized source over a certain range of parallactic angle. The polarization calibrator is sampled for 5 minutes every 
35 minutes or so.
For our observations, we obtained about 100$^{\circ}$ of parallactic angle coverage for J1924-292
which was selected as polarization calibrator. Using this source, we derived solutions for 
the cross polarization delay, the {\em XY}-phase, and the D-terms.
These solutions were applied to all the clumps in our data, along with  
J1751+0939 to calibrate the bandpass, J1851+0035 to calibrate the phase,
and Titan to calibrate the flux. 
After applying the calibration tables, we imaged the data using the $clean$ CASA task
with the {\em Briggs} weighting scheme, robust number 0.5, for sidelobe robustness and the {\em clarkstokes} deconvolution algorithm
to produce the Stokes images. The final images were produced after three, phase-only, self-calibration iterations
using a final solution interval of 90 seconds.

\subsection{ASTE OBSERVATIONS}

The second and third most massive clumps from W43-Main (W43-MM2 and W43-MM3)
were observed during July 2010 using the Atacama Sub-millimeter Telescope
Experiment (ASTE) of the National
Astronomical Observatory of Japan (NAOJ)
\citep{Kohno2005}.  The telescope is located at 4900 meters of altitude at 
{\em Pampa la bola} in
the Chilean Andes plateau reserve for Astronomical research.
The ASTE telescope is a 10\,m diameter dish equipped with a
wide range of instruments, including a
345\,GHz double side band SIS-mixer receiver. We
simultaneously observed 
HCO$^{+}(J=4\rightarrow3)$ and HCN$(J=4\rightarrow3)$ with
a beam size of $\sim 22^{\prime \prime}$, and a velocity resolution
of 0.1\,\kms\ by setting the XF-type digital spectro-correlator
to a bandwidth of 128 MHz.  The pointing accuracy was
of the order of 2$^{\prime \prime}$ with VY\_CMA used as the pointing
source. The observations were done by performing on-the-fly mapping (OTF) with a
grid spacing of $10^{\prime \prime}$ and operating the telescope remotely
from the ASTE base in San Pedro de Atacama under good weather
conditions  (precipitable water vapor or PWV $<$ 1\,mm). Our reference positions
for MM2 and MM3 were taken from Table 1 in \citet{Motte2003}.
All temperatures are presented as
T$_{\mathrm{mb}}=$T$^{*}_{\mathrm{A}}/\eta_{\mathrm{mb}}$, where
$\eta_{\mathrm{mb}}=0.71 \pm 0.07$. Initial data reduction and
calibration was done using the NEWSTAR package \citep{Ikeda2001}.
The calibrated data were
later exported to be plotted using matplotlib \citep{Hunter2007}.

\section{RESULTS}\label{se:CE}

\subsection{DUST CONTINUUM EMISSION}\label{2se:DUST}

\begin{figure*}
\includegraphics[width=0.95\hsize]{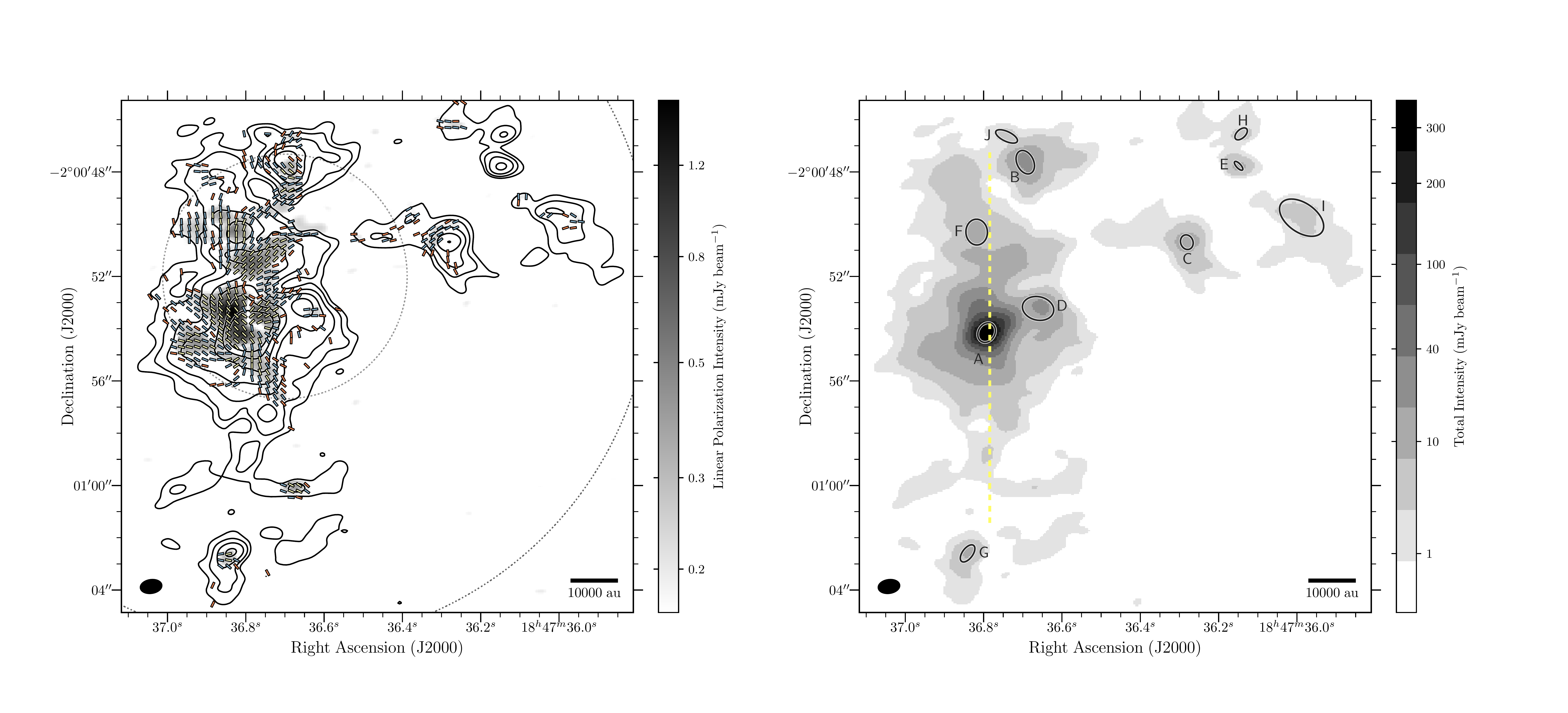}
\caption{
    The Figure shows the Stokes I emission ({\em right})  and the magnetic field map ({\em left}) from the W43-MM2 clump as resolved by ALMA.
    The magnetic field map is composed of a total intensity image (Stokes I) in contours of 4, 8, 16, 32, 64, 128, and 256
$\times \sigma_{I}$, where $\sigma_{I} = 245 \mu$Jy beam$^{-1}$, the polarized intensity image in grey-scale, with
$\sigma_{\mathrm{poli}} = 58 \mu$Jy beam$^{-1}$, and the magnetic field map in pseudo vectors of 2.5 (red), 3 (blue), and
$> 5 \sigma_{\mathrm{poli}}$ (yellow), which was constructed by rotating the polarization position angle (EVPA) by 90$^{\deg}$. 
We are plotting two circles as black dotted lines to indicate the 1/3 FWHM and FWHM primary beam (field of view). The 1/3 FWHM
covers the area where the ALMA polarization accuracy, or minimum detectable polarization, is 0.1\%.
The total intensity Stokes I emission map is shown to the right in a grey-scale of mJy beam$^{-1}$ as indicated
by the colorbar.
    Overlaid are the cores extracted and indicated as ellipses in black. These ellipses represent the deconvolved sized obtained from the
    Gaussian fits. The yellow, segmented, line corresponds to the main axis of the filament. In both maps, the bar at the bottom right corner
gives the spatial scale in au.
}
\label{mm2}
\end{figure*}

\begin{figure*}
\centering
\includegraphics[width=0.95\hsize]{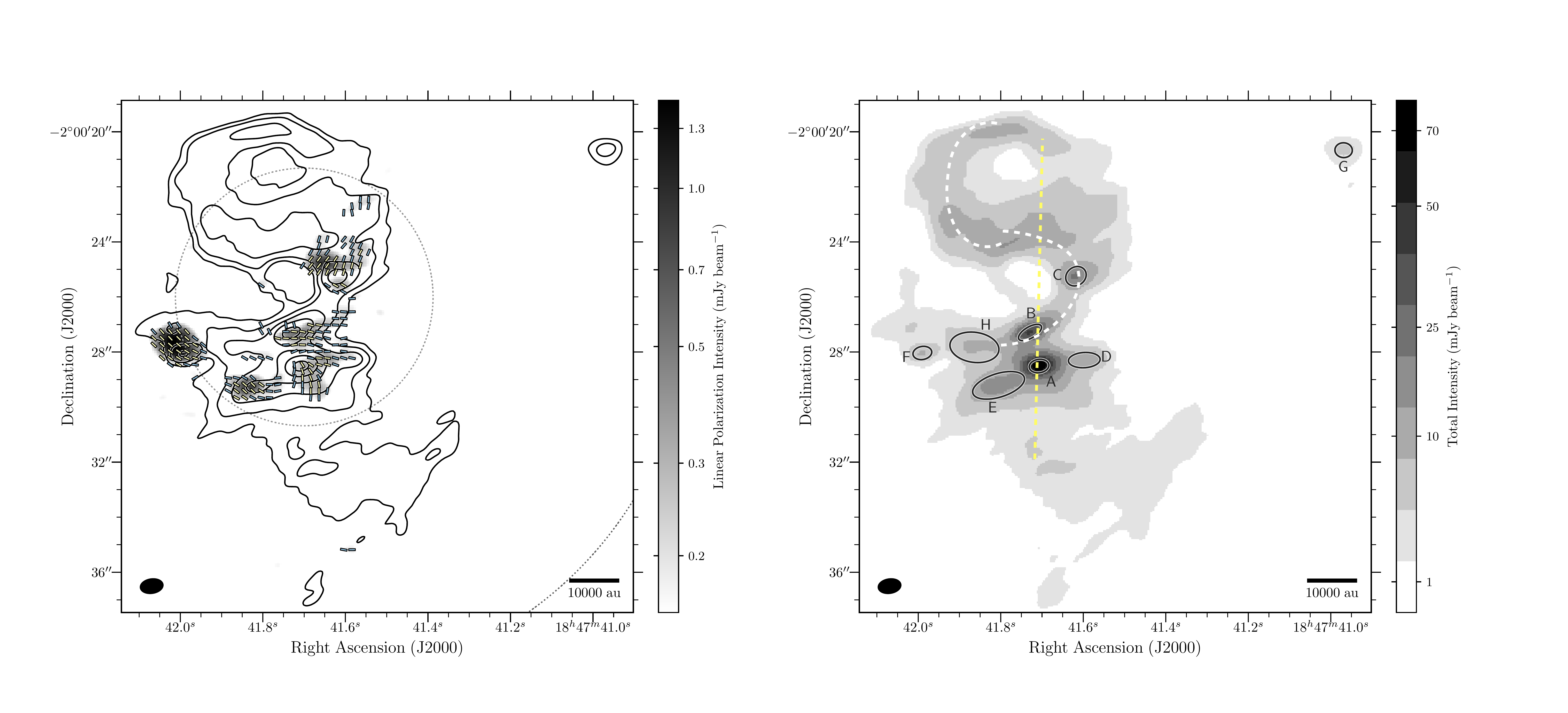}
\caption{
\label{mm3}
The Figure shows the Stokes I and magnetic field maps for the W43-MM3 clump as shown by Figure \ref{mm2}.
To the {\em left}, the Stokes I is represented in contours of 8, 16, 32, 64, 128, and 256 $\times \sigma_{\mathrm{I}}$,
where $\sigma_{\mathrm{I}} = 240 \mu$Jy beam$^{-1}$, the polarized intensity image in grey-scale, with
$\sigma_{\mathrm{poli}} = 52 \mu$Jy beam$^{-1}$, and the magnetic field map in pseudo vectors of 2.5 (red), 3 (blue), and
$> 5 \sigma_{\mathrm{poli}}$ (yellow).
The total intensity Stokes I emission map is shown to the {\em right} in a grey-scale of mJy beam$^{-1}$ as indicated
by the colorbar.  Overlaid are the sources extracted as ellipses, in black, representing the deconvolved sized obtained from the
    Gaussian fits.  The yellow, segmented, line corresponds to the main axis of the filament while the white, segmented, lines
indicate the cavities, or shell, seen in the dust continuum map. 
}
\end{figure*}

 Continuum emission results from the W43-MM2, MM3, MM4, MM6, MM7, and MM8 clumps are presented in this section and shown
in Figures \ref{mm2}, \ref{mm3}, \ref{mm4}, \ref{mm6}, \ref{mm7}, and \ref{mm8}.
The ALMA data uncovered filamentary structure inside all of these clumps. Additionally, we resolved them into
two populations, where the MM6, MM7, and MM8 clumps (population 1) present major fragmentation when compared to 
the MM2, MM3, and MM4 clumps (population 2) . This later population also appears to have a dominant region in the emission, which
we later identified as a massive core. It is also relevant here to note, that there is a fundamental difference in
the strength and distribution of the polarized emission detected between the two populations. In the population 1, the polarized intensity
is weaker and compact while in the population 2 the polarized emission is strong and widely distributed over significant
parts of the clumps.

As in \citetalias{Cortes2016}, we used the {\em getsources} algorithm \citep{Menshchikov2012} to extract core candidates from the continuum emission.
The extracted core position and size was used as a seed for a 2D Gaussian fitting by means of the CASA task {\em imfit} used
over the primary beam corrected maps.
From this, we obtained accurate fluxes, positions, and sizes. However, only cores where the Gaussian fitting converged 
were considered here.
Although using this additional step (the 2D Gaussian fitting)
may overestimate the core flux as the deconvolved Gaussian area may be more extended than the original size obtained from
{\em getsources}. Our methos provides an increased degree of confidence in the likelihood of the extracted cores given
the convergence criteria required to fit the Gaussian. We chose to do this because we did not meet
the expected sensitivity  and the resulting  beam shape was more elongated than expected due the to the low elevation 
reached by then end of the session.
Later, we calculated the core parameters by first computing the column density and the mass as:

\begin{equation}
	N_{\mathrm{H}_{2}} = \frac{S_{\nu}}{B_{\nu}(T) \Omega \kappa_{\nu} \mu m_{\mathrm{H}}} \\
\end{equation}
\begin{equation}
	M = \Omega \mu m_{\mathrm{H}} N_{\mathrm{H}_{2}} d^{2},
\end{equation}

\noindent where $S_{\nu}$ is the total flux obtained from the Gaussian fitting, $B_{\nu}(T)$ is the Planck function
 and $T$ is the dust temperature. The source size is given
by the solid angle $\Omega$ also obtained from the Gaussian fitting, $\kappa_{\nu} = 0.01$ cm$^{2}$ gr$^{-1}$ 
is the emissivity of the dust grains at 230 GHz, which includes a gas to dust ratio of a 100 \citep{Ossenkopf1994}.
The $\mu = 2.33$ is mean molecular weight and $m_{\mathrm{H}}$ is the mass of Hydrogen in grams. Finally,
$d$ corresponds to the distance to W43-Main, estimated to be 5.5 kpc \citep{Zhang2014a}.
The column density calculation assumes that the dust emission is optically thin. Although this
assumption might break down to the most dense cores, 
it is still a reasonable approximation, in average, for most of them. The temperature range used in the mass calculation was taken from
the Spectral Energy Density (SED) fitting by \citet{Motte2003, Bally2010} done over MM2, MM3, and MM4, the most massive clumps. 
For MM6, MM7, and MM8, we assumed T = 25 K which is the mean value between the temperatures determined by the SED fitting
and representative of the ambient molecular gas temperature.

\subsubsection{W43-MM2 CONTINUUM}
\label{3se:mm2-cont}

 The W43-MM2 (hereafter MM2) clump is the second largest clump from the \citet{Motte2003} continuum survey. 
The most up-to-date mass estimation puts MM2 in the 3.5 to 5.3 $\times 10^{3}$ \Msun\,  where the
estimate was done by \citet{Bally2010} assuming an overall dust temperature of 23 K.
In fact, this mass estimate puts MM2 as the most massive clump in W43-Main.
The presence of OH and CH$_{3}$OH  maser emission, but the lack of cm and 24 $\mu$m emission indicates that it is a 
high mass star forming clump in an early evolutionary stage. In addition, no molecular outflow has been detected so far.
Figure \ref{mm2} shows the ALMA 1 mm total intensity continuum map as well as the polarization map
(see section \ref{2se:MF}).
The clump appears as a fragmented filament to length-scales of 0.01 pc where the emission seems dominated by a single source, 
which  seems to be the case for the population 2 or the most massive clumps presented here. 
We labeled this source as core {\em A} to follow the nomenclature from paper I,
with a total recovered flux of 880 mJy and an estimated mass of 426 M$_{\odot}$. This core
has $\sim$ 63\% of the total flux from all the detected cores in MM2. It also appears to be monolithic
with no obvious indication of further fragmentation, though higher resolution observations with ALMA
will indicate if the core has sub-structure. We treat the mass estimation of 426 M$_{\odot}$ as an upper limit
given that it is unlikely that MM2-$A$ is as cold as 23 K. \citet{Sridharan2014} detected a high temperature
hot core in MM1-$A$ which had a similar initial mass estimate. However and given that we do not have 
spectral line data from MM2-$A$ at sufficient resolution, we can only speculate about the core evolutionary stage.
Table \ref{mmTab} present the source extraction parameters for this clump.
The mass estimate done here assumed a $T_{\mathrm{d}} = 23$ K for all the cores
in the clump.

\begin{figure}[!ht]
\includegraphics[width=0.99\hsize]{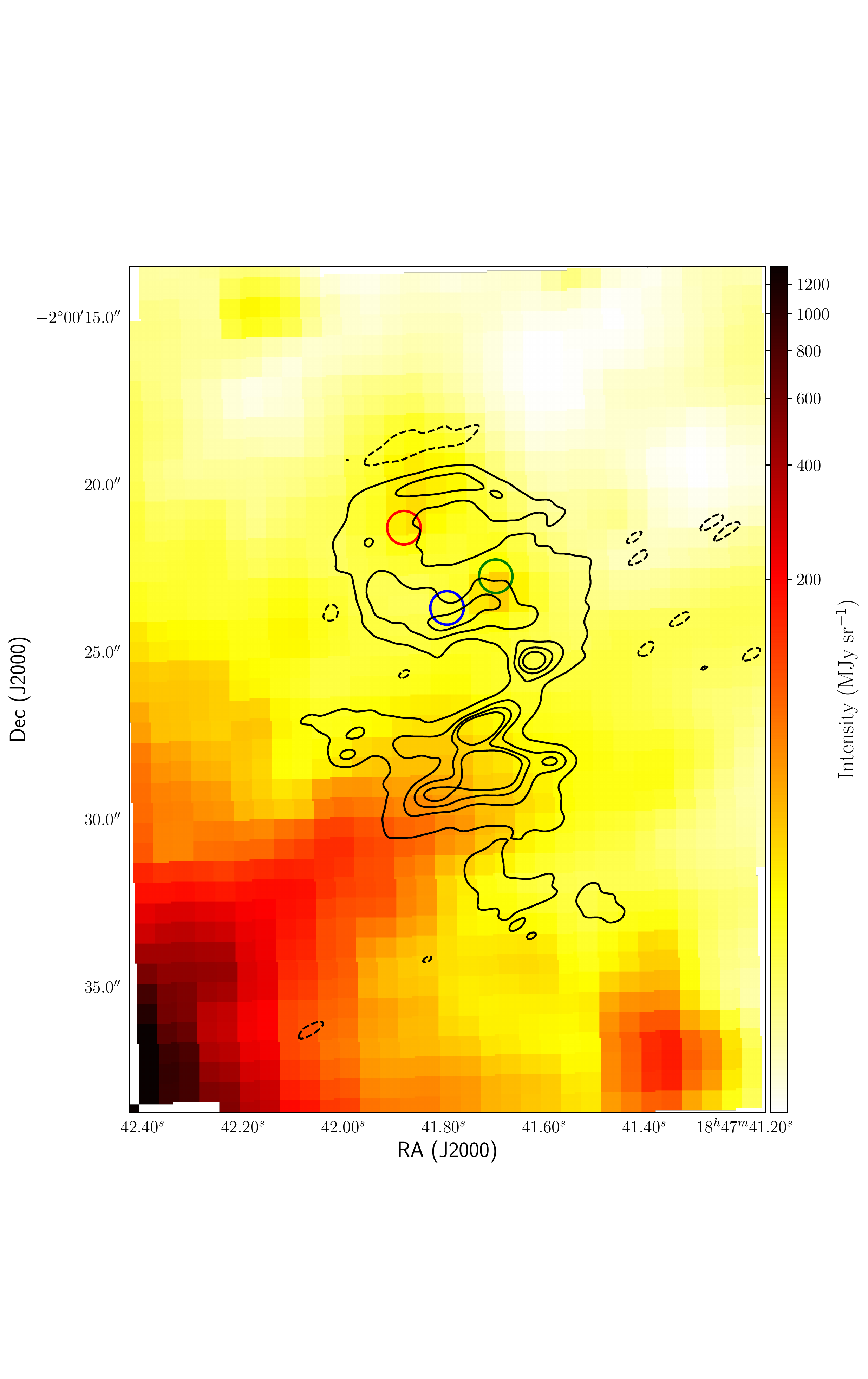}
\caption{
    The Figure shows the dust emission in contours from the MM3 clump. Superposed in color, is an archival
IRAC, 3.6 $\mu$m, SPITZER image from the same region. To the south-east a bright source is shown, which appears to be
a foreground star. To the North, 2 stars, which  seem to coincide with the dust shell seen in MM3, are shown indicated
by  red and green circles. The estimated
surface brightness is 31 $\pm$ 6 MJy sr$^{-1}$ for the red star and 27 $\pm$ 5 MJy sr$^{-1}$ for the green star. 
Using a blue circle, we also indicate the position of G30.720-0.083 from the VLA data.
}
\label{mm3_spitzer}
\end{figure}

\subsubsection{W43-MM3 CONTINUUM} 
\label{3se:mm3-cont}

 W43-MM3 (hereafter MM3) is located in the southern part of W43-Main and to the East of MM2 with an
estimated total mass between 1.5 to 2.3 $\times 10^{3}$ M$_{\odot}$ \citep{Bally2010}. 
The MM3 clump has been identified to contain the G30.720-0.083 radio continuum source
showing emission from 0.6 to 21 cm, with a peak flux of 115 mJy at 6 cm, which is the brightest UC 
\HII\ region in W43 \citep{Bally2010}. 
The MM3 is also seen as an Infrared Dark Cloud (IRDC)
in absorption at 70 $\mu$m and as bright clump at 160 $\mu$m as shown by Herschel data \citep{Bally2010}.
From the ALMA data (see Figure \ref{mm3}), the MM3 clump appears as a fragmented filament, with 8 extracted cores, where
core $A$ is the dominant one with total flux of 150 mJy at 1 mm and an estimated mass of 
59.4 M$_{\odot}$.  The MM3 filament shows a double cavity, or maybe a dust shell, feature to the North of core $A$, 
with an "S" type morphology as seen from North to South (as indicated in Figure \ref{mm3}).  
Furthermore, the cavity to the north is not associated with 
any millimeter detected source.

To ascertain the nature of this shell, we obtained SPITZER IRAC archival images towards MM3.
Figure \ref{mm3_spitzer} shows the MM3 ALMA contours superposed to the IRAC map, where two sources, or stars, 
appear to coincide with the shell. The stars are indicated with red and green circles. Moreover, 
the G30.720-0.083 UC H {\small II} region ($\alpha, \delta =$ 18:47:41.808,-02:00:23.76) 
is also located within the shell and close to one of the IRAC sources \citep{Zoonematkermani1990}. 
The resolution of the VLA data used to detect G30.720-0.083 is about $5^{\prime \prime}$ which is 
about the size of the shell. Despite the uncertainty introduced by the coarse VLA resolution, 
it is plausible that the UC H {\small II} coincides with 
one of the IRAC sources, and thus the UC \HII\ region might explain the origin of the shell.
We should note that the $8\sigma$ to $16\sigma$ contours around the shell are closer toghether than the
more significant contours which may indicate spherical symmetry.
Alternatively, morphologies like this have been observed as the result of
outflow emission. Outflows can carve cavities around the envelope of a star as seen in a number of
regions. \citet{Hull2016} and \citet{Maury2018} presented a clear example in Serpens SMM1 
and B335. In the SMM1 core a high velocity jet of ionized gas and
a highly collimated molecular outflow are carving a cavity around the young, Class 0, proto-stellar source powering the jet.
In B335 the East-West outflow has carved a cavity which wall are clearly visible in their ALMA data.
This scenario might explain the southern most cavity as cores $B$ and $C$ are geometrically close to power the outflows. However, 
we have no data to look for outflow emission from the $B$ and $C$ MM3 cores. Furthermore, the
northern most cavity is difficult to explain in this way, as we do not have an explicit core detection nearby.
For this clump, the core mass distribution goes from 6.6 to 59.4 \Msun\, where we used $T_{\mathrm{d}} = 27$ K uniformly.

\begin{figure*}
\centering
\includegraphics[width=0.95\hsize]{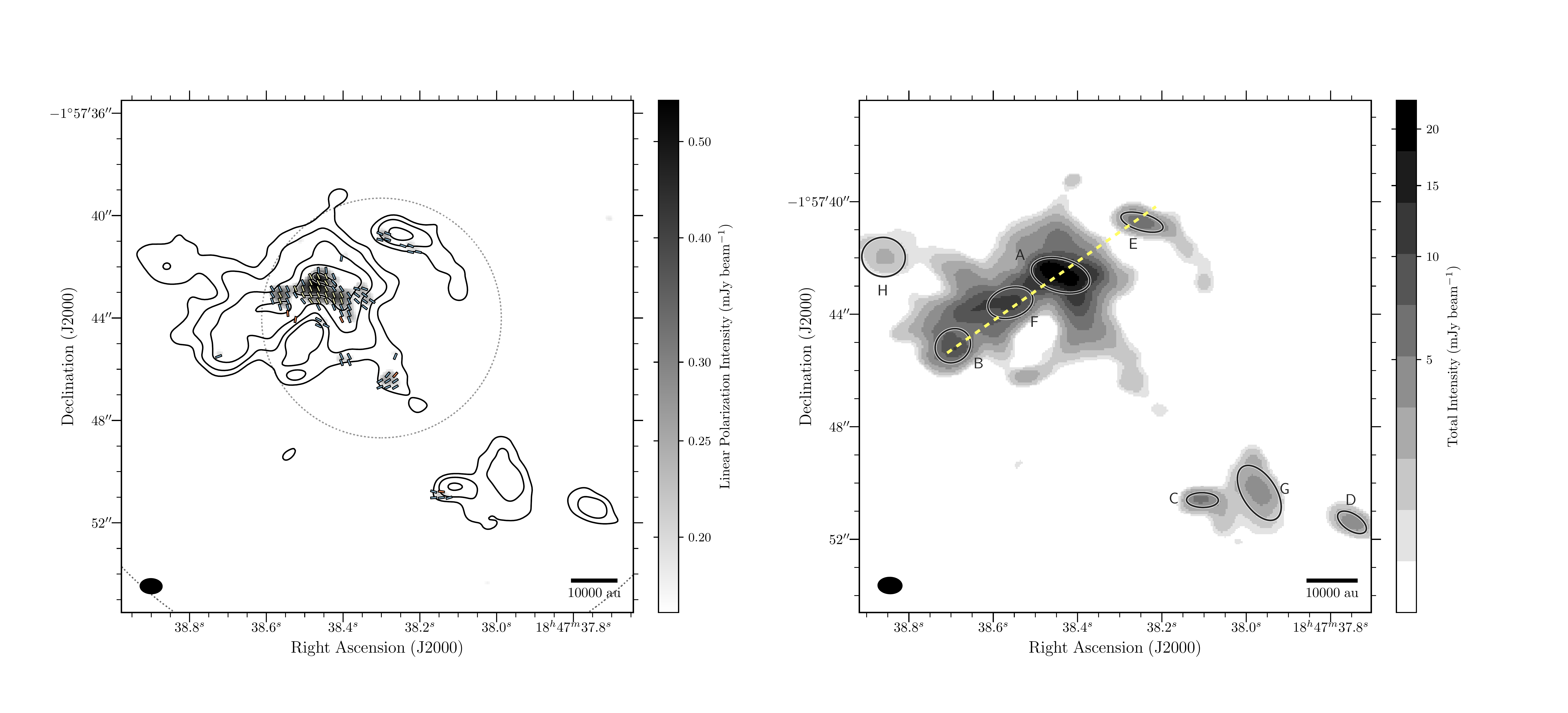}
\caption{
\label{mm4}
The Figure shows the Stokes I and magnetic field maps for the W43-MM4 clump as shown by Figure \ref{mm2}.
To the {\em left}, the Stokes I is represented in contours of 4, 8, 16, 32, 64, and 128 $\times \sigma_{\mathrm{I}}$,
where $\sigma_{\mathrm{I}} = 324 \mu$Jy beam$^{-1}$, the polarized intensity image in grey-scale, with
$\sigma_{\mathrm{poli}} = 56 \mu$Jy beam$^{-1}$, and the magnetic field map in pseudo vectors of 2.5 (red), 3 (blue), and
$> 5 \sigma_{\mathrm{poli}}$ (yellow). The total intensity Stokes I emission map is shown to the {\em right} in a grey-scale of mJy beam$^{-1}$ as indicated
by the colorbar.
    Overlaid are the sources extracted as ellipses, in black, representing the deconvolved sized obtained from the
    Gaussian fits.
  }
\end{figure*}

\subsubsection{W43-MM4 CONTINUUM}
\label{3se:mm4-cont}

 The W43-MM4 (hereafter MM4) is also associated with a radio continuum source.
\citet{Balser2001} suggested the presence of an UC \HII\ region at the center of MM4
while from \citet{Zoonematkermani1990} coincidence was found with strong free-free emission 
($\sim$ 300 mJy), typical from UC \HII\ regions. However, the resolution of both detections
is coarse ($\sim 9^{\prime\prime}$ and $5^{\prime \prime}$), which makes it difficult to
correlate with our higher resolution ALMA data. This clump is also one of the closest,
from our sample, to the W-R/OB cluster powering up W43 with projected distance of about 2 pc.
This distance is close enough to consider the clump inside the shock and the ionization
fronts from the W43 giant \HII\ region\footnote{A quick calculation for the
Stromgren radius for the W43 H$_{\mathrm{\tiny II}}$ region give us $R_{s} = 4.3$ pc assuming the the number of Lyman continuum photons
per star to be $6.0 \times 10^{49}$ (with $\sim 100$ stars, or 10$^{51}$ Lyman continuum photons per second in total) and an electron density of $n_{e} = 10^{3}$ cm$^{-3}$ \citep{Blum1999}}. However, SiO(2-1) maps 
of W43-Main presented by \citet{Nguyen-Luong2013} show insignificant emission towards MM4. Although the authors attribute the
SiO emission to low velocity shock produced by colliding flows, we would have expected an enhancement in the SiO 
emission  and the development of photo-ionized regions (PDRs) from MM4 given it proximity to the W43 \HII\ region.
 
 \citet{Bally2010} estimated an MM4 clump  mass between 0.9 to 1.3 $\times 10^{3}$ \Msun\ assuming an overall 
$T_{\mathrm{d}} = 28$ K. This temperature is consistent with estimations done by \citet{Nguyen-Luong2013} based
on the UV radiation field illuminating this clump.  
Figure \ref{mm4} shows the ALMA
map of MM4, where the emission appears distributed around the main peak with a tail towards the south east. 
Source $A$ dominates the emission with a flux of 160 mJy, though we did not resolve it as seen in Figure \ref{mm4}.
Other 7 sources are also detected mainly around core $A$, the dominant source in MM4 and likely candidate to be a UC \HII\ region.
Three additional cores are detected towards the South-West, $C$, $G$, and $D$, which appears to be connected to the main
filament, though it is interesting to see that no sources were detected within the bridge that joins the tail with the 
main filament. The core mass estimates for MM4 are listed in Table \ref{mmTab}.

\begin{figure*}
\centering
\includegraphics[width=0.95\hsize]{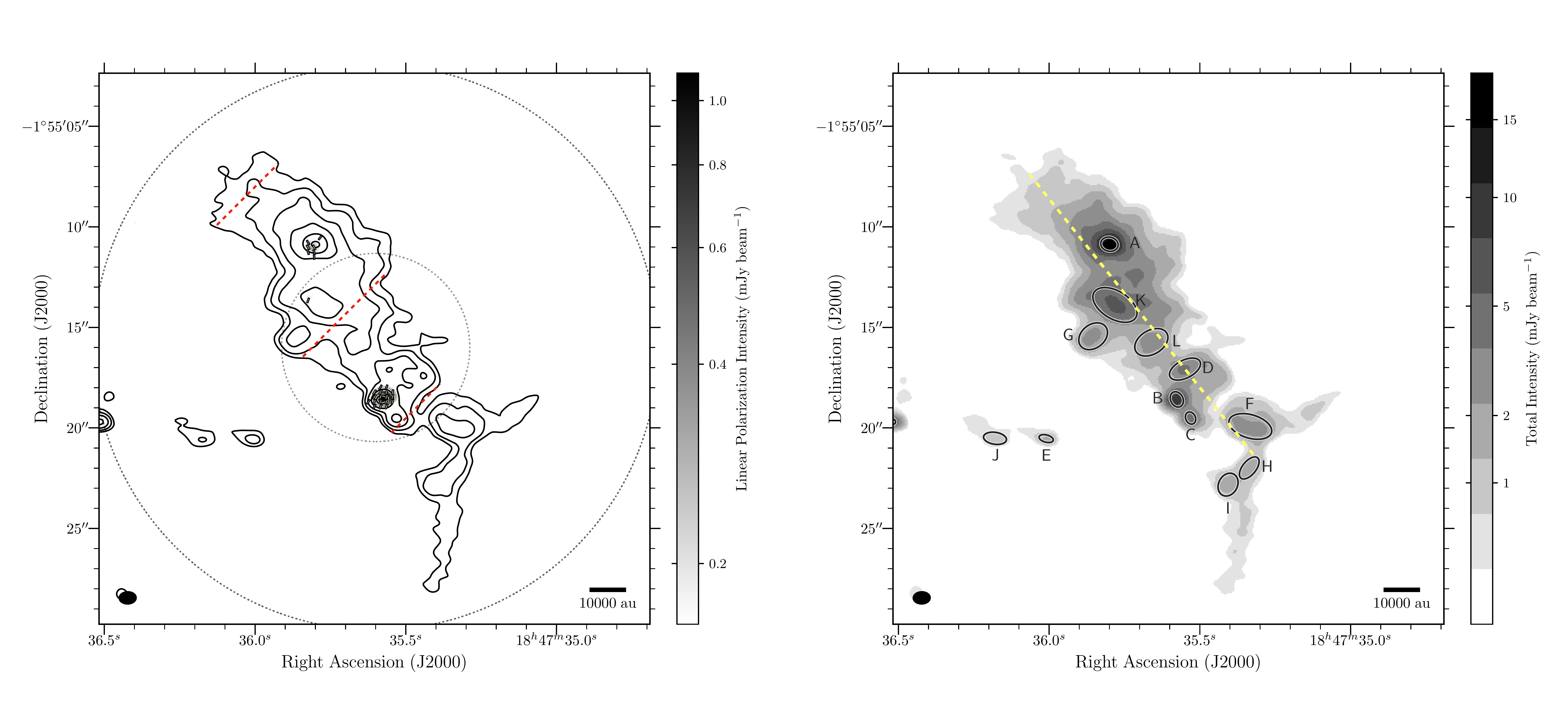}
\caption{
\label{mm6}
The Figure shows the Stokes I and magnetic field maps for the W43-MM6 clump as shown by Figure \ref{mm2}.
To the {\em left}, the Stokes I is represented in contours of 4, 8, 16, 32, 64, and 128 $\times \sigma_{\mathrm{I}}$,
where $\sigma_{\mathrm{I}} = 131$ $\mu$Jy beam$^{-1}$, the polarized intensity image in grey-scale, with
$\sigma_{\mathrm{poli}} = 54 \mu$Jy beam$^{-1}$, and the magnetic field map in pseudo vectors of 2.5 (red), 3 (blue), and
$> 5 \sigma_{\mathrm{poli}}$ (yellow). The segmented red lines shows the cuts used to estimate the filament's width.
The total intensity Stokes I emission map is shown to the {\em right} in a grey-scale of mJy beam$^{-1}$ as indicated
by the colorbar.  Overlaid are the sources extracted as ellipses, in black, representing the deconvolved sized obtained from the
    Gaussian fits. The yellow segmented line indicates the main axis of the filament.
  }
\end{figure*}

\begin{figure*}
\centering
\includegraphics[width=0.95\hsize]{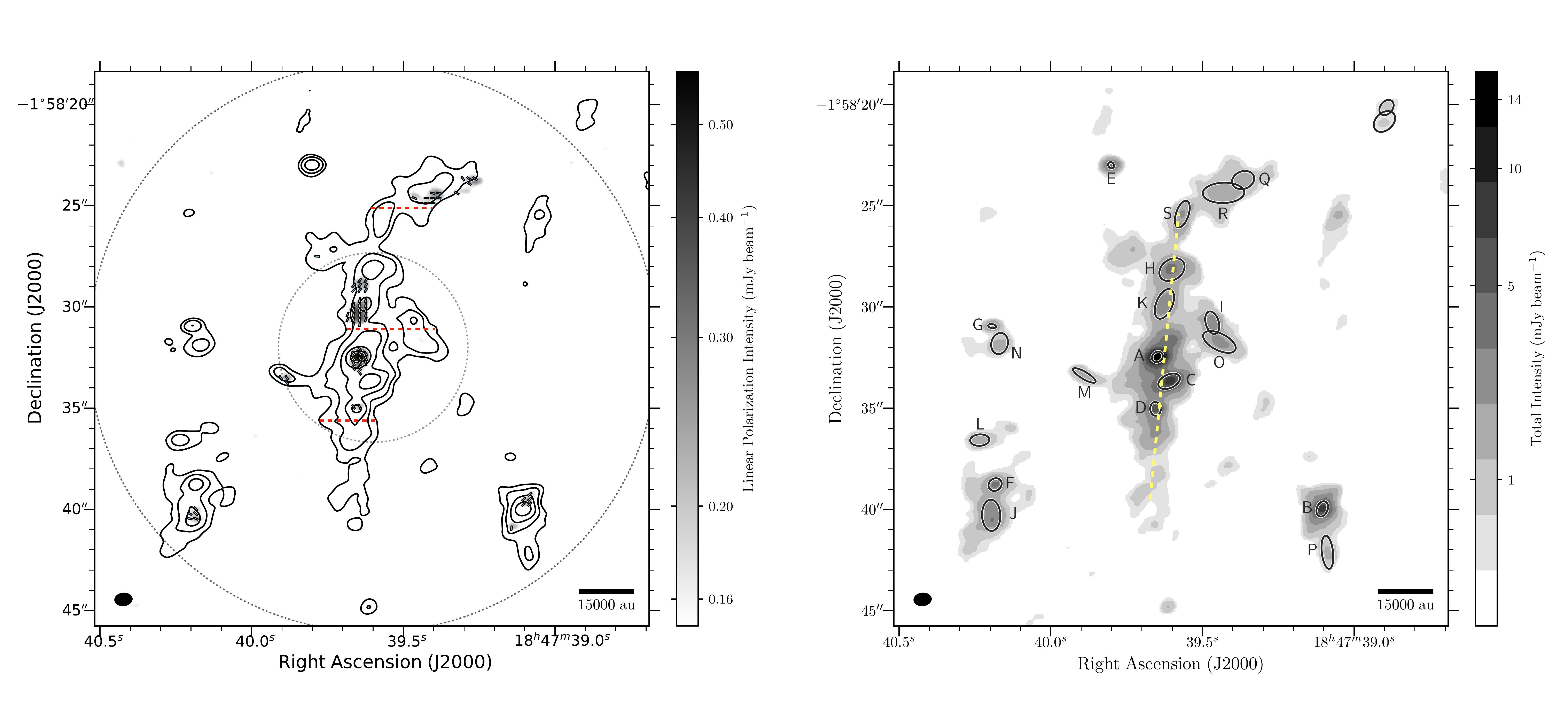}
\caption{
\label{mm7}
The Figure shows the Stokes I and magnetic field maps for the W43\_MM7 clump as shown by Figure \ref{mm2}.
To the {\em left}, the Stokes I is represented in contours of 4, 8, 16, 32, 64, and 128 $\times \sigma_{\mathrm{I}}$,
where $\sigma_{\mathrm{I}} = 150$ $\mu$Jy beam$^{-1}$, the polarized intensity image in grey-scale, with
$\sigma_{\mathrm{poli}} = 50 \mu$Jy beam$^{-1}$, and the magnetic field map in pseudo vectors of 2.5 (red), 3 (blue), and
$> 5 \sigma_{\mathrm{poli}}$ (yellow). The segmented red lines shows the cuts used to estimate the filament's width.
The total intensity Stokes I emission map is shown to the {\em right} in a grey-scale of mJy beam$^{-1}$ as indicated
by the colorbar.  Overlaid are the sources extracted as ellipses, in black, representing the deconvolved sized obtained from the
    Gaussian fits. The yellow segmented line indicates the main axis of the filament.
  }
\end{figure*}

\begin{figure*}
\centering
\includegraphics[width=0.95\hsize]{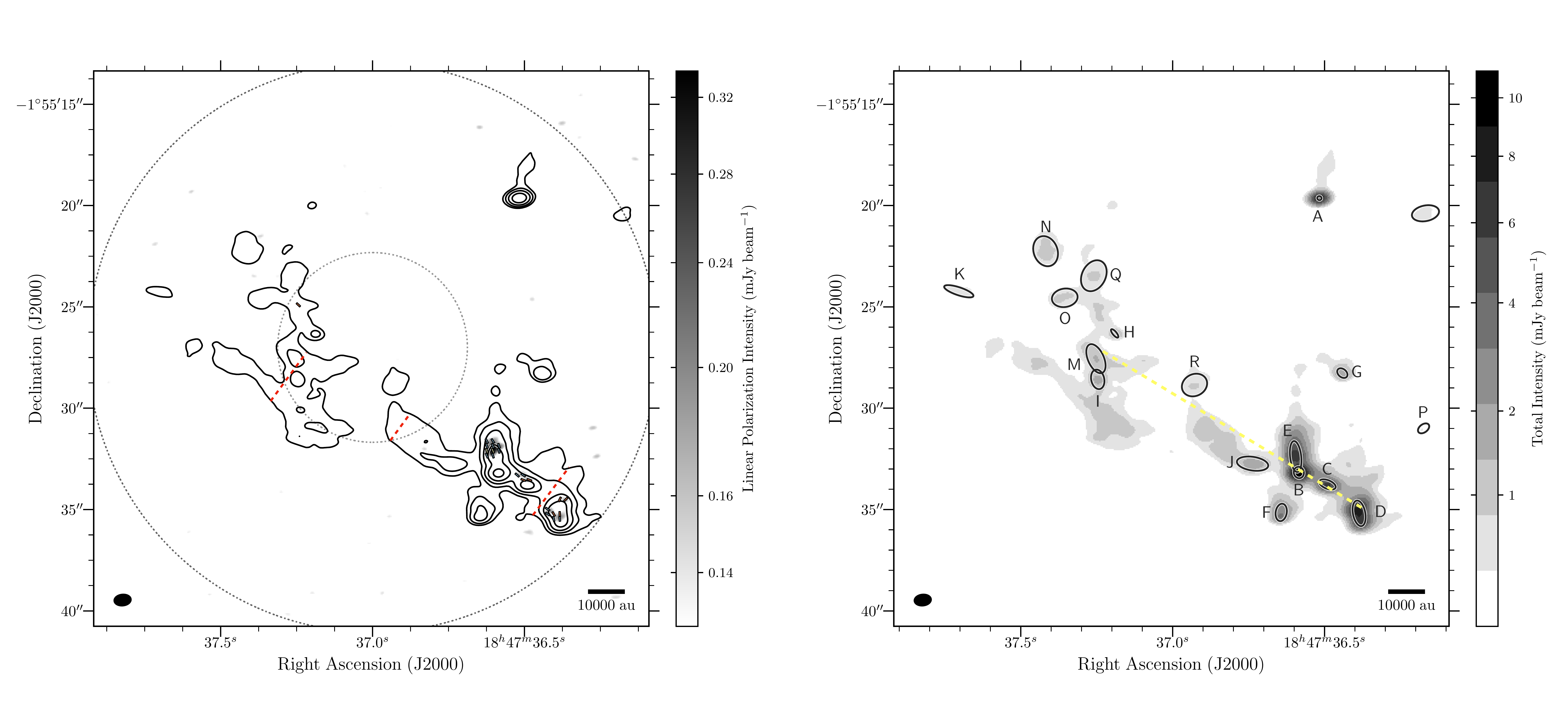}
\caption{
\label{mm8}
The Figure shows the Stokes I and magnetic field maps for the W43-MM8 clump as shown by Figure \ref{mm2}.
To the {\em left}, the Stokes I is represented in contours of 4, 8, 16, 32, 64, and 128 $\times \sigma_{\mathrm{I}}$,
where $\sigma_{\mathrm{I}} = 136 \mu$Jy beam$^{-1}$, the polarized intensity image in grey-scale, with
$\sigma_{\mathrm{poli}} = 51 \mu$Jy beam$^{-1}$, and the magnetic field map in pseudo vectors of 2.5 (red), 3 (blue), and
$> 5 \sigma_{\mathrm{poli}}$ (yellow). The segmented red lines shows the cuts used to estimate the filament's width.
The total intensity Stokes I emission map is shown to the {\em right} in a grey-scale of mJy beam$^{-1}$ as indicated
by the colorbar.  Overlaid are the sources extracted as ellipses, in black, representing the deconvolved sized obtained from the
    Gaussian fits. The yellow segmented line indicates the main axis of the filament.
  }
\end{figure*}

\subsubsection{W43-MM6, MM7, and MM8 CONTINUUM}
\label{3se:mm678-cont}

 The initial mass estimate from \citet{Motte2003} put these clumps
in the $<$ 1000 \Msun\ regime, where the estimates are $\sim$ 500 \Msun\ for MM6, $\sim$ 870 \Msun\ for MM7, and $\sim$ 390 \Msun\ for MM8.
Current data from W43-Main suggests no radio continuum and/or infrared/far-infrared detections sources associated with these clumps. 
Thus, they seem to be in early stage of evolution. As these clumps have not been extensively studied, we lack detailed SEDs for them.
Therefore, we are assuming a dust temperature of T$_{\mathrm{dust}} = 25$ K for all three of them as it seems a representative value
considering their fluxes and the temperatures assumed for the other clumps.
The source parameters for the cores extracted from these clumps are listed in Table \ref{mmTab}.
 
 The MM6 clump shows a clear filamentary structure with 10 cores detected along the major axis
and 2 cores to the East. Figure \ref{mm6}, shows the MM6 ALMA map and the detected cores.
From a total of 12 detected cores, the brightest one, labeled $A$,  has an integrated flux density of 52 mJy and 
and deconvolved size of 0.02 pc. From Table \ref{mmTab}, we see that core $K$ has a higher integrated flux than $A$. However,
this is because the deconvolved size from the Gaussian fit is larger and therefore, its integrated flux. 
Now and by looking at the peak flux, core $A$ is clearly the dominant source in the clump, with 20.5 mJy beam$^{-1}$
compared to 6.5 mJy beam$^{-1}$ from $K$. 
The core estimated masses are within the $< 25$ \Msun\ regime for MM6, and 
 and their mass distribution appears to be uniform even when considering the most massive cores $A$ and $B$. This is a departure from 
 what is seen in the other W43-Main clumps, where the brightest cores in MM1 \citep[see \citetalias{Cortes2016} and ][]{Motte2018a}, MM2, MM3, 
and MM4 have a significant dominance in the core mass distribution of their respective clumps. Given the clear filamentary 
morphology of MM6, we can directly calculate its width\footnote{The filament width was calculated as an average of 3 measurements, 
top, middle, and bottom (shown as segmented red lines on top of the contour maps for each clump), 
which was done following the main axis of the filament. 
The distance was taken from a line placed from lowest contour to the lowest contour from
their respective map while crossing the main axis.},
which in average is closer to 4.3$^{\prime \prime}$ or 0.12 pc assuming 5.5
kpc as the distance to the Sun. Now, the maximum recoverable angular scale for our data  
is about 5.12$^{\prime \prime}$ or 0.14 pc. Thus, the filament width is within the length-scales 
that ALMA is sensitive to and thus, it does not seems to be a cutoff given by the array configuration.
Although such filament widths were initially seen towards low mass star forming regions \citep{Arzoumanian2011},
data obtained from single dish mapping suggests that a similar filament length-scale might be present also in
high mass star forming regions \citep{Andre2016}.

In the MM7 clump, shown by Figure \ref{mm7}, we detected 11 cores along the central filament with 9 additional cores 
to the East and to the West respectively. 
The cores $A$ and $B$ dominate the emission, but with a peak flux distribution similar to what we see in MM6 and
flatter than what is seen in the other most
massive clumps. In total, we extracted 20 cores from MM7 which is about a factor of two more than what we have obtained in
the most massive clump from our sample (MM2).  
This implies a larger fragmentation level, but with a more uniform mass distribution (from 1.7 to 11.8 \Msun\ ).
 The core $A$ estimated mass, of 11.7 \Msun\, is about a factor of six less than the MM3 and MM4 dominant core,
and factor 40 from the MM2 core $A$. However, this mass seems sufficient to form a high mass star especially 
if accretion is still ongoing.
The MM7 clump also shows a clear filamentary structure with an average width of about $3.5^{\prime \prime}$ or 0.09 pc.

 The MM8 clump has the lowest mass in our sample. Here, we detected a total of 18 cores with a peak flux distribution 
similar to MM7 and MM6.
From the detected cores, three dominate the peak flux, cores $A$, $B$, and $D$, which is in line to what is seen in
the MM6 and MM7 clumps (population 1).
In fact, in MM8 we see an increase in the number of low mass cores where some of them have less than 1 \Msun. 
It is worth to remember that 
these sources are all significant detections with fluxes that are a least $10\sigma$ over the noise.
The ALMA map for MM8 with the identified cores is shown in Figure \ref{mm8}.
The MM8 clump also shows a filamentary structure, but in this case the peak of the clump emission was offset
from the phase center by $\sim 20^{\prime \prime}$.
The filament width is, in average, about $2.4^{\prime \prime}$ or 0.06 pc, which is similar to MM7.

\vspace{0.5cm}

 In general, we find that all the clumps in the W43-Main molecular complex show a filamentary structure. 
Here, we define a filament as emission
coming from the clump that shows a geometrically elongated  morphology with a preferred direction. 
Additionally, we found that three of these clumps (MM2, MM3, and MM4) have one core which dominates the flux distributions and therefore,
the mass. The other three clumps (MM6, MM7, and MM8) have a flatter mass distribution with 2 to 3 cores dominating the emission.
We also find, in most the clumps, single and isolated cores which do not seem to be directly associated to the main filament
(such cores $G$ in MM3 and $A$ in MM8).
Although fragmentation in these filaments is evident along the major axis, additional fragmentation is seen, some times in the
orthogonal axis projection (such as MM4), as well as in  filaments parallel to the main one (such as MM7). In these cases, we
see more than one core being detected. As we will see later, this difference is also correlated with a difference in the
amount of polarized emission detected from these filaments. The filaments dominated by a massive core are highly polarized, but the
filaments with a flatter core mass distribution are virtually un-polarized.

\begin{figure*}
\includegraphics[width=0.9\hsize]{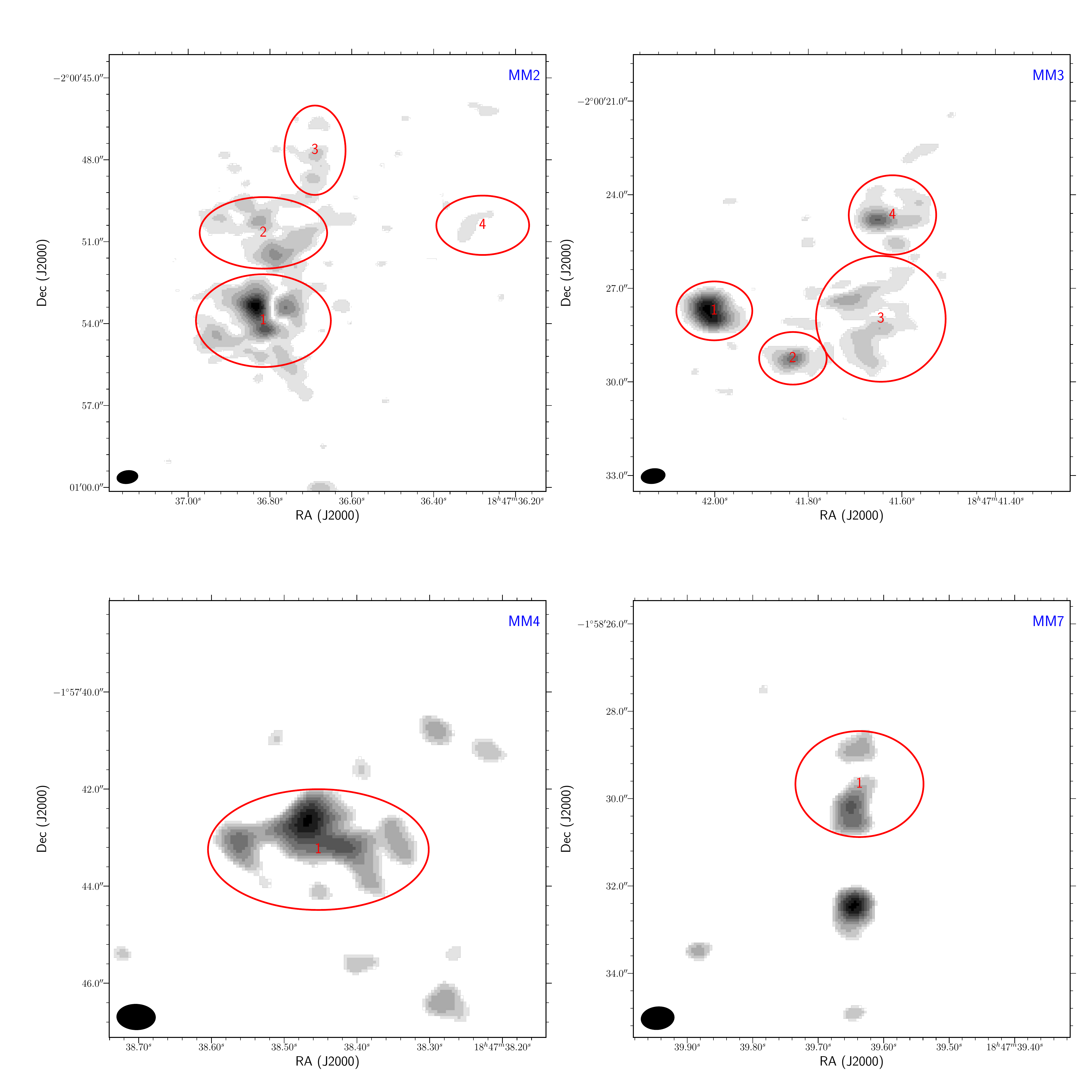}
\caption{
    The Figure shows a panel of polarized intensity maps for the MM2, MM3, MM4, and MM7 clumps.
The polarized intensity emission is shown at the 3$\sigma$ level for each clump.
Superposed are the numbered regions, as red ellipses, indicating the areas  defined for the B$_{\mathrm{pos}}$
estimation in each clump. The clump name is indicated in blue by top-right label.
\label{mmRegions}
}
\end{figure*}

\subsection{THE MAGNETIC FIELD STRENGTH FROM DUST CONTINUUM}\label{2se:MF}

We derived the magnetic field morphology onto the plane of the sky from the polarized emission
assuming grain alignment respect to the magnetic field for all clumps in our sample.
We also estimated the magnetic field strength onto the plane of the sky, or B$_{\mathrm{pos}}$, {by using
the Davis-Chandrasekhar-Fermi (DCF) method \citep{Davis1951,Chandrasekhar1953} and its variants (see
appendix \ref{ape1})}, 
as previously done in \citetalias{Cortes2016}, over selected regions for
each clump (see Figure \ref{mmRegions}). To do this, we used a number of assumptions, most 
notably that the velocity dispersion can be extrapolated from large scales by assuming
a Kolmogorov like power spectrum and the dispersion in the polarization position angle
(also called the electric vector position angle, or EVPA) is due
to perturbations from the gas non-thermal motions.  The details are discussed in appendix \ref{ape1}.

\subsubsection{THE W43-MM2 FIELD}

 The magnetic field morphology onto the plane of sky for the  MM2 clump is shown in Figure \ref{mm2}. 
The polarized emission covers most of the Stokes I map at the $3\sigma$ level  from North to South. 
The inferred magnetic field pattern is quite complex, showing areas on the map where the field appears to be coherent over 
significant portions of the filament. Over these areas, the  EVPA
 dispersion is small which is suggestive of strong fields. However, we have chosen larger regions to estimate the field strength in order to obtain a sufficient number of independent points  and also to derive a  representative value for the ambient magnetic field around the cores in a given region. 
 
To West of the main filament and over the MM2-$C$ core, the field morphology suggests continuity with the field over the main filament,  
despite the gap seen in the total intensity.
It is important to state that most of the polarized emission used to infer the field morphology
and estimate its strength in this clump, is within the 1/3 of the ALMA, 12 meter antenna, primary beam (inner dotted
circle in Figure \ref{mm2}). It is within this region, that ALMA meets the specification of a minimum detectable
amount of linear polarization of 0.1\%. Outside this region, the specification is not guaranteed, but in band 
6 the performance degradation is no worse than 0.6\% at the 50\% level of the FWHM with an error in the EVPA no larger than $2^{\circ}$. Thus, we are confident that the polarization morphology over MM2-$C$ is representative of the magnetic field morphology.

 To estimate the field strength throughout the MM2 clump, 
we have defined four regions based on the projected field apparent connectivity (see Figure \ref{mmRegions}).
Region 1 was defined over sources $A$ and $D$ where the field shows a well connected radial pattern. 
Towards core $A$ in region 1, the field pattern is mostly radial quite similar to the morphology 
seen towards MM1-A \citepalias{Cortes2016}.
In that case, the interpretation was that gravity dominates and the field is being dragged towards the main core.
We will explore this possibility in the discussion section (see section \ref{se:discussion}).
To the North and over core $F$ we defined Region 2, where the field is also continuous, well connected, and clearly 
distinct from the field morphology seen in region 1.
Region 3 is defined over cores $B$ and $J$, where the field pattern seem isolated and with a different 
morphology respect to the field pattern seen in region 2.
Region 4 is defined to the West over core $C$ where the field is isolated from the main filament.
However and at the $2.5\sigma$ level, there seems to be continuity between the field in this region and 
the filament main field.  Figure \ref{mmRegions} shows the regions on top of the polarized intensity map.
In all the four regions, the angular scale for the $3\sigma$ polarized emission is about $5^{\prime \prime}$. 
We used this angular scale for the calculation of $\delta \phi$ and $\sigma_{\mathrm{v}}$.
Table \ref{fieldValues} shows the estimated B$_{\mathrm{pos}}$ for all the cores belonging to the chosen region.

 We obtained field estimations from 0.1 to 3.2 mG throughout the clump (see Table \ref{fieldValues}). This range of field
estimations is smaller to the values obtained towards W43-MM1 where the spread and morphology
the polarized emission is similar.
In MM1 the emission covered the entire filament, or $\sim 13^{\prime \prime}$, which is about the extension of MM2
($\sim 14^{\prime \prime}$) if we consider all the polarized emission. 
The differences in the field estimations here compared to \citetalias{Cortes2016} come from the values of $\Delta V$
where the Kolmorogov extrapolation give smaller line-widths than the 3 \kms\ used there. Also,
the regions used to calculate $\delta \phi$ in \citetalias{Cortes2016} traced field patterns with smaller dispersion values
in the EVPA. In MM2, the situation is more complex as the field is clearly connected throughout the filament 
which complicates the definition of these regions. It is important to point out that the DCF method assumes that the EVPA dispersion is the product of perturbations in the field lines by the gas non-thermal motions
(a more detailed discussion can  be found in \ref{ape1}). However and in region 1, over the MM2-$A$ core, the EVPA
dispersion seems to be produced by infalling gas perturbations of the field lines rather than by turbulence (see section \ref{2se:lineEm}
for our gas infalling results). 
One way to see this, is by noting that the Nort-East and North-West pseudo vectors, towards the center of MM2-$A$, seem to be almost orthogonal to each other. It is unlikely that turbulence will induce such level of change in the field lines and therefore it has come from gravity.
Although in MM1-$A$ we have a similar situation, the geometrical distribution of the polarized emission seemed
to favor a bimodal distribution between two distinct sub-regions, but where one of them is more extended. 
As both regions have small EVPA dispersion, one dominated which decreased the overall value of $\delta \phi$
\citepalias[see Figure 2 in ][]{Cortes2016}. This resulted in an increase in the estimated field strength when compared to MM2-$A$, though both regions have similar field morphologies.  
Therefore, this bias in the estimation of $\delta \phi$, likely, yielded magnetic field strengths estimations which
are lower limits from the ``true'' magnetic field strength around the MM2-$A$ and $D$ cores. 
The same might be said from the estimation done towards MM1-$A$ in \citetalias{Cortes2016}.

Although the field lines around the
main core (region 1) have a clear radial pattern, the field to the North is more difficult to interpret.
Here and to the East, the pattern appears remarkably smooth and uniform with a clear direction along the major axis
of the filament for the field over MM2$-F$, but turning to $\sim 45^{\circ}$ between MM2-$B$ and MM2-$D$. 
This uniformity strongly suggests that magnetic tension along the filament is dynamically significant.
The width of that emission is about half the filament width, or $3.2^{\prime \prime}$, which is
about 0.08 pc. 
However and to the West, the field lines deviate $\sim 40^{\circ}$ and are even orthogonal to the
main axis in some parts. Continuity in the field lines is present as the field, smoothly, change direction.
This change is significant and it happens across an important section of the filament. In Figure \ref{mm2} we see
that the width of the $5\sigma$ pseudo-vectors that change direction in region 2 is about $1.4^{\prime \prime}$,
or 0.04 pc. Thus, the field appears to have a coherent and smooth direction over many core length-scales, which
suggests that the field is not affected by the gravitational pull of these cores. 
In region 3, the field lines approximate a radial pattern over core $B$. However, the significance of the 
emission is not as good as over core $A$ and thus this interpretation is inconclusive. To the East of core $B$
we have almost negligible polarized emission at a Stokes I level where we do detect polarized emission elsewhere
(region 4). At the eastern edge of region 3, the field lines appear to follow the main axis, but they
smoothly deviate below core $B$ to be orthogonal to the main axis of the filament. It is not clear how
the field connects from the eastern edge of region 3 with the field in region 2 given the deviation seen.
In region 4 the polarized emission is not as significant and the required number of points to estimate the field
was met by considering the $2.5\sigma$ pseudo-vectors. Nevertheless, the  field morphology seem consistent
with the more significant points. The inferred field lines in this region appear to connect with the 
field in the main filament. It is clear that the field evolves from being orthogonal to the main axis of the 
filament to an smooth alignment with the main axis.

\subsubsection{THE W43-MM3 FIELD}

 The magnetic field morphology over the MM3 clump, shown in  Figure \ref{mm3}, is distributed 
mostly throughout  the center of the filament. 
It shows four distinct regions that include all the detected cores in the clump, and where the field appears to be well ordered.
Although the polarized emission appears to be more compacted than in MM2, there is indication of continuity, 
from region to region, in the derived field morphology. 
In all of these four regions, we have a sufficient number of independent points to estimate B$_{\mathrm{pos}}$.
These distinct regions are indicated by the cores associated to the field pattern which can be 
identified by looking at Figure \ref{mm3} and at Table \ref{fieldValues}. 
The estimates for B$_{\mathrm{pos}}$ and the parameters
from the polarization map are also listed in Table \ref{fieldValues}.
Most of the polarized emission is located around cores $A$, $B$, and $D$ (region 3) at the center of the filament, 
and over core $F$ where the polarized emission is the brightest in the clump.
Over MM3-$A$, the field morphology appears similar to
what we have seen in MM2 and MM1, where the field appears to be dragged by gravity. Following the discussion for the MM2 field strength estimation over region 1,  the EVPA dispersion seen over
region 3 in MM3 appears to also be result of  perturbations by infalling gas over the 
field lines rather than turbulence. Therefore, we also consider the estimations here to be 
lower limits with estimated values for B$_{\mathrm{pos}}$ to be between 0.1 and 1.4 mG.

 The polarized intensity map from the MM3 clumps (Figure \ref{mm3}), shows strong and compact emission 
(up to 30\% of Stokes I) around and over core $F$. The inferred magnetic field morphology is uniform and ordered with 
a small dispersion in the field lines orientation (i.e. equivalent to the EVPA dispersion, or $\delta \phi = 10.3^{\circ}$).
A similar situation is seen over core $C$ and to a lesser degree over $E$, 
where the polarized emission is compact but not as strong as
what is seen over core $F$. However, we do not detect polarization emission over the shell associated with
the UC \HII\ region. Given the size and flux obtained from the shell, we would have expected to 
detect polarization at similar levels respect to the other regions in the clumps (see the regions defined in
Figure \ref{mmRegions}). We will explore possible explanations to this in section \ref{sse:poli}.

We obtained significant values for B$_{\mathrm{pos}}$ in the order of  milli-Gauss from all
3 estimation methods for all the regions in MM3. The range of the fractional polarization is also significant.
We obtained from 0.5\% to 30.7\% where most of the emission is on-axis with the exception of region 1, which is
located at the edge of the 1/3 of the primary beam (see inner dotted circle in Figure \ref{mm3}) and 
slightly off-axis. The 0.5\% corresponds to region 3 and over core $A$, which 
is expected given the increase in the Stokes I emission due to the larger column density. 
This tendency of lower polarization fraction with increasing column density has been commonly observed \citep{Fissel2016}.  
In contrast, the highest amount of fractional polarization comes from compact 
and bright spots of polarized intensity. It is not clear why we see such a compact clusters of polarized emission 
(see also MM6 and MM7 results), but it might corresponds to places where the dust its well illuminated by the
radiation field produced by the proto-stars, or some other source, which might significantly 
 increase the grain alignment. 

\subsubsection{THE W43-MM4 FIELD}

 In the case of the MM4, the polarized intensity is confined to center of the clump and around core $A$.
Polarized emission is also detected to the North-West over core $E$ and to the South-West, but there we did not detected
cores with sufficient significance. All the detections are on-axis as indicated by the inner dotted circle in Figure \ref{mm4}.
The inferred magnetic field morphology is remarkably uniform suggesting, almost, parallel field lines over cores $A$ and $F$.
The extension of the detected field is about $4^{\prime \prime}$, or 0.11 pc. Although it is difficult to conclude whether MM4 is 
a filament given its compact morphology, it is possible to define a major axis from South-East to North-West
(see Figure \ref{mm4}), which seems to be orthogonal to the 
main field direction. Over core $E$ and to South-West tail from core $A$ (a tail which seems orthogonal to the defined
main axis), the field lines are also parallel but the number of independent points is too small to reliably
estimate the field. Nonetheless over core $E$, the EVPA uniformity is significant with a small dispersion estimate
($\delta \phi \sim 5^{\circ}$), which suggests a strong field.
Towards the cores $C$, $G$, and $D$, we did not detected significant polarization. Only a compact amount
of polarized intensity, at the level of 10\%, is detected at the edge of core $C$.
We estimated B$_{\mathrm{pos}}$ towards the center of the MM4 clump where we obtained
strengths on the order of milli Gauss ($< 10$ mG, see Table \ref{fieldValues}). 

 The position of the associated UC H {\footnotesize II} region in W43-MM4 is not obvious from our map.
As a difference with MM3, we did not find IRAC counterparts, but nevertheless this suggests that the 
MM4 clump is in a more evolved state of evolution than some of the other clumps (e.g. MM2). 
Given that the sizes of UC \HII\ regions are expected to be small and $< 0.1$ pc \citep{Tan2014},
the UC \HII\ region is, likely, unresolved in our data. 
Although,  an \HII\ region  will expand more easily along the field lines, it is 
difficult to ascertain if the UC \HII\ region has a significant impact in the magnetic field, as the field coherent scale is larger
than the expected UC \HII\ region size. Nevertheless, here we found striking differences in the polarized emission 
between two clumps where UC \HII\ regions have been detected.  In MM3, we found no polarized emission at all around 
the area where the UC \HII\ is detected, but in MM4 we found significant amount of polarized emission.
A possiblity is that in MM4 we are still seeing an unperturbed envelope of gas and dust while in MM3 a cavity
has already been formed by the UC \HII\ region. 
Accretion from infalling motions is difficult to establish in MM4.
Although the line profile from the HCO$^{+}(3 \rightarrow 2)$ suggest self-absorption, we do not have confirmation from 
H$^{13}$CO$^{+}(3 \rightarrow 2)$ as the line appears to be optically thick \citep[see Figure 4 from ][]{Motte2003}.

When comparing to other objects, the MM4 clump seems similar to 
W3-Main IRS 5 clumps which has been resolved into, at least, five cores, though the column densities are smaller \citep[$\sim 10^{-23}$ cm$^{-2}$ from][]{Rivera-Ingraham2013}. 
\citet{Hull2014} imaged W3-Main in 1 mm polarization with CARMA obtaining a sufficient
magnetic field map. There map shows a field morphology over IRS 5 which resembles an hourglass shape, which is different from 
the MM4 field morphology.
However and to the West of IRS 5, the field morphology seem relatively smooth with parallel field lines to the South.
\citet{Hull2014} associated this emission to the free-free emission from the W3-B \HII\ region, which is also 
associated with an infrared source (IRS 3), resolved into a  type O6 star (IRS 3a). As free-free emission is very weakly polarized ($< 1\%$),
the polarization pattern seen in both W3-Main and MM4 cannot be explained in this way. Thus, grain alignment seems the 
only possibility to explain the polarized emission which might be coming from the envelope around the UC \HII\ region. 
The line of magnetic force tend to remain straight when densities irregularities develop \citep{Spitzer1978}, which
might be interpreted as 
field lines becoming more ordered through compression \citep[as seen in the low-mass star forming core B335, ][]{Maury2018}.

\subsubsection{THE W43-MM6, MM7, AND MM8 FIELDS}

 The MM6, MM7, and MM8 clumps show significantly less amount of polarized emission than the previous clumps. 
Although the MM6 and MM8 clumps are in the lower range of masses in our sample, 
the dust continuum emission from some of their cores is 
comparable to cores in the other clumps where we did detected polarized emission. 
However, the amount of polarized emission seen in their maps 
appears negligible (see Figures \ref{mm6}, \ref{mm7}, \ref{mm8}, and \ref{sipol}). 
Thus, we did not attempt to estimate B$_{\mathrm{pos}}$ for these
sources. What seems to be common in all these clumps (including the whole sample), is the presence of strong and compact
sources of polarized intensity. This is clearly seen in MM6 core $B$, where the polarized emission is confined, just,
to that core. We will explore this further in the discussion (see section \ref{sse:poli}).
 On the other hand, the MM7 clump has a sufficient number of points to estimate B$_{\mathrm{pos}}$ over core $K$
(if we consider the $2.5\sigma$ values, see Figure \ref{mm7}). Here, we obtained magnetic field strengths around 10 mG
considering the average from all three methods. The maximum fractional polarization is also high and about 18\% which suggests
an increase in grain alignment efficiency.

\begin{figure}
\centering
\includegraphics[width=0.9\hsize]{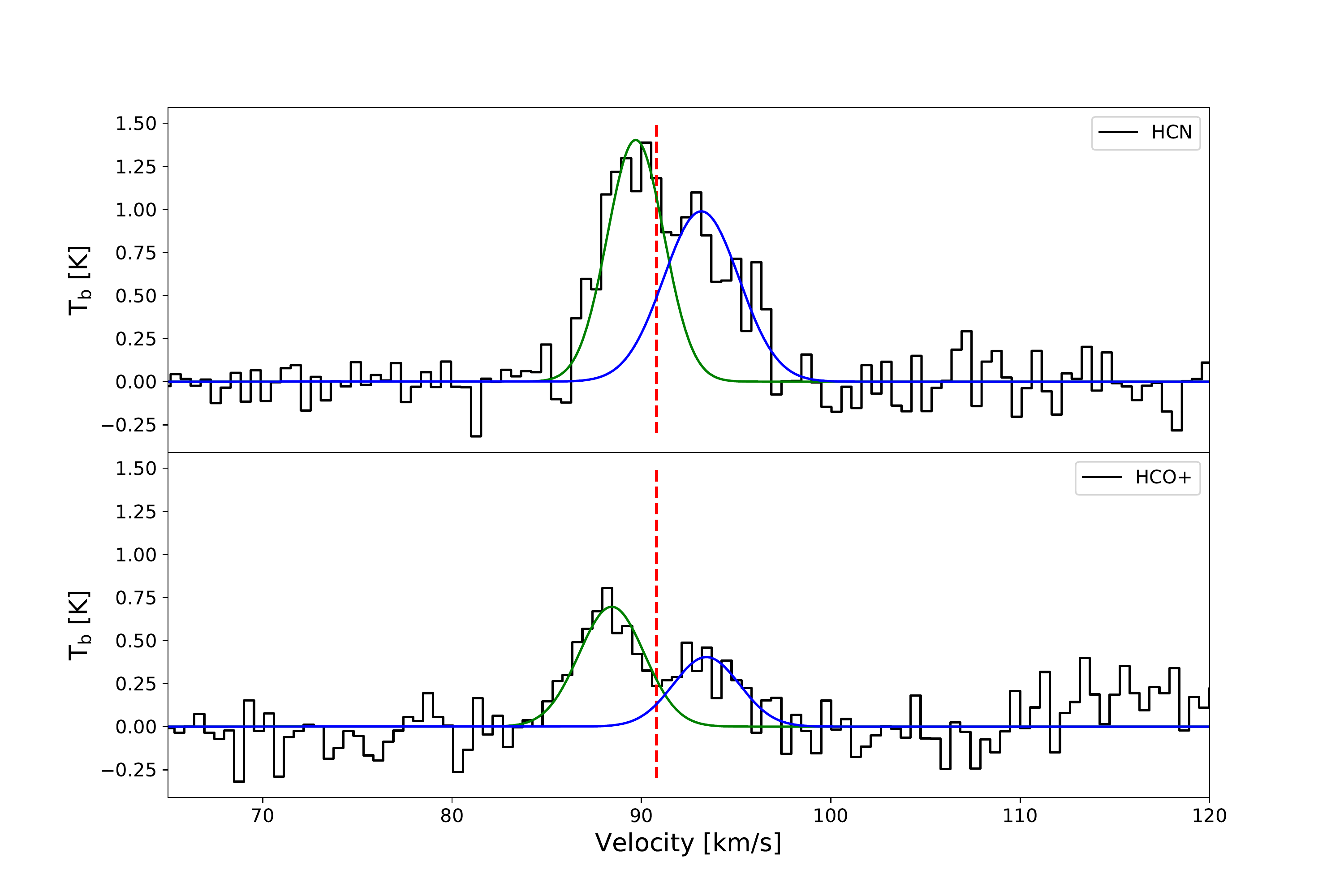}
\includegraphics[width=0.9\hsize]{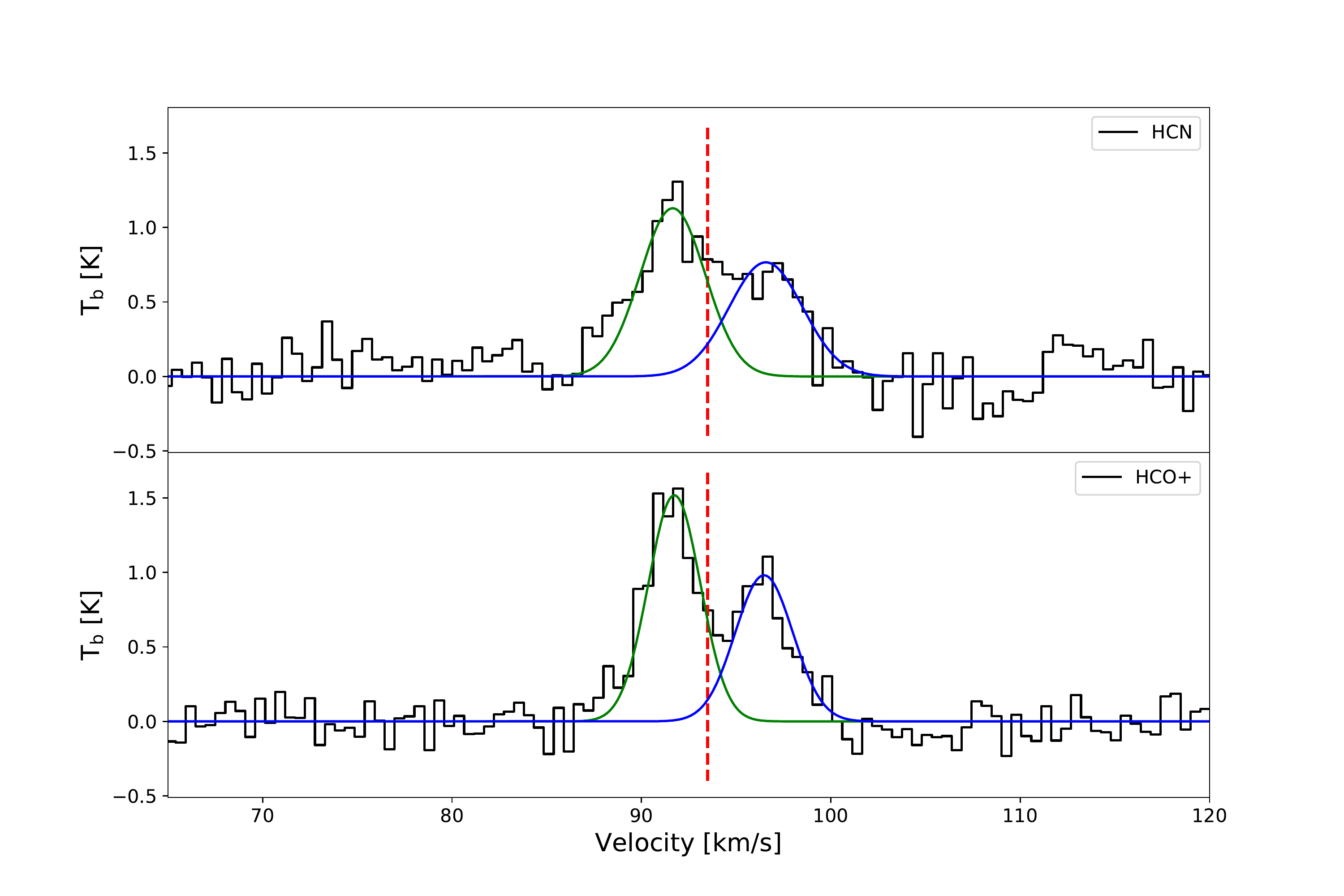}
\caption{\label{mmLineCenter} The Figure shows the HCN$(J=4 \rightarrow 3)$ and HCO$^{+}(J=4 \rightarrow 3)$ spectra from the MM2 ({\em upper panel}) and the
MM3 ({\em lower panel}) clumps. The spectra corresponds to observations of the clump phase centers done with  ASTE.
The red segmented line corresponds to the
V$_{\mathrm{lsr}}$ of those clumps as obtained from the H$^{13}$CO$^{+}(J=3 \rightarrow 2)$ spectral from \citet{Motte2003}.
In blue and green are the Gaussian fits done over line components.}
\end{figure}

\subsection{LINE EMISSION} \label{2se:lineEm}

\begin{figure}
\centering
\includegraphics[width=0.95\hsize]{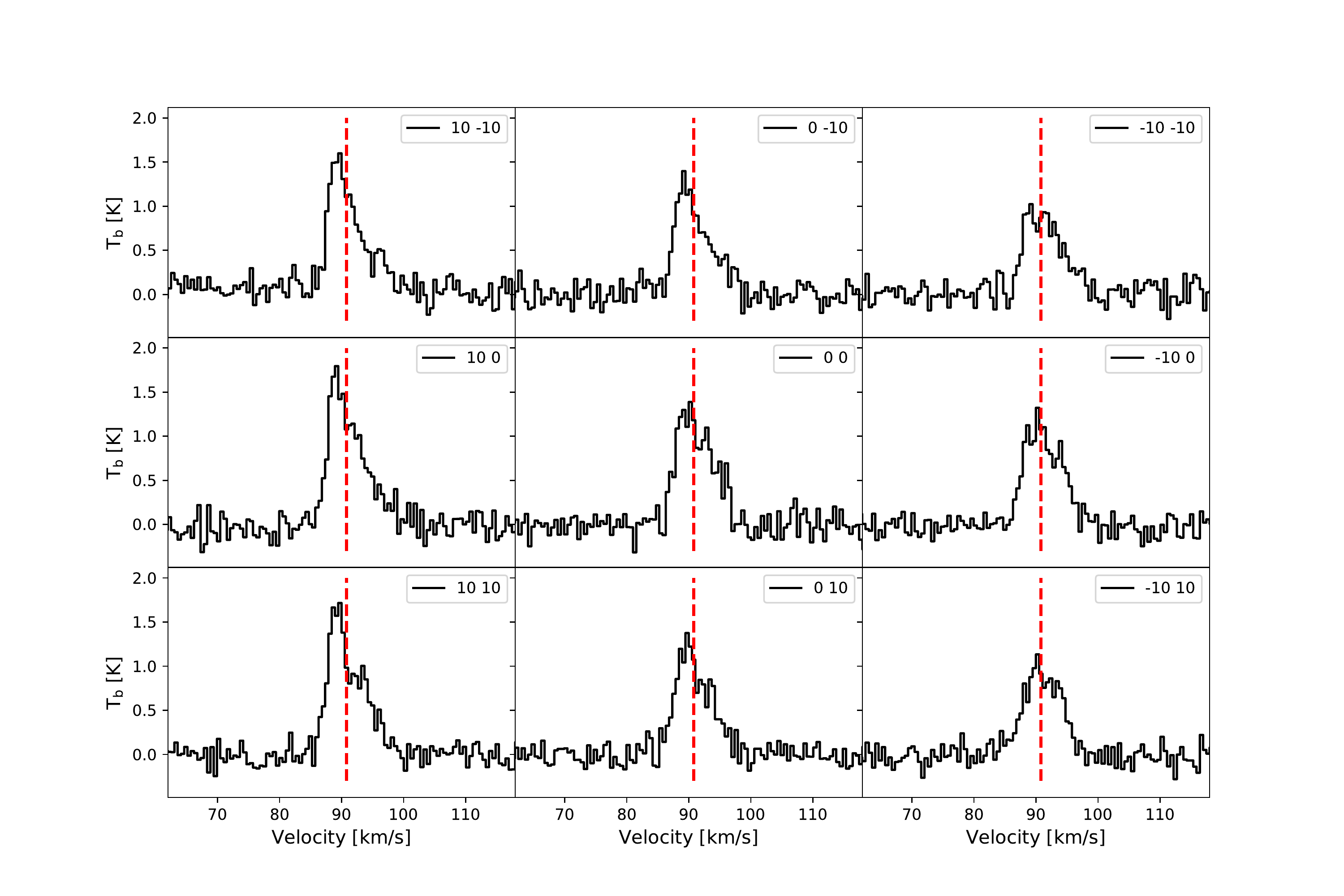}
\caption{Here we show a $3\times3$, $10^{\prime \prime}$ spaced, HCN$(J=4 \rightarrow 3)$ panel from the MM2 clump. In red is the V$_{\mathrm{lsr}}=90.8$
km s$^{-1}$. The labels corresponds to the offsets respect to the center coordinates, which is also the sampling rate used for the
OTF mapping done with ASTE over the MM2 clump.
\label{mm2Panel}
}
\end{figure}

\begin{figure}
\centering
\includegraphics[width=0.95\hsize]{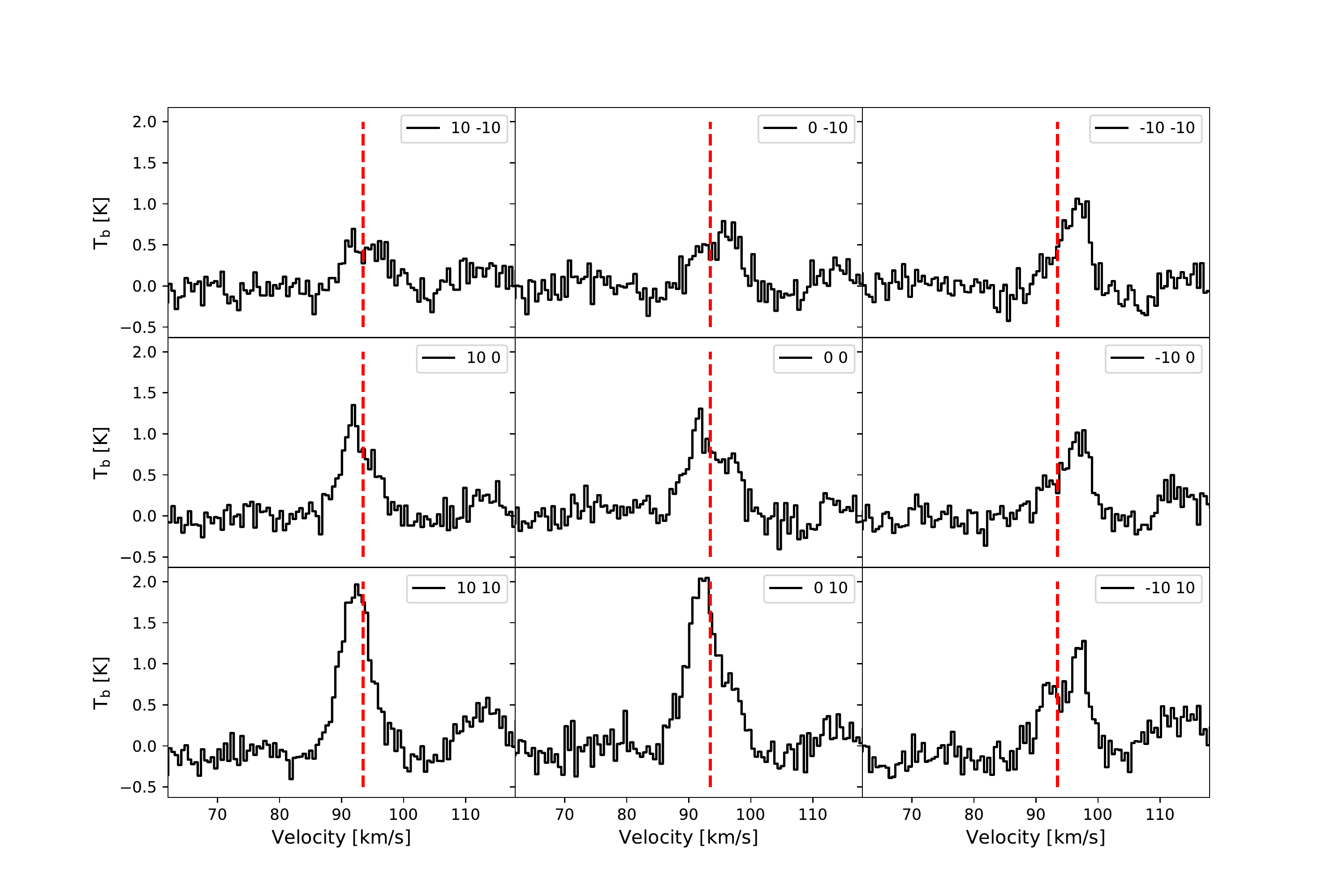}
\caption{ Here we show a $3\times3$, $10^{\prime \prime}$, HCN$(J=4 \rightarrow 3)$ panel from the MM3 clump. In red is the V$_{\mathrm{lsr}}=93.5$
km s$^{-1}$. The labels corresponds to the offsets respect to the center coordinates, which is also the sampling used for the
OTF mapping done with ASTE over the MM2 clump. Although the profiles suggest a blue asymmetry in most of the spectra, to the
West there is a clear red asymmetry, which may suggests rotation at the clump scales. Additionally, there is some indication
of high velocity emission about 115 km s${-1}$ in some of the spectra.
\label{mm3Panel}
}
\end{figure}

\subsubsection{LINE EMISSION FROM W43-MM2}

\begin{deluxetable}{c c c c c}
\tablecolumns{5}
\tablewidth{0pt}
\tabletypesize{\scriptsize}
\tablecaption{The Table presents the line parameters for ASTE  molecular line observations.\label{obspar} }              
\tablehead{
\colhead{Line} & 
\colhead{Transition} & 
\colhead{Frequency} & 
\colhead{Map size} & 
\colhead{RMS-noise} \\     
\colhead{}     &
\colhead{}     &
\colhead{[GHz]}   &  
\colhead{[arcsec$^2$]}  & 
\colhead{[K]} 
}
\startdata 
HCO$^{+}$  & $(J=4\rightarrow3)$ & 356.7342880 & $200^{\prime \prime} \times 160^{\prime \prime}$ & 0.83 \\   
HCN        & $(J=4\rightarrow3)$ & 354.5054773 & $200^{\prime \prime} \times 160^{\prime \prime}$ & 0.83 \\   
\enddata
\end{deluxetable}   

 Molecular line emission from high density tracers was also mapped towards the MM2 and MM3 clumps.
Figure \ref{mmLineCenter} shows the single dish, ASTE, 
HCN($J=4 \rightarrow 3$) and HCO$^{+}(J=4 \rightarrow 3)$ spectra towards the center of MM2,
along with their respective Gaussian fits.
Both molecules show a dip in their line profiles at $\sim 91$ km/s which is suggestive of self-absorption. 
In the absence of molecular emission from an optically thin species tracing similar densities (such as H$^{13}$CO$^{+}$($J=4 \rightarrow 3$) ),
we use the published H$^{13}$CO$^{+}$($J=3 \rightarrow 2$) from \citet{Motte2003} spectra to confirm the self-absorption. 
A line center velocity of $V = 90.8$ km/s was derived from the H$^{13}$CO$^{+}$($J=3 \rightarrow 2$) spectrum,
which is consistent with the dip in the HCN($J=4 \rightarrow 3$) and 
HCO$^{+}(J=4 \rightarrow 3)$ spectra \citep[see Figure \ref{mmLineCenter} here and Table 2 in ][]{Motte2003}. 
In fact, this velocity is used as the $V_{\mathrm{lsr}}$ of MM2.
Now, our Gaussian fitting results show two components centered at 89.7 and 93.2 \kms\ 
with amplitudes of 1.4 and 0.99 K  for the HCN line; while for the HCO$^{+}$ line the two components
are found at 88.4 and 93.4 \kms\ with amplitudes 0.7 and 0.4 K respectively.
Comparing the two components respect to the V$_{\mathrm{lsr}}$, the blue asymmetry is clear as the blue component 
from both lines has a much higher intensity than the red component. 
This type of profile is suggestive of infalling motions \citep{Leung1977}.
Figure \ref{mm2Panel} shows HCN($J=4 \rightarrow 3$) spectra around the 
reference coordinates covering and area of $44^{\prime \prime} \times 44^{\prime \prime}$ sampled every 
$10^{\prime \prime}$.
The blue asymmetry is present in all spectra throughout MM2, we will explore the implications of this
blue asymmetry further on (see section \ref{2se:infall}).

\subsubsection{LINE EMISSION FROM W43-MM3}

 Figure \ref{mmLineCenter} also shows the HCN($J=4 \rightarrow 3$) and HCO$^{+}$($J=4 \rightarrow 3$) 
spectra towards the same coordinates used as phase center for MM3.
As with MM2 the emission is also suggestive of self-absorption, though the HCN line does not present a strong
dip in between components and it is also broader and noisier when compared to the HCO$^{+}$ spectrum.
For MM3 we also used the H$^{13}$CO$^{+}$($J=3 \rightarrow 2$) results from \citet{Motte2003} giving us a V$_{\mathrm{lsr}} = 93.5$,
consistent with the dip seen the spectra.
We also did a $44^{\prime \prime} \times 44^{\prime \prime}$ grid around the MM3 reference coordinates to look
for blue asymmetry in the spectra as it was done with MM2. However in this case, we found that only spectra 
$\delta \alpha = 0^{\prime \prime}$ and $\delta \alpha =  10^{\prime \prime}$ shows blue asymmetry while
the spectra with $\delta \alpha = -10^{\prime \prime}$ show red-asymmetry in both HCN($J=4 \rightarrow 3$) and 
HCO$^{+}$($J=4 \rightarrow 3$) emission (see Figure \ref{mm3Panel}. This may indicate rotation in the MM3 clump at
large scales.

\subsection{INFALLING MOTIONS}
\label{2se:infall}

\begin{deluxetable}{c c c}
\tablecolumns{3}
\tablecaption{\scriptsize Normalized velocity difference. The asymmetry in line profiles is presented by
    the calculation of $\Delta V_{\mathrm{be}}$, as shown in Eqn. \ref{dv}, where the optically thin 
    species correspond to H$^{13}$CO$^{+}$.  The values used to compute the blue excess
        are taken from Gaussian fits to the corresponding spectra. \label{dvbe}  }
\tablehead{                           
\colhead{Clump} &                     
\colhead{Line} &
\colhead{$\Delta V_{\mathrm{be}}$} \\     
}
\startdata
   MM2 & HCN$(J=4\rightarrow3)$              & -0.3  \\   
   MM2 & HCO$^{+}(J=4\rightarrow3)$          &  -0.6 \\   
   MM3 & HCN$(J=4\rightarrow3)$              & -0.4  \\   
   MM3 & HCO$^{+}(J=4\rightarrow3)$          &  -0.4 \\   
\enddata 
\end{deluxetable} 

 Historically, asymmetries found in the spectra of molecular lines have been used to probe for infalling
motions in star forming cores. \citet{Leung1977} suggested that an asymmetry in the line profile towards the blue 
may indicate the presence of infalling motions. Thus, the low excitation and red-shifted infalling layers of gas in the 
front part of the cloud absorbs some of the emission from the rest of the gas. 
This red-shifted self-absorption is what makes the spectrum show a brighter blue peak. 
To quantify this asymmetry, we computed the normalized velocity difference from our data as
explained in appendix \ref{ape2}.

We calculated $\Delta V_{\mathrm{be}}$
for both HCN$(J=4\rightarrow3)$ and HCO$^{+}(J=4\rightarrow3)$ spectra coming from the center of both MM2 and MM3 (see Table \ref{dvbe}),
as well as the H$^{13}$CO$^{+}$($J=3 \rightarrow 2$) previously used to check for self-absorption.
The $\Delta V_{\mathrm{be}}$ values obtained for both MM2 and MM3 from the HCN and HCO$^{+}$ emission are all suggestive of infalling motions.
These values are less pronounced respect to results obtained towards MM1. \citet{Cortes2010} computed the $\Delta V_{\mathrm{be}}$
using same molecular transitions obtaining values $\sim -1$ \citep[see Table 3 in ][]{Cortes2010}.

To characterize these infalling motions, we applied a simple infalling model as also described in appendix \ref{ape2}.
We first fitted a double Gaussian to the HCN and HCO${^{+}}$ spectra from
the center of both MM2 and MM3 clumps.  The Gaussian fitting is shown in Figure \ref{mmLineCenter}.
The model was then adjusted to the resulting Gaussian spectra by minimizing
the $\chi^{2}$ function through the Levenburg-Marquardt algorithm \citep{Press2002}.
The fit parameters are presented in Table \ref{infParam}.
The infalling model has accretion onto a proto-stellar object as an implicit assumption.
The two layers which defines the model are assumed to have an
infalling direction onto a single core. By modeling infalling motions in this way
it is possible to estimate the infalling speeds by averaging both the front and rear components
as listed in Table \ref{infParam} for both MM2 and MM3.
For MM2 and MM3, the infalling speeds obtained from HCN and HCO$^{+}$ are comparable within 70\%.
Using these infalling speeds and assuming
spherical geometry we estimate the mass infalling rate for these clumps by calculating,

\begin{equation}
\dot{M} = \frac{dM}{dt} \sim \frac{M}{t} = \frac{\rho V v_{\mathrm{in}}}{R} = \frac{4}{3}\pi n_{\mathrm{H_{2}}}\mu 
m_{\mathrm{H}} R^{2} v_{\mathrm{in}}
\end{equation}

\noindent where $\mu$ = 2.35 is the mean molecular weight, R = 0.3 pc is the geometric radius,
 and $n_{\mathrm{H_{2}}}$ is the gas number density taken from Table \ref{mmTab}. The results are listed in
Table \ref{infParam}.
The blue asymmetry in the line shape
is better manifested in the HCO$^{+}$ emission; and thus, we are only considering those results
to estimate $\dot{M}$. Therefore,  we obtained a mass infall rate of $\dot{M} = 1.8 \times 10^{-2}$ \Msun/yr for MM2
and $\dot{M} = 6.3 \times 10^{-3}$ \Msun/yr for MM3. In principle, we expect that the HCO$^{+}$ emission is diluted
in the ASTE $22^{\prime \prime}$ beam and thus to estimate the infalling radius, we used the dust emission
as a proxy. Although it is not clear that the dense molecular gas will follow the spatial distribution
of the dust emission, this will still give a better estimate than deriving the radius from the ASTE beam.
The interferometric maps of MM2 and MM3 present here have a maximum angular scale of about 0.3 pc and thus, these
ALMA data is insensitive to length-scales larger than this. Given this limitation, we have assumed 0.15 pc as
the radius for the infalling mass rate estimation. This is reasonable given the geometric assumptions used.
\citet{Cortes2010} estimated the mass infall rate for MM1 assuming a radius derived from interferometric
observations done with the BIMA array at a much coarser resolution. Here we also refine this calculation using the
same assumption for the radius, and also updating the mass density to $n_{\mathrm{H_{2}}} = 10^{6}$ cm$^{-3}$,
which give us a mass infall rate for MM1 of  $1.3 \times 10^{-2}$ \Msun/yr. These infall rates are consistent
with previous estimates for MM1 \citep{Herpin2009}, and results obtained towards other high mass star forming regions
by similar analysis \citep{Saral2018} and by radiative trasfer modeling and SED fitting \citep{Fazal2008}. 

\section{DISCUSSION}\label{se:discussion}

\subsection{THE DYNAMICAL EQUILIBRIUM OF CORES IN THE W43-MAIN CLUMPS}

\subsubsection{Initial considerations}

 A total of 81 cores have been detected from our sample of clumps in W43-Main. 
Most of these cores are located along the main axis of their respective filaments, but some exceptions,
corresponding to  isolated cores, are seen outside the main filaments.
The estimated core masses range from a few to 427 \Msun\, where only 2, marginally, sub-solar cores were 
detected\footnote{The significance of these sub-solar cores is $\sim 10\sigma_{M}$ given a mass sensitivity of $\sigma_{M} = 0.08$ \Msun\ .}.
The spatial distribution of cores in all clumps is suggestive of clusters or associations. However, a significant number of cores, 54 or 66\%, 
have masses below 10 \Msun\, which is suggestive of low mass star formation. A similar situation has also been seen in
W43-MM1 \citep[see \citetalias{Cortes2016} and ][]{Motte2018a}. Additional examples from other high-mass star forming regions have 
been presented in the literature
\citep{Lu2018,Sanhueza2017,Cyganowski2017,Zhang2015,Frau2014}. This indicates that low mass stars are also being formed along
high-mass stars and therefore, low mass and high mass star formation processes might be coupled.

 Understanding the dynamical equilibrium of proto-stellar cores is required
to assess whether these cores are collapsing and forming stars or if they are still unbound.
We do this by  considering
the contributions from thermal motions, non-thermal motions (turbulence), and magnetic fields,
as the most important physical parameters against gravity. 
Rotation at the core scales might be relevant, especially if flattened structures (disks?) have formed. However, it has been found 
that in low mass star forming clouds, rotation does not seems to provide significant support against gravity  at the length-scales
traced by our data \citep{Tobin2012}. Therefore, and given the lack
of detailed kinematical information from our observations, we do not consider rotation here. 
Additionally, previous studies suggest that magnetic fields play a role in the level of fragmentation within a core, although this is still debateable \citep{Palau2013,Palau2015}.

 We understand that a core is collapsing, if the estimated mass is larger than some critical mass by the
application of virial equilibrium. The considerations 
behind determining this
critical mass has been the subject of a number of works in the past \citep{Spitzer1978,Shu1992,Bertoldi1992}. 
It is often found that the determination of the virial mass was done only by considering the gas kinematics, 
thermal and non-thermal (turbulence),
and, sometimes, by considering ``reasonable'' estimations of the  magnetic field strength. 
However, in this work we do have quantitative information about the
magnetic field and therefore, we can include more accurate estimations of its strength in the calculation.
We start by computing what in the literature is referred to as the virial mass, or M$_{\mathrm{vir}}$.
As this mass is a representation of the energy contained in both thermal and non-thermal motions, 
we refer to it here a the kinetic mass, which we calculate as,

\begin{equation}
M_{\mathrm{kin}} = 3k\frac{R\sigma_{\mathrm{v}}^{2}}{G},
\end{equation}

\noindent  where $R$ is the core radius, $\sigma_{\mathrm{v}}$ is the gas velocity dispersion along the line of sight 
which is derived from the molecular line-width as 
$\sigma_{\mathrm{v}} = \Delta \mathrm{V}_{\mathrm{FWHM}}/2\sqrt{2\ln{2}}$, $G$ is the
gravitational constant, and $k$ is a correction factor introduced by \citet{MacLaren1988} to account for the a power law density profile
$\rho \sim R^{-a}$. 
Assuming a constant density in the condensation, or $a$ = 0 and k = 5/3, the kinetic mass can be rewritten as,

\begin{equation}
M_{\mathrm{kin}} = 1.17 \left( \frac{R}{\mathrm{mpc}} \right) \left( \frac{\sigma_{\mathrm{v}}}{\mathrm{km\ s}^{-1}} \right)^{2} \Msun \\
\end{equation}

\bigskip
 The gas kinetic energy is represented by the non-thermal motions seen from the molecular line-widths, 
which are here midly larger than the thermal sound speed. 
Here, we use the same line-widths used for the estimation of B$_{\mathrm{pos}}$, but at core scales of 
1$^{\prime \prime}$ which are about the average of our core sizes (see section \ref{2se:MF} and appendix \ref{ape1}). 
For completion, we also calculated the Jeans mass, M$_{\mathrm{J}}$, for each core.
The computed values for $M_{\mathrm{J}}$ and $M_{\mathrm{kin}}$ are shown in Table \ref{mmTab}.
The kinetic mass already includes the contributions from thermal and  non-thermal energy into the balance.   
To assess the support provided by motions against gravity, we use the kinetic mass to define a kinetic 
virial parameter, 
$\alpha_{\mathrm{kin}}=M_{\mathrm{kin}}/M_{\mathrm{gas}}$ \citep{Bertoldi1992}. 

To quantify the contribution from the magnetic field in the balance against gravity, we define the magnetic mass as 
$M_{\mathrm{\Phi}} = \Phi/2\pi G^{1/2}$ \citep{Nakano1978,Crutcher2004}, where $\Phi$ is the magnetic flux and $G$ is 
the gravitational constant. We used this definition to be consistent with the mass-to-magnetic flux ratio definition (see
paragraph below). Having defined $M_{\mathrm{kin}}$ and $M_{\mathrm{\Phi}}$, we can now calculate 
a total virial parameter as,

\begin{equation}
\alpha_{\mathrm{total}} = \frac{M_{\mathrm{kin}} + M_{\mathrm{\Phi}}}{M_{\mathrm{gas}}},
\end{equation}

 The $\alpha_{\mathrm{total}}$ includes the contribution of the magnetic field in the ratio by adding the magnetic
field mass. In this way, we account for all the relevant physical processes that oppose gravitational collapse.

 To explore the magnetic-only contribution we will compute the mass to magnetic flux ratio, as done before 
in W43-MM1 \citepalias{Cortes2016}. We do this to compare the relevance of the magnetic field with respect to gravity.
 The mass to magnetic flux ratio is defined following \citet{Crutcher2004},

\begin{equation}
\lambda_{B} = \frac{\left<(M/\Phi)_{\mathrm{observed}}\right>}{(M/\Phi)_{\mathrm{crit}}} = 7.6\times10^{-21}\frac{N(\mathrm{H_{2}})}{3\mathrm{B}_{\mathrm{pos}}},
\end{equation}

\noindent where N$(\mathrm{H_{2}})$ is the molecular hydrogen column density in cm$^{-2}$
calculated for each source independently, 
B$_{\mathrm{pos}}$ is the magnetic field strength in $\mu G$ assumed to be a global physical parameter for a given region (see
section \ref{2se:MF} for a description of the chosen regions),
where the factor of 3 in the denominator corresponds to 
an statistical geometrical correction for $(M/\Phi)_{\mathrm{observed}}$. The geometrical shape of a core depends on the
strength of the magnetic field, which will tend to flatten the core along its minor axis. In the limit case
of a strong magnetic field, the core might converge to a pseudo-disk \citep{Mouschovias1987}.
Table \ref{fieldValues} presents the derived values for $\lambda_{B}$
using all three B$_{\mathrm{pos}}$ estimations. Although the nomenclature might appear confusing as $\lambda_{B}$ is 
essentially $M_{\mathrm{gas}}/M_{\mathrm{\Phi}}$, which is the inverse of a magnetic $\alpha$ parameter, we have decided to keep it
in order to be consistent with the literature.


\begin{figure}
\centering
\includegraphics[width=0.98\hsize]{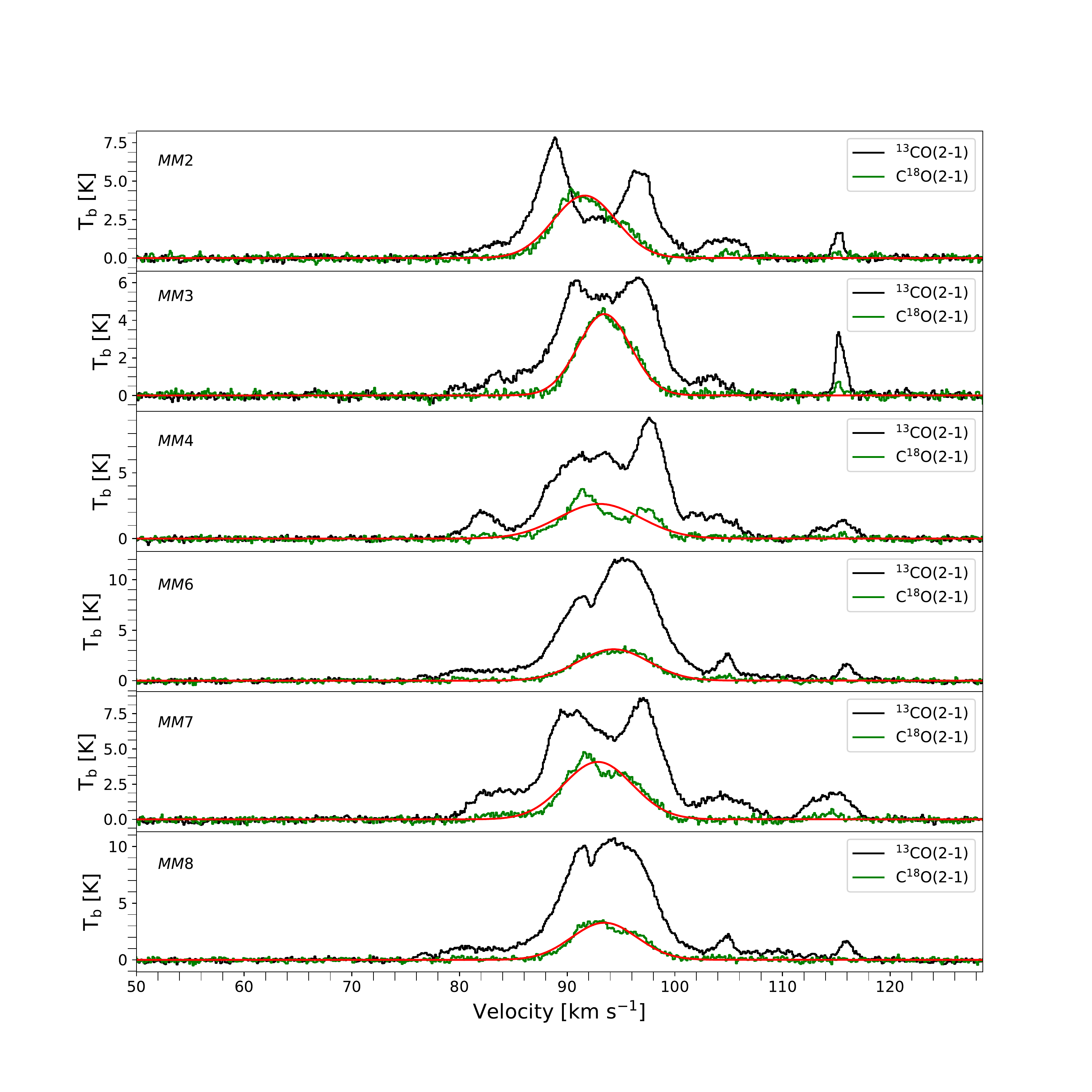}
\caption{The panel shows $^{13}$CO$(J=2\rightarrow1)$, in black, and C$^{18}$O$(J=2\rightarrow1)$ spectra in green. 
The data has been extracted from the public survey done by \citet{Carlhoff2013} using circular regions of 10$^{\prime \prime}$
in diameter center at the coordinates of the most massive cores, from each clump, detected in this work.
Superposed in red is the Gaussian fit to the C$^{18}$O$(J=2 \rightarrow 1)$ line. The name of each clump is indicated
by the legend box at each plot.
\label{coLine}
}
\end{figure}

\subsubsection{Caveats}

Before stating the result of our computations, it is necessary to factor in the uncertainties in the calculation.
We are deriving velocity dispersions by assuming a Kolmogorov-like power spectrum which we believe is representative of the
non-thermal motions induced by the turbulence cascade from the large scales, though its nature is uncertain.  
Furthermore, this assumption makes the calculation of the kinematic mass (see section \ref{virialAnalysis}) independent
of the temperature.
However, turbulence can be injected at smaller scales
by outflows, shocks and ionization fronts, and winds coming from the newly born stars; all of which we are not considering in this analysis.
Nevertheless, magnetic fields have a dampening effect on turbulence at small scales, which might decrease the
contribution to the turbulent energy close to the core scales.
This is also suggested by MHD simulations where the velocity dispersion in
gravitationally bounded cores was found to be sonic or mildly supersonic \citep{Hennebelle2018}.
Therefore, the injection of additional energy into the turbulence at these length-scales should happen continuously to
alter the energy cascade  and to modify the velocity dispersion significantly.
Also of importance are the uncertainties in the mass
estimation.  The mass estimates depends on the dust temperature, a proxy for the gas temperature, which is not well constrain
for these clumps.
High mass proto-stellar cores are seen to develop hot-cores quite early in their evolution, though the exact process is unclear.
In W43-Main, the binary in MM1-$A$ already has developed a hot-core \citep{Sridharan2014}.
An increase in the inner core temperature will decrease the mass estimate and
therefore, increase the $\alpha_{\mathrm{total}}$ ratio, which might potentially alter the conclusions.
However, this is difficult  to address without better spectroscopic millimeter, or sub-millimeter, data from
these cores. Additionally, all three  B$_{\mathrm{pos}}$ estimations require the volume number density as an input which we computed
by assuming spherical geometry. This assumption might bias the B$_{\mathrm{pos}}$ computation
towards larger values, especially if the geometry in the line-of-sight has a more flatter shape.

\subsubsection{The virial equilibrium}
\label{virialAnalysis}

 We first compare turbulence and magnetic fields independently against gravity. 
For thermal energy, all cores in our sample show Jeans masses which are, 
 orders of magnitude lower than the core estimated mass as shown in Table \ref{mmTab}. 
It is clear at this point, that thermal pressure, on its own, does not contribute significantly 
as a support mechanism against gravitational collapse  at this stage in the evolution of our cores.

\begin{figure*}
\centering
\includegraphics[width=0.45\hsize]{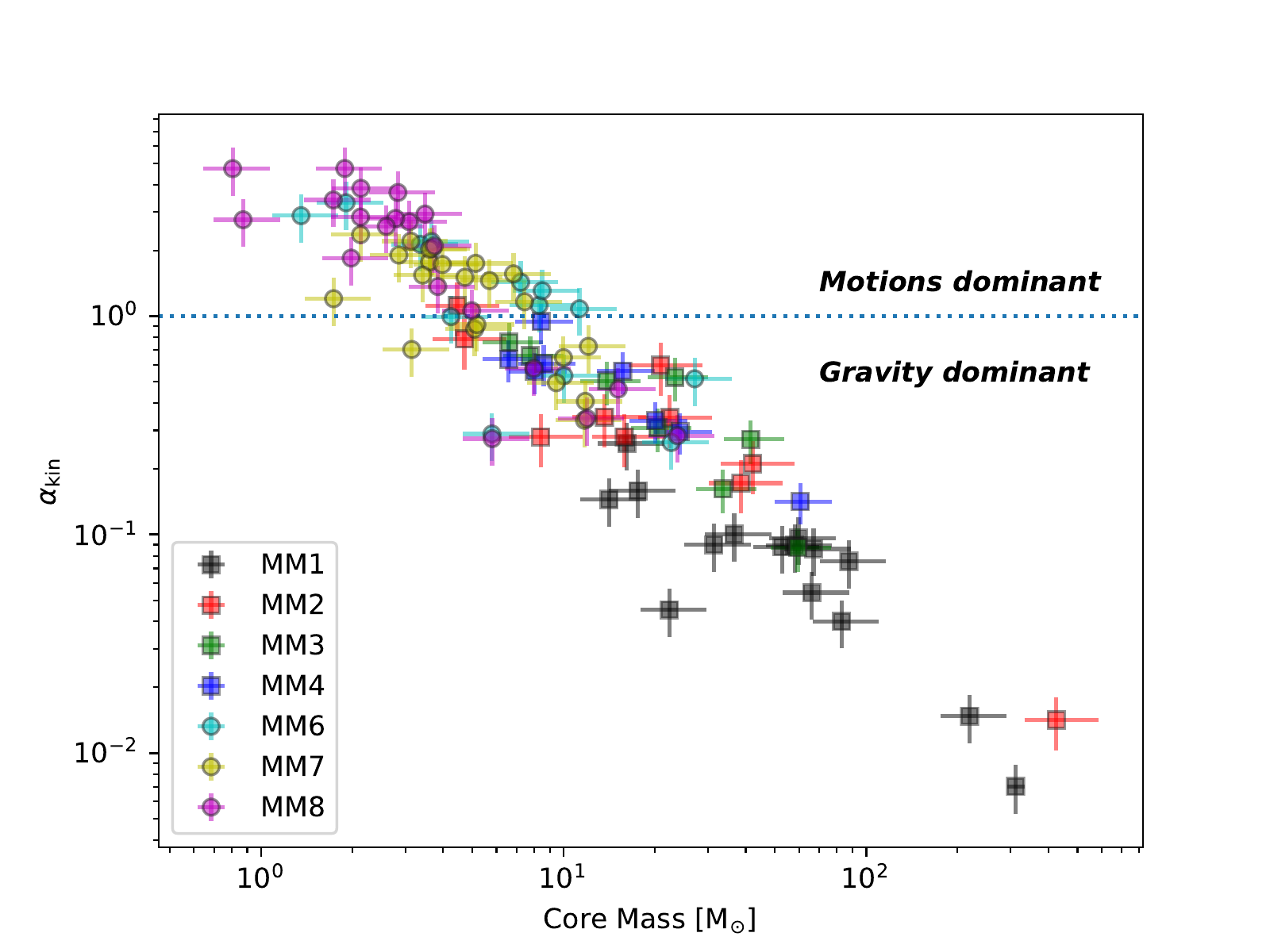}
\includegraphics[width=0.45\hsize]{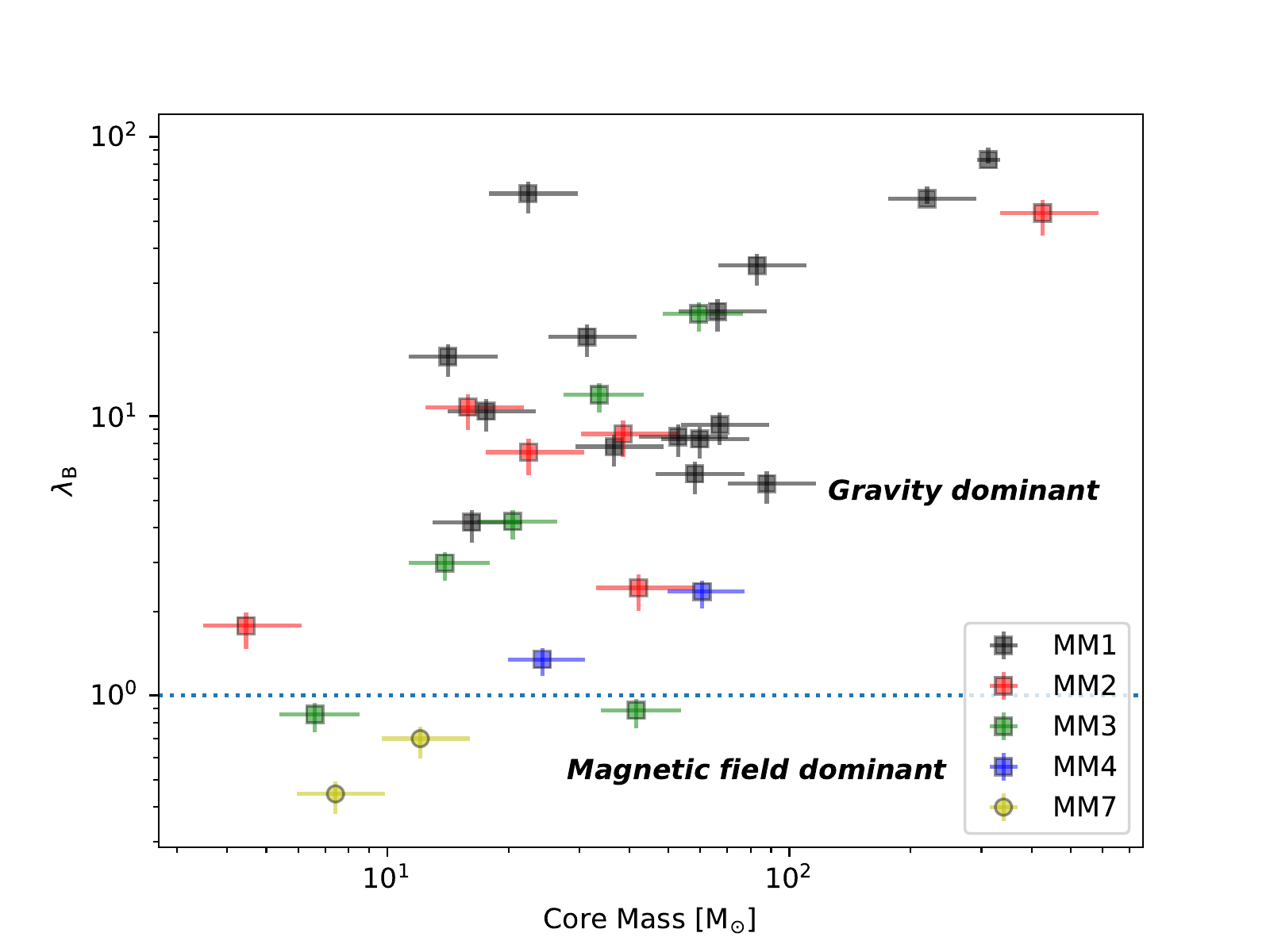}
\caption{Here we show the $\alpha_{\mathrm{kin}}$ virial parameter ({\em left}) and the 
mass to magnetic flux ratio, or $\lambda_{\mathrm{B}}$ ({\em right}),  
as a function of core mass in M$_{\odot}$. The cores are colored according to their clumps as indicated by 
the label. Error bars are displayed for the mass, $\lambda_{\mathrm{B}}$, and $\alpha_{\mathrm{kin}}$  values
computed by using a range of
three temperatures for all cores in each clump (see section \ref{virialAnalysis} discussion).
The blue dotted line corresponds to the critical threshold in the virial equilibrium.
\label{jtr}
}
\end{figure*}

When considering support only from kinetic energy, we found that 43, out of 81, cores have
$\alpha_{\mathrm{kin}} < 1$ (as shown in Figure \ref{jtr}). This corresponds to
most of the cores from MM2, MM3, and MM4, and with some additional cores from the other clumps. 
Values of $\alpha_{\mathrm{kin}} < 1$ suggests that turbulence alone cannot support
these cores against gravity.
An extreme case is MM2-$A$ which is 2 orders of magnitude away from virial equilibrium.
In contrast, the cores with $\alpha_{\mathrm{kin}} > 1$ belong, mostly, to MM6, MM7, and MM8 with just one core from MM2,
suggesting that these cores are gravitationally unbound. 
This set of cores belong to 
the low-end range of the mass distribution and also to the clumps with largest degree of fragmentation (the MM7 and MM8 clumps).

 Considering that we have detected large scale infalling motions from large scales, 
we would have expected to see most of these cores with $\alpha_{\mathrm{kin}} < 1$. 
However, in $\sim$ 47\% of our cores the energy from turbulence alone seems sufficient to support the 
cores against gravity. Although transient cores, due to the turbulence energy cascade, cannot be immediately ruled out, 
the large column densities obtained (mostly in excess of $10^{24}$ cm$^{-2}$)
suggests that these cores are likely
pressure confined, and thus the most likely scenario is that these
cores are still accreting gas, but have not yet reached a critical mass to ensue gravitational collapse. 

\begin{figure*}
\centering
\includegraphics[width=0.45\hsize]{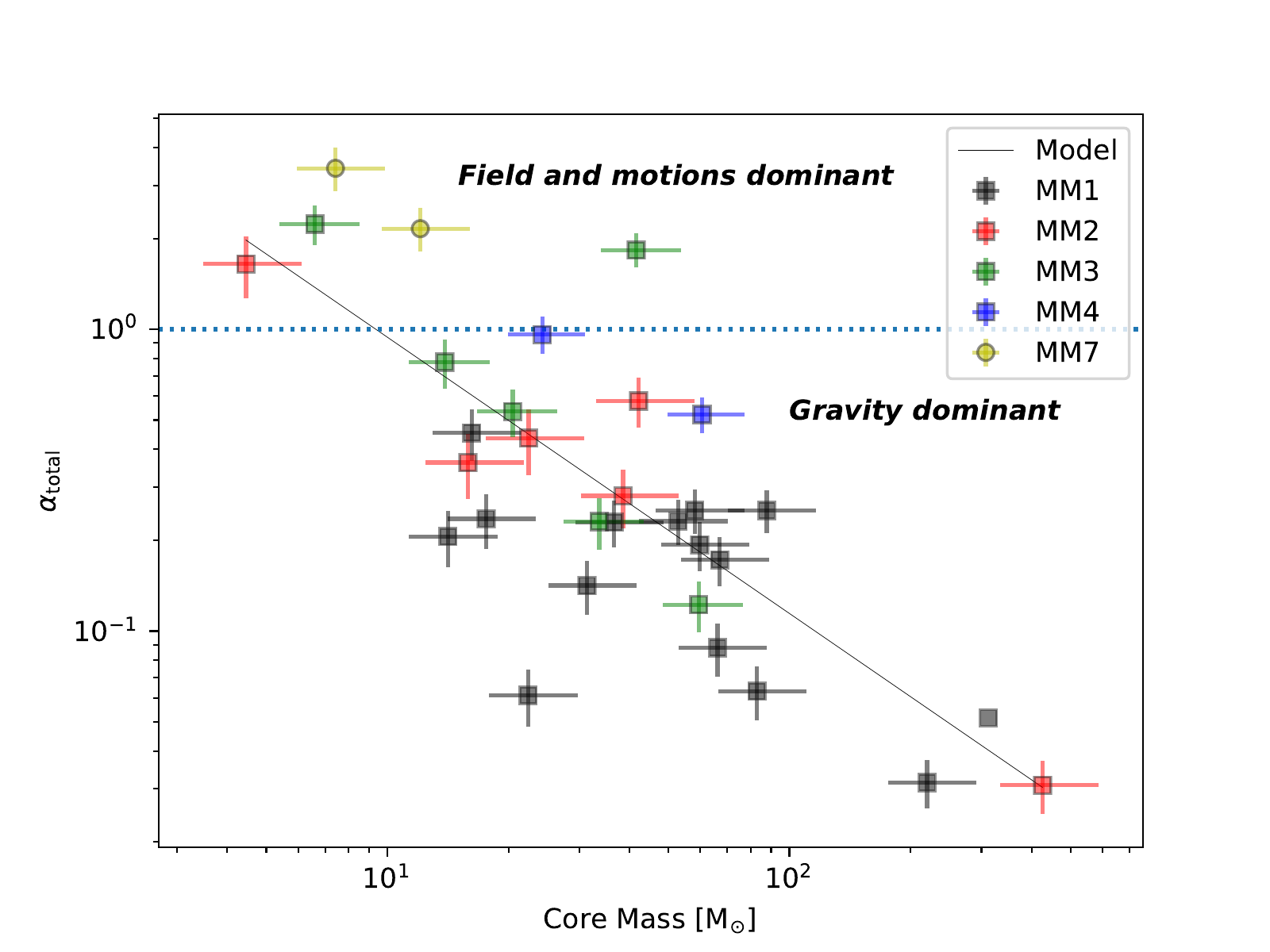}
\includegraphics[width=0.45\hsize]{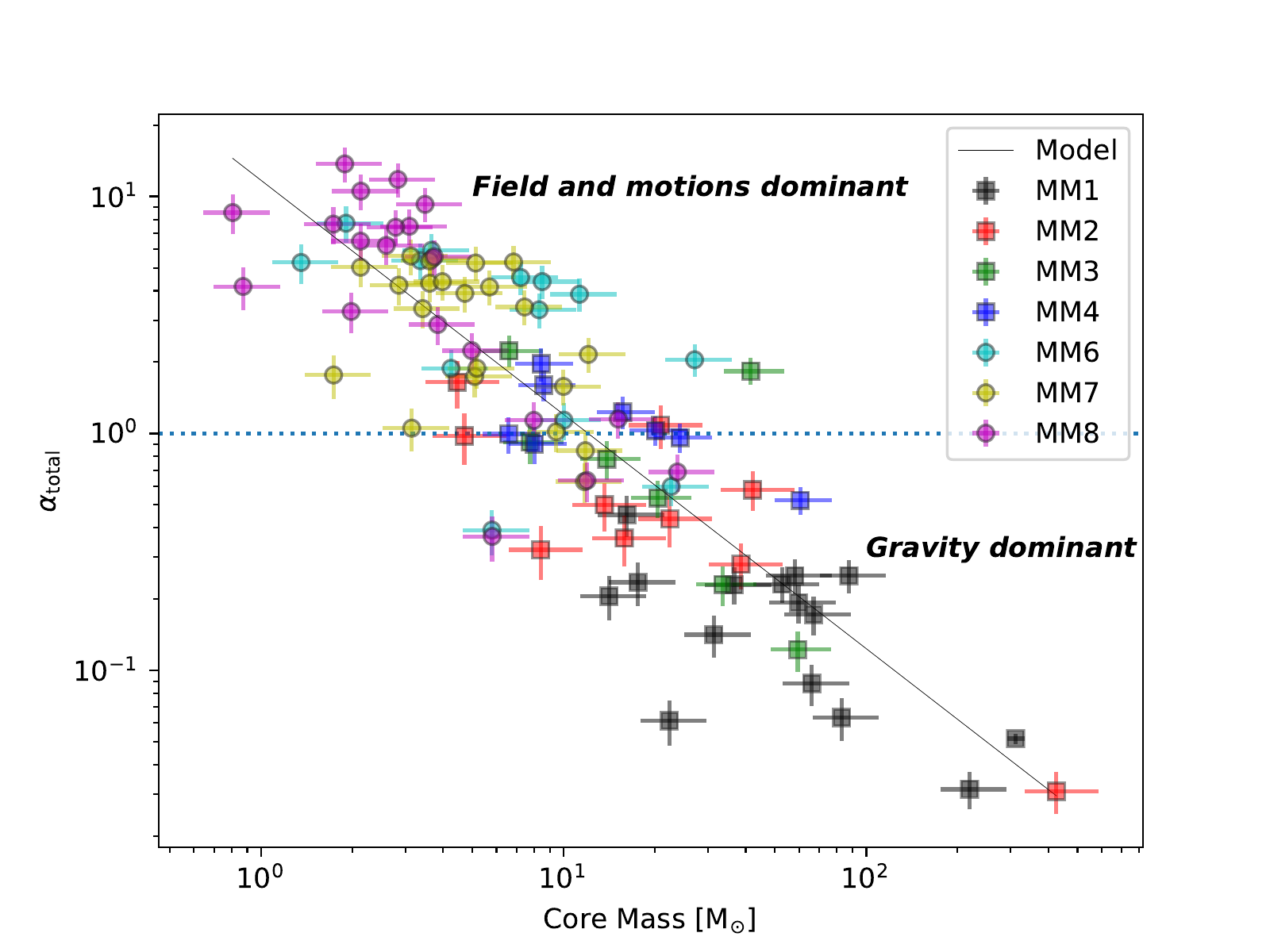}
\caption{
Here we show the total virial parameter, or $\alpha_{\mathrm{total}}$ ({\em left}), for 
the sample of cores where we have B$_{\mathrm{pos}}$ estimations directly derived from the data, 
and the $\alpha_{\mathrm{total}}$ parameter when all cores are considered ({\em right}). Both panel show
plots as a function of core mass in M$_{\odot}$. The cores are colored according to their clump as indicated by  
the label. 
Error bars are displayed for the mass and $\alpha_{\mathrm{tot}}$ estimates 
computed by using a range of
three temperatures for all cores in each clump (see section \ref{virialAnalysis} for discussion).
The blue dotted line corresponds to the critical threshold in the virial equilibrium.
The black line corresponds to the best power law fit to $\alpha_{\mathrm{total}}$.
\label{lr}
}
\end{figure*}

 To consider the magnetic-only contribution to the virial analysis, we calculated $\lambda_{B}$
only for the cores with associated B$_{\mathrm{pos}}$ estimates (see Figure  \ref{jtr} and Table \ref{fieldValues}).
The computation of $\lambda_{B}$ was done by using the average of all three B$_{\mathrm{pos}}$ estimates.
Our results show that most of the cores from MM2, MM3, MM4, and MM7, with B$_{\mathrm{pos}}$ estimates,  are super-critical, 
particularly MM2-$A, B$, and MM3-$A, B$. Two cores from MM3 (MM3-$E$ and MM3-$F$) are sightly sub-critical while
two from MM7 which appear to be sub-critical (MM7-$H$ and MM7-$K$). However and for MM2-$A$, MM2-$D$, MM3-$A$,
MM3-$B$, and MM3-$D$ the values for $\lambda_{B}$ are upper limits given that the magnetic field estimations are
lower limits (see section \ref{2se:MF}).
For the super-critical cores, the interpretation is that the magnetic field, by itself, does not 
provides sufficient support against gravity and therefore they should be out of virial equilibrium
unless additional sources of energy (turbulence) can further contribute to the core support.
For the sub-critical cores, the results suggests that the magnetic field alone might
be supporting these cores against gravity.
Although, the two cores in MM3 have large differences in the mass estimates, the magnetic field morphology
that surrounds them shows almost no dispersion in the EVPA resulting in strong B$_{\mathrm{pos}}$ estimates
(of $\left<\mathrm{B}_{\mathrm{pos}}^{\mathrm{F}}\right> = 13$ mG and 
$\left<\mathrm{B}_{\mathrm{pos}}^{\mathrm{E}}\right>  = 17$ mG). 
\citet{Frau2014} investigated  NGC 7538, IRS 1-3, with the SMA, where they found a number of low mass cores
along the spiral arms of the filament which also appear to be sub-critical. These core also are in the low-end
of their mass distribution. 

Finally, we consider both turbulence and magnetic energy, in the dynamical equilibrium of the cores in W43-Main, 
by computing the total virial parameter $\alpha_{\mathrm{total}}$ as a function of the
estimated core mass (see left panel in Figure \ref{lr}). From a total of 17 cores,
we found 5, or 29\%, with $\alpha_{\mathrm{total}} > 1$.  
The remaining 12 cores, or 71\% of the total, have $\alpha_{\mathrm{total}} < 1$ and should be out of virial equilibrium. 
Here, out of virial equilibrium implies that, as  we are
accounting for all the relevant physical parameters at these length-scales, these cores should be self-gravitating
and forming stars. The final multiplicity inside each core is unclear from these data and only higher resolution observations
might provide answers to that question. 
Although the number of bound cores dominates the distribution, the evidence from B$_{\mathrm{pos}}$ estimates coming from 
real data as well as reasonable velocity dispersions, strongly suggests that the unbound cores are real.
As previously stated, we also note that all of these core are likely to be pressure confined given the high column densities
observed towards these clumps. Thus, it is highly unlikely that the unbound cores are transient structures due to the turbulent cascade, and the
more likely explanation is  that these cores are still accreting and have not
yet accumulated sufficient mass to become self-gravitating.

To increase our statistics, we estimated $\alpha_{\mathrm{total}}$ for the totality of our cores 
by assuming  ``reasonable'' B$_{\mathrm{pos}}$ values.
We did this by using proximity to regions with B$_{\mathrm{pos}}$ estimates and, 
in the case of MM6 and MM8, we use the average obtained from MM7 given the similarities in their fragmentation.
The result of adding all of our cores is shown in Figure \ref{lr} (right panel).
The Figure shows two distinctive populations of bound and unbound cores. 
The bound cores belong to the population of the most massive clumps
with direct evidence for infalling motions from large scales, though there a number of cores in ``critical'' state. 
The unbound cores
correspond to 70\% of our sample, which although quite uncertain, strongly suggests that there is a population of
cores which are not gravitationally bound. For this population of cores, the support against gravity comes from a
combination of magnetic and turbulent energy. Although it is quite possible that the ``reasonable'' B$_{\mathrm{pos}}$
estimate used for MM7 and MM8 is large (we will discuss this possibility in section \ref{sse:poli}), 
any additional source of support will increase the cores unbound state already seen
from turbulence alone.

As previously discussed, plausibly, a main uncertainty in our calculations comes from the temperature assumptions.     
Although the temperature used for the column density estimation at each clump comes from SEDs derivations, 
these represent the large scale structure within the clump (given the resolution of the single dish 
telescopes used to derive such SEDs). Thus, the internal core structure, particularly in the most massive sources,
might have temperatures which are higher than the nominal ones used here. 
To account for variation in the core temperatures, we calculated
error bars for the mass, $\alpha_{\mathrm{kin}}$, $\lambda_{B}$, and $\alpha_{\mathrm{total}}$ by using a $\pm 5$ K
temperature deviation relative to the nominal SED temperature of the clump. Using $\pm 5$ K is a reasonable assumption for 
the bulk of the cores in our sample, although the more massive cores might have higher temperatures (as previously discussed).
The spread shown by the uncertainty in the 
temperature might sligthly move the values, but the overall trend remains as shown in Figures \ref{jtr} and \ref{lr}.

From ALMA observations of IRDC G28.34, \citet{Zhang2015} found 10 gravitationally bound cores
and one clump, G28-P1, was found to be unbound. However, they did not have observational information about the
magnetic field and thus, they  assumed a value for B = 270  $\mu$G in their analysis. 
\citet{Zhang2015},
stated correctly that magnetic fields can substantially increase the value of the virial mass, which is exactly
what we see in W43-Main. The values that we have estimated for the magnetic field are, at maximum, close to 2 orders of magnitude
over the values used for G28.34,  but biased toward the lower end as we used the region number density which is lower than 
the core density (see appendix \ref{ape1}). The column densities derived toward the G28.34 cores, though not listed, appear to
approach our results as the number density values are between $10^{6}$ and $10^{7}$ cm$^{-3}$. Given that density regime,
it is likely that the field strength will be on the order of milli Gauss for G28.34, as expected from observations \citep{Crutcher2012}.
This will certainly increase the virial mass for those cores which might resemble what we see in W43-Main.

Having uncovered these two populations of cores in W43-Main, 
it should be noted that these populations will likely produce low and high mass stars. 
Thus, as some of these cores are gravitationally bound while the others are not, 
the time-scale for star formation in these two populations has to be different.
For the bound cores collapse cannot proceed faster than the free-fall time, which sets a lower limit. However,
the unbound cores might be accreting at time-scales set by the infall rate, which is substantially lower.
This essentially suggests that the evolutionary time-scale for star-formation in high mass filaments is not uniform.

 \citet{Bertoldi1992} derived a power law for the virial parameter in the form of 
$\alpha = b\left(\frac{M_{o}}{M_{\mathrm{obs}}}\right)^{a}$ for magnetized star forming cores. To compare, 
we fitted our data with such  power law obtaining $b=1.06$ and $a=-0.92$ for the initial 17 cores 
and $b=1.04$ and $a=-0.88$ for the whole sample (see Figure \ref{lr}). \citet{Bertoldi1992} derive $b = 2.12$ and 
$a=2/3$ for magnetized critical clumps, and $b=2.9$ and $a=2/3$ for pressure confined clumps. 
However, in their analysis they considered $M_{O} = M_{\mathrm{J}}$ which is different from what we are using in 
$\alpha_{\mathrm{total}}$. In our case, $M_{O} = M_{\mathrm{\Phi}} + M_{\mathrm{kin}}$, which might explain the difference.
\citet{Frau2014} also fitted a power law to their 
data, finding $b = 2.26$ and $a = -0.68$ in agreement with \citet{Bertoldi1992}, though they do not comment about
the usage of the Jeans mass. 

\subsubsection{Comparison between motions and magnetic fields}

\begin{figure}
\centering
\includegraphics[width=0.95\hsize]{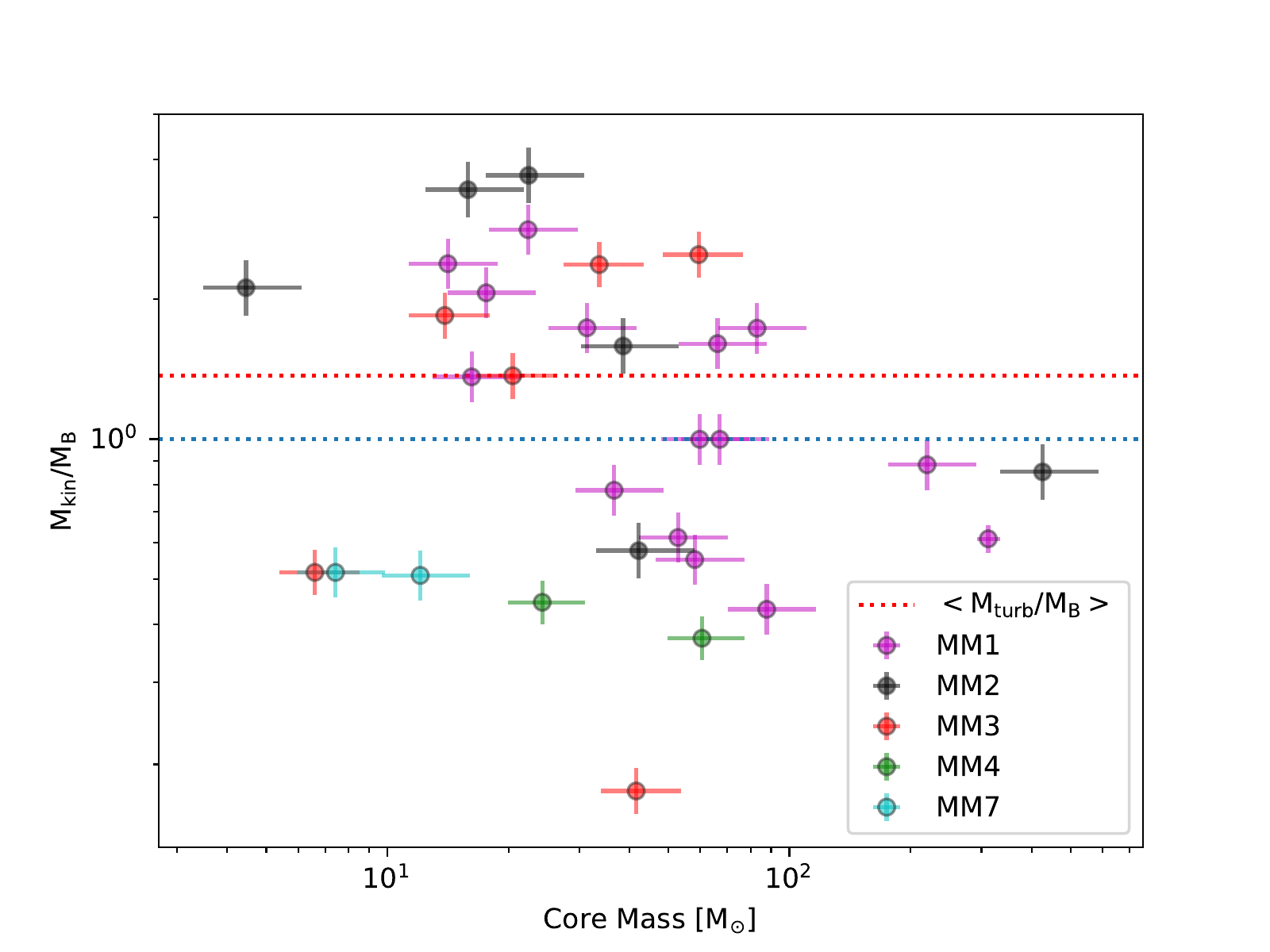}
\caption{In this Figure we show the ratio between the turbulence mass,
or M$_{\mathrm{kin}}$, and the magnetic mass, or M$_{\mathrm{B}}$,
as a function of core mass in M$_{\odot}$. 
Error bars are displayed for the mass and and $M_{\mathrm{kin}}/M_{\mathrm{B}}$ ratio 
computed by using a range of
three temperatures for all cores in each clump (see section \ref{virialAnalysis} for discussion).
 The blue dotted line corresponds to the equipartition threshold between both physical parameters . 
The red dotted line marks the average for the ratio between M$_{\mathrm{kin}}$ and M$_{\mathrm{B}}$.
\label{tb2MB}
}
\end{figure}

 The virial parameter $\alpha_{\mathrm{total}}$, provides an assessment of the core equilibrium, but not information about
the relevance of each of the parameters that counteract gravity. To explore the relative importance of kinetic energy
and magnetic energy, we computed the ratio between the kinetic mass and the magnetic field mass (see Figure \ref{tb2MB}).
The mean value for the turbulent to magnetic field mass ratio is 
$\left<M_{\mathrm{kin}}/M_{\mathrm{\Phi}}\right> = 1.43$,
which suggests that turbulent energy dominates over magnetic energy.
Now, the Alfven speed, $V_{\mathrm{a}} = B/\sqrt{4\pi\rho}$, has a range between $ 0.19 < V_{\mathrm{a}} < 2.9$
km s$^{-1}$ which is on average larger than the velocity dispersion, $\sigma_{B}$, 
as $\left<V_{\mathrm{a}}\right>=1.1 > \left<\sigma_{B}\right> = 0.9$ km s$^{-1}$, if our assumptions are correct. This
also suggests that magnetic tension in these filaments is strong and, might, damp the turbulent cascade at these
scales, unless turbulence gets replenished.
One way in which turbulence can be replenished is through outflow feedback. Figure \ref{coLine} shows extracted 
$^{13}$CO$(J=2 \rightarrow 1)$ and C$^{18}$O$(J=2 \rightarrow 1)$ spectra from the W43-Main large scale CO mapping by 
\citet{Carlhoff2013}. 
The spectra was extracted from the clumps phase center position and by using pencil beam regions at the resolution of the IRAM telescope 
($\sim 10^{\prime \prime}$).
The spectra shown in Figure \ref{coLine} suggests that the $^{13}$CO emission is optically thick at 1 mm, where there is also
substantial structure in velocity for the points considered.
Gaussian fits were done to the C$^{18}$O spectra (shown in red in Figure \ref{coLine}), which also suggests that, in
some clumps, C$^{18}$O might also be optically thick.
The emission at the lines-wings from $^{13}$CO  strongly suggests the presence of outflows in some of these clumps and
it is particularly clear in MM8 and MM6. 
Although, $^{13}$CO and C$^{18}$O traces
more diffuse gas at 1 mm (the $^{12}$CO$,J=2 \rightarrow 1,$ has a critical density $\sim 10^{4}$ cm$^{-3}$), 
than our dust emission, they are still a sufficient proxy for outflows. 

\begin{figure}
\centering
\includegraphics[width=0.99\hsize]{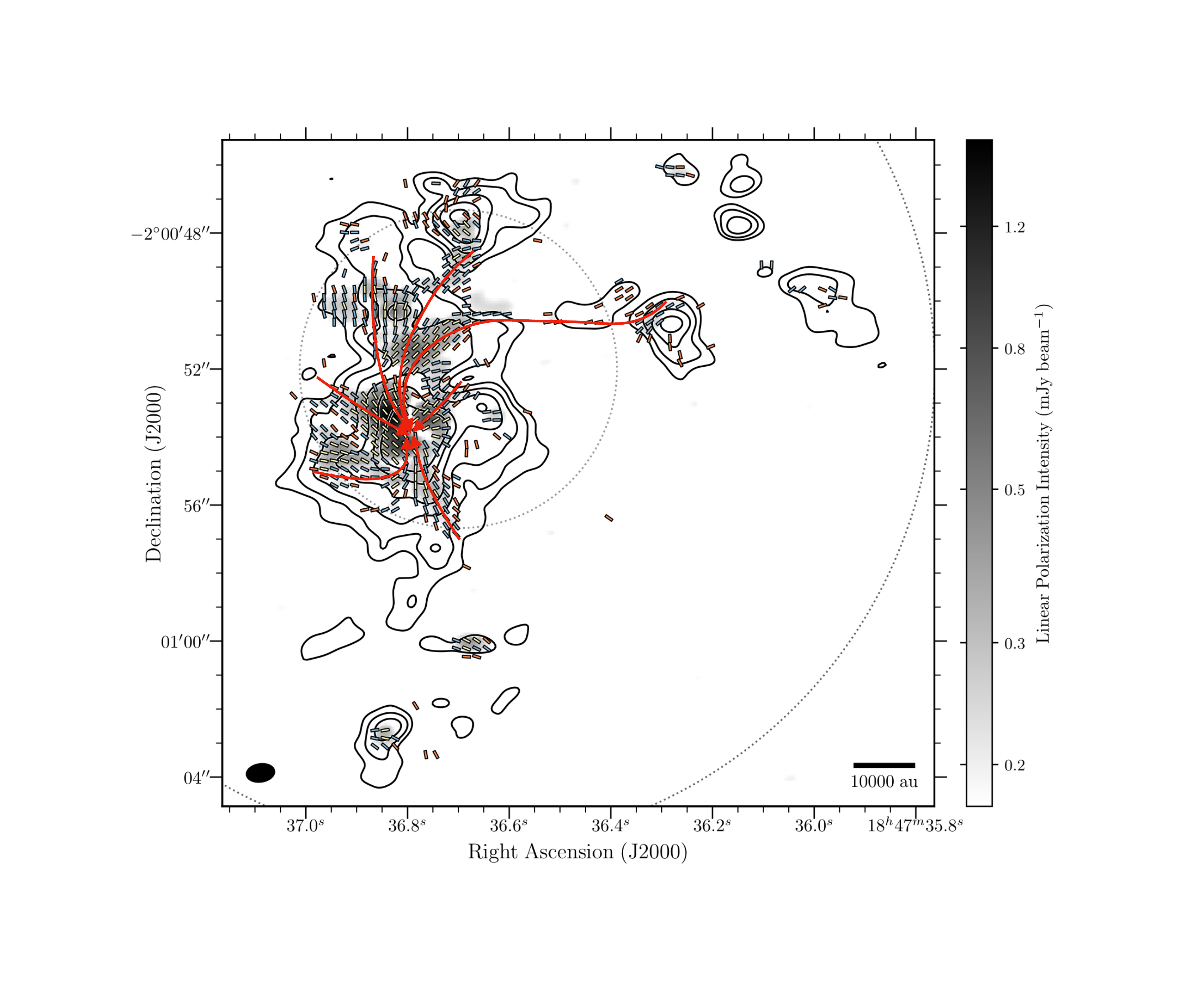}
\caption{Here we show the magnetic field map in MM2 as shown by Figure \ref{mm2}. Superposed is a field line
scheme, as red arrows, to illustrate the proposed field connectivity throughout the clump. Here we suggests that
the magnetic field is connected and likely dragged into the gravitational potential well of MM2-$A$.
\label{mm2Field}
}
\end{figure}

\subsubsection{On the nature of MM2-$A$}

 The MM2-$A$ core accounts for 73\% of the total estimated mass for all cores in MM2. The MM3-$A$ corresponds to 30\%,
and MM4-$A$ is 44\% of the total mass for all the cores in both clumps. Although, these are the most massive cores in 
our sample, MM2-$A$ is, by far, the most massive one. In fact, MM2-$A$ might be the most massive proto-stellar core 
in W43-Main now that MM1-$A$ has been resolved into a binary \citep{Motte2018a}. 
Having said that and given that at the 0.01 pc scales MM2-$A$ appears to be an unresolved and round proto-stellar core despite the elongated 
beam shape, it becomes interesting to understand how massive the resulting 
star can be. To estimate the final mass for a star in MM2-$A$, we follow \citet{Sanhueza2017} which  
made use of \citet{Larson2003} relation between
the maximum stellar mass and the mass of the whole cluster, or

\begin{equation}
\left(\frac{M_{\mathrm{max}}}{\Msun}\right) = 1.2\left(\frac{M_{\mathrm{cluster}}}{\Msun}\right)^{0.45},
\end{equation}

\noindent  here we derive the $M_{\mathrm{cluster}}$ from the MM2 total mass by making used of the current
W43-Main star formation rate. 
For the MM2 mass, the current best estimate is given to be 
$\left<M_{\mathrm{MM2}}\right> = 4.4 \times 10^{3}$ M$_{\odot}$ \citep{Bally2010}. Thus and assuming a star formation efficiency 
of 25\% \citep{Motte2003}, we obtain $M_{\mathrm{cluster}} = 1.1 \times 10^{3}$ \Msun\ which gives 
$M_{\mathrm{max}} = 28 \pm 14$ M$_{\odot}$. 
The uncertainty in the mass estimates comes, mostly, from the temperature determination, which 
can introduce a 50\% error in the estimation. Given the mini-starburst nature of W43-Main, the star formation rate may be higher 
than 25\% as estimated in other high-mass star forming regions \citep{Lada2003,Alves2007}. Thus, for a conservative increase to 
30\% in the star formation rate, we obtained a $M_{\mathrm{max}} = 30 \pm 15$ M$_{\odot}$. 
A 30 \Msun\ star will corresponds, likely, to an O7 star which is consistent with the spectral type of 
most of the stars in the W43 \HII\ region \citep{Blum1999,Heap2006}.
However and as previously stated,
the main caveat in the MM2-$A$ mass estimation is the temperature assumption. As with MM1-$A$, it is possible that
MM2-$A$ has already developed a hot core inside which will certainly increase the temperature also decreasing the
mass estimate. Following \citetalias{Cortes2016}, we also consider here T=70 K to estimate the mass for MM2-$A$
which  yields $M_{A} = 118$ M$_{\odot}$. This mass estimate corresponds to 27\% of the original estimate. Thus
if we assume the same decrease in the maximum mass previously obtained, we get  a $M_{\mathrm{max}} \sim 8$ M$_{\odot}$.
Although an 8 \Msun\ star is still a massive star, it is unlikely that MM2-$A$ will result in such star.
The current evidence suggests that accretion is still ongoing and despite the temperature uncertainty, the
final mass of the star might be close to the original estimate, or 30 M$_{\odot}$. It may very well be, that the
accretion time scale is larger than previously considered given the presence of strong magnetic fields.

 Pre-stellar cores are the smoking gun for the core accretion model
in high-mass star formation \citep{McKee2003}. These cores should be massive, cold condensations (T $<$ 20 K),
where the emission is purely thermal. Thus, a high-mass pre-stellar core has not developed hot cores or outflows, and
corresponds to the initial stage for high-mass star formation in the core accretion model.
The MM2-$A$ does not have a counterpart at 24 $\mu$m emission and radio continuum, which indicates
a proto-stellar cores in its earliest stages of evolution. Its massive gravitational potential well seems to
be dragging the field throughout the clump and into core. We elaborate further this proposal by sketching
the magnetic field connection in Figure \ref{mm2Field}. The red arrows indicate clump-scale field lines as they
are dragged into MM2-$A$. It is unlikely that a massive core that can drag the field will be in a
pre-stellar phase. Although at a coarse resolution and lower sensitivity, the SMA has detected perturbed 
magnetic fields in both high and intermediate star forming regions. \citet{Juarez2017} obtained 
a bimodal magnetic field distribution towards NGC 6334 V, which seem to converge towards a massive hot core.
Also a similar situation is seen in DR21(OH) core, where the field also seemed dragged by gravity towards the core,
though in this case the field appears to have a toroidal morphology as traced by SMA.
Additionally, outflow emission from MM2-$A$ can be explored from the $^{13}$CO  and C$^{18}$O
spectra shown in Figure \ref{coLine}. The
$^{13}$CO spectrum from MM2-$A$ at 0.27 pc scales ($10^{\prime \prime}$ angular scales) appears to show 
excess emission in both lines-wings with resembles an outflow.
The C$^{18}$O spectrum show no excess emission in the line-wings, but the line-shape is not Gaussian and suggests
some self-absorption, which makes the emission likely optically thick. Therefore, it is likely that MM2-$A$ 
is not pre-stellar, but it might be still quite early on in its evolution, and thus, further studies at higher
resolution will, hopefully, uncover its nature.


\subsection{The polarized intensity distribution}\label{sse:poli}

\begin{figure*}
\centering
\includegraphics[width=0.95\hsize]{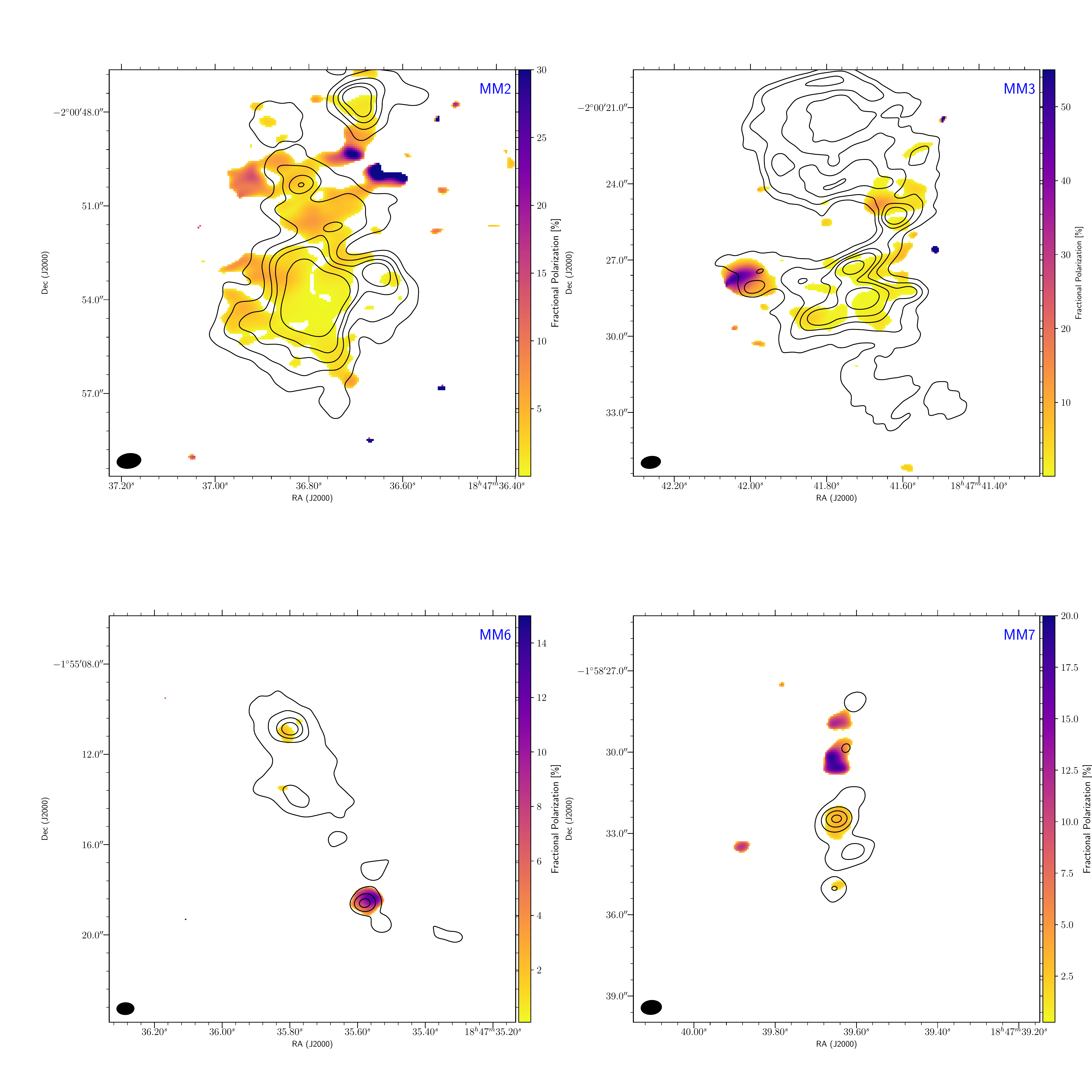}
\caption{\label{sipol} 
The Figure shows the fractional polarization for the MM2, MM3, MM6, and MM7 clumps as indicated by the blue label 
in upper right corner of each panel.
The fractional polarization map was calculated as the ratio between the polarized intensity map and the
Stokes I map, but using only the values greater or equal to $3\sigma_{\mathrm{poli}}$. Superposed to each 
fractional polarization map are  Stokes I contours (10, 20, 33.3, 50, 100) $\times$ $\mathrm{min}_{\mathrm{poli}}$, 
where $\mathrm{min}_{\mathrm{poli}} = 5\times\sqrt{\sigma_{\mathrm{poli}}}$. The values for $\sigma_{\mathrm{poli}}$
are for each clump, $\sigma_{\mathrm{poli},mm2} = 5.9\times10^{-5}$, $\sigma_{\mathrm{poli},mm3} = 5.2\times10^{-5}$,
$\sigma_{\mathrm{poli},mm2} = 5.3\times10^{-5}$, and $\sigma_{\mathrm{poli},mm2} = 5.6\times10^{-5}$ all in
units of Jy beam$^{-1}$.
}
\end{figure*}

A visual examination of the magnetic field maps from our clumps reveals
differences in the amount and distribution of the polarized emission. 
For example, the MM2 clump has the most intense and spread out amount of polarized emission,
with a peak intensity of 1.6 mJy beam$^{-1}$ and an extension of 0.27 pc along the main axis of the 
filament. In contrast, the MM6 clump presents only compact and almost 
unresolved emission at two places in the filament (cores $A$ and $B$ in Figure \ref{mm6}). 
The peak polarized emission in MM6 is 1.1 mJy beam$^{-1}$, which is of the same order as the peak emission in MM2. 
The MM6 filament is almost twice the size of MM2 (with an extension of 0.42 pc along the main axis), 
but it is almost devoid of polarized intensity even
at places where, from the Stokes I emission, we would have expected a detection (when compared to MM2). 
At the Stokes I lowest contour, which in both clumps is about 1.3 mJy beam$^{-1}$, we see significant polarized emission 
in MM2 but no emission in MM6. This is also the case at some of the highest contours in Stokes I in MM6. 

To explore this apparent discrepancy further, we produced fractional polarization maps of the most polarized clumps 
(MM2 and MM3) and the ones with the least amount of polarized emission (MM6 and MM7). The maps are shown in
Figure \ref{sipol}. Although the MM8 clump is also not significantly polarized, we did not consider it here as it
has the lower mass in our sample and thus, the lack of polarized emission,  could just be a sensitivity issue.
The maps shown in Figure \ref{sipol} were constructed by considering only the
$3\sigma_{\mathrm{poli}}$ or greater polarized intensity ( where
$\sigma_{\mathrm{poli}} = \sqrt{\sigma_{\mathrm{Q}}^{2} + \sigma_{\mathrm{U}}^{2}}$), 
which we used to create masks that were applied to  the fractional polarization maps. 
Over the fractional polarization images, we superposed Stokes I emission 
from each clump by defining sensitivity contours of expected fractional polarization.
To construct these contours, we assumed a minimum polarized intensity 
$\mathrm{min}_{\mathrm{poli}} = 5 \sigma_{\mathrm{poli}}$, which we later used to 
compute Stokes I levels of 10, 5, 3, 2, and 1\% expected polarization.  

The results suggests that, indeed, there are regions in MM6 and MM7, and also MM3, where we should have 
detected polarized emission, but we did not as seen by checking the expected polarization contours in 
Figure \ref{sipol}.
In MM6, the filament shows an unpolarized extension of about 6$^{\prime \prime}$ around
core MM6-$A$ and in MM3 the whole shell appears to be unpolarized. For MM7 the map suggests that we should have been
able to detect more polarized emission at the 15\% level around core MM7-$A$, but we only detect emission
at the 5 to 1\% levels over the core itself.
The origin of this lack of polarized emission can not be completely explained from these data.
However, there are some possibilities which we can explore.
The clumps where the polarized emission is significant are also the ones with the least number of cores detected (or with a lesser
fragmentation degree). In MM2, MM3, and MM4 we are detecting about 8
cores in each clump while for MM6, MM7, and MM8 we are detecting significantly more (13 cores in MM6, 21 in MM7, and 19 in MM8). 
It is clear that the fragmentation degree is higher in 
clumps where less polarization is detected. 
Although MM2 and MM3 are the most massive clumps in our sample, the MM4, MM6 and MM7 clumps are comparable in mass. 
This suggests that sensitivity to detected polarized emission is not an issue. 

A plausible explanation might be related to grain alignment efficiency. Besides the differences in the 
amount of polarized emission seen between the clumps, we see regions of strong and compact polarized dust emission
in almost all clumps. Noteworthy is the case of core MM3$-F$ where the dust appears to be highly polarized up to 51\%, though
this region is at the edge of the lowest Stokes I contour used in our maps (see Figure \ref{mm3}).
The current theoretical consensus suggests that grains are aligned by
radiative torques produced by radiation fields \citep{Lazarian2014,Lazarian2007}. 
Thus, it is possible that we are seeing the result of inner structures within these clumps with
preferred directions which allow efficient illumination of the dust in these regions.
This type of preferential illumination might increase the
efficiency in the alignment between the grains and the magnetic field. 
Now, it is harder to explain what we see in the regions where we do not see
polarized emission just from a decrease in the grain alignment efficiency. 
Specially considering the UC \HII\ region which we is associated with the shell in the MM3 clump (see section \ref{3se:mm3-cont}).
If preferential illumination increases grain alignment efficiency, the radiation field from this star should have yield 
detectable polarized emission from the MM3 shell. However, the shell is almost not polarized. 
In contrast, the MM4 UC \HII\ region seems to be located within highly polarized emitting dust.
\citet{Tang2009} mapped the UC \HII\ region G5.89-0.39 with the SMA, where they found significant linear
polarization coming from a dust ridge around the UC \HII\ region. It is interesting to note that this ridge is located
at one side respect to the star, which migth also be the result of preferential illumination by the star radiation field.
In our case, we see a small amount of polarized dust emission at the East side of the MM3 shell. The SMA
data cannot be directly compared with ours, but it is clear that the effect of an UC \HII\ region on the polarized
dust emission is not understood and deserves more investigation.

 Differences in clump fragmentation have also been observed in numerical simulations \citep{Hennebelle2018,Lee2017}.
These studies have suggested that magnetic fields play a significant role
in the fragmentation of filamentary structures in star forming regions. The simulations found that in
non-magnetized filaments the number of cores produces are in excess of what observations tell us. By introducing
magnetic fields in the simulations, excessive fragmentation is suppressed and the obtained core mass functions
are closer to what is seen in the data. 
Our B$_{\mathrm{pos}}$=1.7 mG estimate for MM7 is near the average value of the whole sample ($\left<\mathrm{B}_{\mathrm{pos}}\right>$=1.8 mG),
which makes it difficult to conclude unless the field outside the densest region is
tangled along the line of sight which will produce depolarization. 
If this is the case, the magnetic field might very well be a regulator of the number of proto-stellar cores formed and their initial masses.
Hence, weak fields remain a possibility and more observation are needed to understand these difference better.

\section{SUMMARY AND CONCLUSIONS}\label{se:sum}

 In this work, we have presented high spatial resolution ($\sim 0.01 $ pc) polarized dust emission observations of
six the most  massive clumps from W43-Main, MM2, MM3, MM4, MM6, MM7, and MM8 obtained with ALMA. The W43-MM1 results 
were presented in \citetalias{Cortes2016}, but their results were used in the virial analysis, see section \ref{virialAnalysis}.
The  ALMA data revealed clumps with filamentary morphologies, where a total of 81 cores were extracted
(96 when considering the MM1 cores).
We also derived detailed magnetic field morphologies from these clumps by assuming magnetic grain alignment between the 
ambient magnetic field and the dust grains. From these morphologies, we were able to estimate the
strength of the field and to evaluate the dynamical equilibrium of the cores detected here. Additionally, we
obtained single dish observations of the high density molecular gas tracers HCN$(J=4 \rightarrow 3)$ and 
HCO$^{+}(J=4 \rightarrow 3)$, and archival $^{13}$CO$(J=2 \rightarrow 1)$ and C$^{18}$O$(J=2 \rightarrow 1)$ observations
to investigate the large scale kinematics of these clumps. Detailed results are as follows,

\begin{enumerate}

\item We have found that the clumps mapped with ALMA can be divided into two population of filaments. One where the number of 
extracted cores is larger suggesting increased fragmentation, and the other where the fragmentation is 
lower and the core distribution seems dominated by a single-massive core.
We identify the first population with the MM6, MM7, and MM8 clumps and the second with the MM2, MM3, and MM4 clumps. 
The  MM2-$A$ proto-stellar core is the most massive in our sample, with an estimated mass of 425.7 M$_{\odot}$.
The core remains unresolved, but we our calculations suggest that its evolution will result in a 
30 \Msun\ O7 type star. 

\item Two of our clumps, MM3 and MM4, are associated with UC H {\small II} regions. The MM3 clump
shows two cavities, or a shell, which are not directly  associated with any of the cores detected here. However,
the G030.720-00.082 UC H {\small II} region is located within this shell and it is also seems to be associated with 
the IRAC sources also located within the shell. As such, it is plausible that this shell is the result of
the expanding shock and ionization fronts from  UC H {\small II} in MM3.
 
\item From the dust polarized emission, we derived the magnetic field morphologies from MM2, MM3, MM4, and MM7.
The field morphology from these clumps is quite ordered and spread out over significant portions of the filaments.
Towards most of the massive cores, the field lines appear to have a projected radial patterns 
suggesting that gravity dominates over the field. However, over the filaments and away from the massive cores, we
found that the field pattern appears to be along the filament with low EVPA dispersions, which suggests that
magnetic tension is important.

\item We analyzed single dish spectra from HCN$(J=4 \rightarrow 3)$ and HCO$^{+}(J=4 \rightarrow 3)$ looking for infalling motions.
The blue asymmetry in the spectral profile from both lines towards MM2 and MM3 strongly suggest infalling.
We quantified this asymmetry by computing the normalized velocity difference and fitting a simple infalling model to the line
profile. We found consistent evidence supporting infall from scales of $\sim 0.3 $ pc, obtaining infall rates of
$\dot{\mathrm{M}}_{\mathrm{MM2}} = 1.2 \times 10^{-2}$ \Msun\ yr$^{-1}$  and 
$\dot{\mathrm{M}}_{\mathrm{MM3}} = 6.3 \times 10^{-3}$ \Msun\ yr$^{-1}$, large enough to overcome radiative feedback and form massive stars. 

\item From the dust polarized emission we estimated the magnetic field strength onto the plane of the sky,
or B$_{\mathrm{pos}}$, using the original Davis-Chandrashekar-Fermi method with two additional derivations.
We found average values for the field from 500 $\mu$G to a few mG (with a $\left<\mathrm{B}_{\mathrm{pos}}\right>$=1.8 mG
for the whole sample). 
The strongest value was estimated towards MM3-$F$ where we found $\left<\mathrm{B}_{\mathrm{pos}}\right> = 4.2$ mG.	
We used these estimations to analyze the dynamical equilibrium of the cores by calculating
the total virial parameter, $\alpha_{\mathrm{total}}$, considering both turbulence and magnetic fields.
We found 12, out of 17, bound cores and 5 unbound cores.
We increased our statistics by assuming a B$_{\mathrm{pos}}$
value for the remaining cores. Here, we found that about half of our cores seem to be unbound, and where these cores are
also the least massive ones. Given that these cores are likely pressure-confined, we conclude that they should still be accreting
to become self-gravitating at a later stage. Therefore, this strongly suggests different evolutionary 
time-scales between the bound and unbound cores and thus the star formation time-scales in W43-Main might not be uniform.

\item We finally considered the difference in the distribution and strength of the polarized emission between 
the two population of clumps in our sample.
The MM2, MM3, and MM4 are highly polarized, while the MM6, MM7, and MM8 are not.
We compared the fractional polarization maps with the Stokes I emission using ``expected'' contours where
polarized emission should have been detected (with the exception of MM8 where issue seem to be 
sensitivity).
Thus, we found that in the MM6 and MM7 clumps the Stokes I emission was sufficient for succesful detections of polarized emission. 
But and as previously said, these clumps are largely devoid of it. Also, the MM6, MM7, and MM8
clumps show significantly more fragmentation compared to the other clumps.
Non ideal MHD mumerical simulation have
suggested that the magnetic field has a direct impact in the degree of fragmentation seen in their simulated clumps.
It is plausible that weaker magnetic fields in these clumps may explain this difference, but, certainly, more data is needed to confirm this.

\end{enumerate}

\begin{acknowledgements}
We acknowledge the useful comments made by the anonymous referee, which certainly improved the results shown by this article.\\
The National Radio Astronomy Observatory is a facility of the National Science Foundation operated under cooperative agreement by Associated Universities, Inc.\\
This paper makes use of the following ALMA data: ADS/JAO.ALMA\#.2013.1.00725.S. ALMA is a partnership of ESO (representing its member states), NSF (USA) and NINS (Japan), together with NRC (Canada) and NSC and ASIAA (Taiwan), in cooperation with the Republic of Chile. The Joint ALMA Observatory is operated by ESO, AUI/NRAO and NAOJ. \\
J.M.G. acknowledges support from the Spanish AYA2017-84390-C2-2-R grant and the support from the Joint ALMA Observatory Visitor Program.\\
C.L.H.H. acknowledges the support of both the NAOJ Fellowship as well as JSPS KAKENHI grant 18K13586.\\
Z.Y.L. is supported in part by NASA grant 80NSSC18K1095 and NNX14B38G and NSF grant AST-1815784 and 1716259.
S.P.L. thanks the support of the Ministry of Science and Technology (MoST) of Taiwan through
Grants NSC 98-2112-M-007-007-MY3, NSC 101-2119-M-007-004, and MoST 102-2119-M-007-004- MY3.\\
C.O. acknowledges the NRAO's Office of Diversity and Inclusion for its financial support of the undergraduate research experience program for Chilean students.\\
\software{CASA \citep[v4.7][]{McMullin2007}, NEWSTAR \citep{Ikeda2001}, APLpy \citep{Robitaille2012}}

\end{acknowledgements}

\appendix

\section{THE DCF METHOD}\label{ape1}

 Assuming perfect alignment between the ambient magnetic field and dust grains, the magnetic
morphology onto the plane of the sky can be inferred from the polarized dust emission by rotating
the electric vector position angle (EVPA) by 90$^{\circ}$ \citep[For a review see][]{Lazarian2007}.
However, polarized emission from dust can only provide morphological information about the field,
but not a direct measurement of its strength. Nevertheless, the field strength
 can be estimated from the projected, onto the plane of the sky, field lines pattern by
using the Davis, Chandrasekhar, \& Fermi technique and its derivatives
\citep[here after DCF][]{Davis1951,Chandrasekhar1953}. The usual DCF technique can be expressed as follows,

\begin{equation}
    \label{cf}
    B_{\mathrm{pos}} = 9.3 \frac{\sqrt{n_{\mathrm{H_{2}}}}\Delta V}{\delta \phi}
\end{equation}

 where $n_{\mathrm{H_{2}}}$ is the molecular Hydrogen number density, the $\Delta V$ represent 
the velocity dispersion in the gas, and $\delta \phi$ is the EVPA dispersion,
which representats the perturbations in the magnetic field lines.
A number of extension, or derivatives, to the technique have been suggested to improve and account 
for the caveats of the original DCF technique.
\citet{Heitsch2001} attempted to address the
limitation of the small angle approximation by replacing $\delta\phi$ by $\delta \tan(\phi)$ which is
calculated locally and by adding a geometric correction to avoid underestimating the field in the super-Alfvenic case.
In contrast, \citet{Falceta2008} assumed that the field perturbation is a global property and thus, they
replaced $\delta \phi$ by $\tan(\delta \phi) ~\sim \delta B/B_{sky}$ in the denominator of equation \ref{cf}. These two 
extensions can be written as,

\begin{equation}
B_{\mathrm{pos}} = Q\frac{\sqrt{n_{\mathrm{H_{2}}}}\Delta V}{\tan{\delta \phi}} 
\end{equation}
\begin{equation}
B_{\mathrm{pos}} = Q\frac{\sqrt{n_{\mathrm{H_{2}}}}\Delta V}{\delta \tan{\phi}}\left(1 + 3(\delta \tan{\phi})^{2}\right)^{1/4}
\end{equation}

where $Q$ is an scaling factor.
All the parameters used in the three calculation have large uncertainties, where,
ideally, these parameters should trace similar spatial scales. A main assumption behind the DCF method is that the field perturbations are produced by the gas non-thermal motions (turbulence). in fact, numerical simulations by \citet{Ostriker2001} showed that $\delta \phi < 25^{\circ}$ values are able to reproduce the field strengths obtained from the simulations. However, the ALMA data is showing regions where $\delta \phi > 25^{\circ}$ and where also the field morphology suggests that gravity might be behind the increased $\delta \phi$ values as the infalling gas might be also pulling the field along into the potential well of the core. Therefore, it is likely that the 
field strength is being under-estimated in those regions.

To trace the gas non-thermal motions, data
from an optically thin and  high density molecular line tracer is required. At the clump scales, we have
obtained HCN$(J=4 \rightarrow 3)$ and HCO$^{+}(J=4 \rightarrow 3)$ data for MM2 and MM3, but
the H$^{13}$CO$^{+}(J=3 \rightarrow 2)$ observations from \citet{Motte2003} are also available in the literature.
However, the HCN$(J=4 \rightarrow 3)$ and HCO$^{+}(J=4 \rightarrow 3)$ will be shown to be
to be optically thick (see section \ref{2se:lineEm}) and therefore cannot be used reliably
to trace the gas turbulent motions. But, the H$^{13}$CO$^{+}(J=3 \rightarrow 2)$ emission
is, likely, optically thin for some of our clumps. The resolution of these data is
$\sim 26^{\prime \prime}$, or $\sim 0.7$ pc,
which as a first approximation, can be used to derive $\Delta V$ at the clump scale.  Nonetheless, we can
estimate $\Delta V$ at smaller scales
if we assume a Kolmogorov type turbulent power spectrum. A usual expression for
a Kolmogorov type turbulence power spectrum is given by $\sigma_{\mathrm{v}}^{2} = bL^{n}$.
Here, $\sigma_{\mathrm{v}}$ corresponds to the velocity dispersion
of a molecular line as expected to be from the turbulence energy cascade at lenght-scale $L$. To obtain the
powerlaw we are require to know $b$, a scaling constant, and $n$ the power law exponent.
\citet{Li2008} derived $n=0.36$ from their M17 emission, which is a well studied high-mass star forming region.
Unfortunately, we do not have sufficient single dish data to fit the power law and derive the parameters ourselves. Nevertheless,
the values derived from M17 by \citet{Li2008} are in concordance what is seen in other high-mass star forming regions
\citep{Tang2018}, and an acceptable representation for the non-thermal motions in W43-Main.
In fact, molecular line observations of C$^{18}$O(2-1) at similar length-scales
from G28.34, reveal velocity dispersions $\sigma_{\mathrm{C^{18}O}} \sim 0.7$ km s$^{-1}$ \citep{Zhang2015}.
These values are similar to  our estimate, though C$^{18}$O(2-1) traces
a more diffuse environment than H$^{13}$CO$^{+}$(3-2). The C$^{18}$O(2-1) line-width can be enlarged
by additional turbulence added to the gas by outflows and infall, which is indeed true for W43-Main clumps where there is evidence
for infalling motions and outflows have been detected.
 The H$^{13}$CO$^{+}$($3 \rightarrow 2)$ data show a large spread in the line-widths, from 3.6 to 6.3 km/s for MM4 and MM8
respectively with values in between for the other clumps \citep[see Table 2 in][]{Motte2003}. Starting
from clump-scales, we can obtain the $b$ values for each clump, as $b = \sigma_{\mathrm{H}^{13}\mathrm{CO}^{+}}^{2}/L^{n}$.
Later the estimation for the velocity dispersion, $\sigma_{\mathrm{v}}$, can be obtained by using the relevant angular scale,
where the line-width, $\Delta V$, is given by multiplying by the FWHM value, or $\Delta V = 2\sqrt{2\log{2}}\sigma_{\mathrm{v}}$.
The derived values of $b$ are listed in Table \ref{vdTab}.

 The dispersion of the EVPA, $\delta \phi$, requires closer attention.
We use the standard deviation of the EVPA values from the polarization map as our estimator of the dispersion.
If we assume that the distribution of EVPA values is Gaussian, 10 beam independent points
will give us a dispersion value within 15\% from the true dispersion. Here we defined an independent point
as a Nyquist sampled cell in the map. This criteria gives a minimum threshold to use the
polarization data in the B$_{\mathrm{pos}}$ estimation. However, the exquisite sensitivity of ALMA is giving
us a much larger number of independent points which decreases the uncertainty significantly.
We use this threshold to establish a criteria to define regions where the field shows an apparent coherence.
We do this by requiring continuity in the field lines, or a smooth variation in EVPA, from sub-regions where
we have 10 independent points, the minimum threshold needed to estimate $\delta \phi$. Therefore, a region
where we  estimate B$_{\mathrm{pos}}$ is composed of continuous EVPA measurements. We justify this as  the underlying
assumptions in the DCF method are that the field in frozen in the fluid and the perturbations in the Alfven waves happen
at small-scales. Therefore, we cannot consider the dispersion in the field from a whole filament unless its
dispersion is small.
Based on these considerations,
the clumps where the polarized emission is sufficient to satisfy our constraints are MM2, MM3, MM4, from
population 2  and only MM7 from population 1.
As previously identified, the remaining clumps from population 1 (MM6 and MM8) have well defined regions with polarized emission,
but not sufficient independent points. We will discuss this
later in section \ref{sse:poli}. Thus, we devided the polarized emission in regions \citepalias[as done in ][]{Cortes2016} where the
field appears to be continuous and calculated the required parameters for the region lenght-scale (see section \ref{2se:MF} for details).

Finally, the number density is estimated by computing the column density for the region and assuming spherical geometry. 
In this way, we are accounting for perturbation by the ambient gas in the field lines which provides consistency in the 
calculation. However, this field strength is not necesarilly representative of core field strengths as B$ \sim \sqrt{\rho}$
\citep{Ciolek1995b} or B$ \sim \rho^{2/3}$ \citep{Mestel1966}. Either case, the total field strength might the larger
than the values we are estimating here.

\section{INFALLING METHODS}\label{ape2}

The ratio between the difference in the velocities, associated with the
peak emission between the optically thick and the optically thin lines, respect to the line-width
from the optically thin emission is known as the the normalized line velocity difference \citep{Mardones1997},

\begin{equation}
\Delta V_{\mathrm{be}} = \frac{V_{\mathrm{thick}}^{\mathrm{max}} - V_{\mathrm{thin}}^{\mathrm{max}}}{\Delta V_{\mathrm{thin}}}
\label{dv}
\end{equation}

 where $V_{\mathrm{thick}}^{\mathrm{max}}$ is the velocity from the peak emission of the optically think line,
$V_{\mathrm{thin}}^{\mathrm{max}}$ is the velocity from the peak emission of the optically thin line, and
$\Delta V_{\mathrm{thin}}$ is the line-width from the optically thin molecular tracer. A negative $\Delta V_{\mathrm{be}}$
will indicate blue-shifted gas velocities, which is suggestive of infalling motions. 

The normalized velocity difference is not sufficient to characterize infalling motions.
A more detailed model is needed to derive the flow parameters especially the mass infall rate.
Such model was developed by \citet{Myers1996} and \citet{DiFrancesco2001} where the clump is
approximated by two infalling gas layers,
a front and a rear layer, with a central source simulating the pre-stellar core.
Thus, the observed brightness temperature is quantified by the following equation, where the subscripts “f”, “r”, “C”, and “cmb”
stand for the front layer, the rear layer, the central source, and the cosmic background emission.
The model is described by the following equation;

\begin{equation}
\label{model}
\Delta T_{\mathrm{B}} = (J_{\mathrm{f}} - J_{\mathrm{C}})[1 - e^{-\tau_{\mathrm{f}}}] +
(1-\Phi)(J_{\mathrm{r}} - J_{\mathrm{cmb}})[1 - e^{-(\tau_{\mathrm{r}} + \tau_{\mathrm{f}})}].
\end{equation}

 The main terms in this equation are
  the Planck excitation temperature given by $J_{\mathrm{i}} = 
T_{0}/[\exp{(T_{0}/T_{\mathrm{i}})} - 1]$ with $T_{0} = h\nu/k$, and $T_{\mathrm{i}}$ corresponding to
either $T_{\mathrm{f}}$, $T_{\mathrm{r}}$, $T_{\mathrm{c}}$, and $T_{\mathrm{cmb}}$. Also,
$J_{\mathrm{C}} = \Phi J_{\mathrm{c}} + (1 - \Phi)J_{\mathrm{r}}$ where $\Phi$ is the beam
filling factor of the continuum source, which due to the large beam size of the ASTE telescope is assumed to be 0.
The $\tau_{\mathrm{i}}$ expressions correspond to the optical depths, which
we assumed to be Gaussian.  Thus and following \citet{Myers1996}, the front and rear optical depths are given by

\begin{eqnarray}
\tau_{\mathrm{f}} = \tau_{0} \exp{\left[ \frac{-(v - v_{\mathrm{f}} - v_{\mathrm{lsr}})^2}
{2\sigma^2} \right]} \\
\tau_{\mathrm{r}} = \tau_{0} \exp{\left[ \frac{-(v + v_{\mathrm{r}} - v_{\mathrm{lsr}})^2}
{2\sigma^2} \right]},
\end{eqnarray}

 where $\tau_{0}$ is the peak optical depth for both
the front and the rear layers, $v_{\mathrm{f}}$ and  $v_{\mathrm{r}}$
are the infalling velocity for both slabs and $\sigma$ is the velocity dispersion.

\newpage

\begin{longrotatetable}
\startlongtable
\begin{deluxetable}{c c c c c c c c c c c c c c c}        
\tablecolumns{15}
\tablewidth{0pt}
\tabletypesize{\scriptsize}
\tablecaption{The Table shows the physical parameters for each of the cores that were detected. 
Additionally, we are showing here the Jeans length ($\lambda_{J}$)  and mass (M$_{J}$), 
as well as the kinetic mass (M$_{\mathrm{kin}}$) values for each core\label{mmTab}.}
\tablehead{
\colhead{Clump} &
\colhead{Core} & 
\colhead{R.A.} &
\colhead{DEC} & 
\colhead{Peak} & 
\colhead{Flux} & 
\colhead{Major\tablenotemark{a}} & 
\colhead{Minor\tablenotemark{a}} & 
\colhead{P.A.} & 
\colhead{n\tablenotemark{b}} & 
\colhead{Mass} & 
\colhead{Size\tablenotemark{c}} & 
\colhead{$\lambda_{J}$} &
\colhead{M$_{J}$} &
\colhead{M$_{T}$}  \\
\colhead{ } & 
\colhead{ } & 
\colhead{[J2000]} & 
\colhead{[J2000]} & 
\colhead{[mJy beam$^{-1}$]} & 
\colhead{[mJy]} & 
\colhead{[arcsec]} & 
\colhead{[arcsec]} & 
\colhead{[deg]} & 
\colhead{[$10^{7}$ cm$^{-3}$ ]} & 
\colhead{[\Msun]} & 
\colhead{[mpc]} & 
\colhead{[mpc]} & 
\colhead{[\Msun]} &
\colhead{[\Msun]} \\
}
\startdata
    & A & 18:47:36.79 &  -02:00:54.15 & 354.1 $\pm$ 17.8 & 879.8 & 0.79 & 0.63 & 149.80 & 283.7 & 425.7 & 18.82 & 0.60 & 0.02 & 6.03\\
    & B & 18:47:36.69 &  -02:00:47.63 & 27.4 $\pm$ 2.1 & 79.7 & 0.95 & 0.64 & 25.57 & 19.3 & 38.6 & 20.72 & 2.32 & 0.07 & 6.64\\
    & C & 18:47:36.28 &  -02:00:50.69 & 17.7 $\pm$ 1.2 & 32.8 & 0.57 & 0.47 & 20.16 & 26.4 & 15.9 & 13.87 & 1.98 & 0.06 & 4.44\\
    & D & 18:47:36.66 &  -02:00:53.22 & 21.4 $\pm$ 4.2 & 87.1 & 1.21 & 0.90 & 77.42 & 8.7 & 42.1 & 27.80 & 3.44 & 0.11 & 8.90\\
MM2 & E & 18:47:36.15 &  -02:00:47.76 & 12.9 $\pm$ 0.7 & 17.4 & 0.44 & 0.17 & 43.31 & 94.5 & 8.4 & 7.34 & 1.05 & 0.03 & 2.35\\
    & F & 18:47:36.82 &  -02:00:50.30 & 13.6 $\pm$ 1.4 & 46.4 & 0.99 & 0.82 & 3.02 & 7.2 & 22.4 & 24.03 & 3.79 & 0.12 & 7.70\\
    & G & 18:47:36.84 &  -02:01:2.60 & 14.0 $\pm$ 1.0 & 28.2 & 0.77 & 0.40 & 143.18 & 19.1 & 13.7 & 14.71 & 2.33 & 0.07 & 4.71\\
    & H & 18:47:36.14 &  -02:00:46.54 & 6.1 $\pm$ 0.6 & 9.7 & 0.57 & 0.33 & 132.75 & 13.6 & 4.7 & 11.54 & 2.76 & 0.09 & 3.70\\
    & I & 18:47:35.99 &  -02:00:49.75 & 6.0 $\pm$ 0.7 & 43.3 & 1.90 & 1.12 & 55.18 & 1.6 & 20.9 & 38.98 & 8.11 & 0.25 & 12.49\\
    & J & 18:47:36.74 &  -02:00:46.64 & 4.3 $\pm$ 1.7 & 9.2 & 0.91 & 0.37 & 64.82 & 5.3 & 4.5 & 15.53 & 4.42 & 0.14 & 4.97\\
\\
\hline
\\
    & A & 18:47:41.71 &  -02:00:28.51 & 76.1 $\pm$ 5.5 & 149.9 & 0.72 & 0.49 & 98.23 & 67.4 & 59.4 & 15.77 & 1.34 & 0.05 & 5.19\\
    & B & 18:47:41.73 &  -02:00:27.29 & 39.2 $\pm$ 3.7 & 84.8 & 0.94 & 0.41 & 120.45 & 32.9 & 33.6 & 16.57 & 1.92 & 0.07 & 5.45\\
    & C & 18:47:41.62 &  -02:00:25.25 & 20.8 $\pm$ 1.7 & 51.7 & 0.78 & 0.66 & 125.23 & 12.9 & 20.5 & 19.20 & 3.07 & 0.11 & 6.32\\
MM3 & D & 18:47:41.60 &  -02:00:28.29 & 12.6 $\pm$ 1.5 & 35.0 & 1.15 & 0.55 & 93.33 & 6.4 & 13.9 & 21.31 & 4.36 & 0.16 & 7.01\\
    & E & 18:47:41.81 &  -02:00:29.21 & 18.4 $\pm$ 2.4 & 104.8 & 1.96 & 0.85 & 107.35 & 4.5 & 41.6 & 34.47 & 5.19 & 0.19 & 11.34\\
    & F & 18:47:41.99 &  -02:00:28.04 & 8.7 $\pm$ 0.8 & 16.6 & 0.69 & 0.48 & 100.28 & 8.3 & 6.6 & 15.26 & 3.83 & 0.14 & 5.02\\
    & G & 18:47:40.97 &  -02:00:20.67 & 9.9 $\pm$ 0.7 & 19.6 & 0.63 & 0.54 & 86.17 & 9.2 & 7.8 & 15.55 & 3.64 & 0.13 & 5.12\\
    & H & 18:47:41.86 &  -02:00:27.83 & 9.1 $\pm$ 0.2 & 58.8 & 1.79 & 1.09 & 82.19 & 2.0 & 23.3 & 37.27 & 7.78 & 0.28 & 12.27\\
\\
\hline
\\
    & A & 18:47:38.44 &  -01:57:42.62 & 22.6 $\pm$ 3.1 & 159.6 & 2.13 & 1.18 & 74.89 & 3.6 & 60.6 & 42.16 & 5.92 & 0.22 & 8.59\\
    & B & 18:47:38.70 &  -01:57:45.12 & 11.2 $\pm$ 0.9 & 53.0 & 1.31 & 1.16 & 127.05 & 2.5 & 20.1 & 32.89 & 7.07 & 0.27 & 6.70\\
    & C & 18:47:38.10 &  -01:57:50.61 & 7.0 $\pm$ 0.6 & 17.4 & 1.13 & 0.52 & 88.89 & 3.4 & 6.6 & 20.53 & 6.10 & 0.23 & 4.18\\
MM4 & D & 18:47:37.75 &  -01:57:51.40 & 7.6 $\pm$ 0.3 & 21.1 & 1.15 & 0.59 & 57.31 & 3.3 & 8.0 & 22.03 & 6.14 & 0.23 & 4.49\\
    & E & 18:47:38.25 &  -01:57:40.73 & 6.7 $\pm$ 0.4 & 22.7 & 1.55 & 0.60 & 73.94 & 2.3 & 8.6 & 25.66 & 7.45 & 0.28 & 5.23\\
    & F & 18:47:38.56 &  -01:57:43.59 & 12.2 $\pm$ 0.1 & 64.0 & 1.64 & 1.07 & 106.15 & 2.5 & 24.3 & 35.28 & 7.15 & 0.27 & 7.18\\
    & G & 18:47:37.97 &  -01:57:50.35 & 5.4 $\pm$ 0.1 & 41.3 & 2.20 & 1.19 & 32.91 & 0.9 & 15.7 & 43.15 & 12.04 & 0.45 & 8.79\\
    & H & 18:47:38.86 &  -01:57:41.97 & 3.6 $\pm$ 0.0 & 22.2 & 1.55 & 1.40 & 84.38 & 0.6 & 8.4 & 39.18 & 14.20 & 0.54 & 7.98\\
\\
\hline
\\
    & A & 18:47:35.80 &  -01:55:10.86 & 20.5 $\pm$ 1.5 & 52.0 & 0.87 & 0.70 & 78.28 & 11.1 & 22.7 & 20.89 & 3.19 & 0.11 & 5.99\\
    & B & 18:47:35.58 &  -01:55:18.58 & 9.9 $\pm$ 1.4 & 23.0 & 0.80 & 0.62 & 27.08 & 6.8 & 10.0 & 18.67 & 4.05 & 0.14 & 5.35\\
    & C & 18:47:36.52 &  -01:55:19.69 & 11.7 $\pm$ 0.2 & 13.3 & 0.24 & 0.20 & 15.77 & 129.2 & 5.8 & 5.84 & 0.93 & 0.03 & 1.67\\
    & D & 18:47:35.53 &  -01:55:19.51 & 5.3 $\pm$ 0.1 & 9.7 & 0.63 & 0.48 & 25.24 & 5.9 & 4.2 & 14.74 & 4.36 & 0.15 & 4.22\\
    & E & 18:47:35.55 &  -01:55:17.07 & 4.0 $\pm$ 0.9 & 19.0 & 1.67 & 0.89 & 116.92 & 1.1 & 8.3 & 32.48 & 10.21 & 0.34 & 9.31\\
MM6 & F & 18:47:36.01 &  -01:55:20.53 & 1.8 $\pm$ 0.2 & 3.1 & 0.73 & 0.36 & 75.65 & 2.4 & 1.4 & 13.65 & 6.88 & 0.23 & 3.91\\
    & G & 18:47:35.33 &  -01:55:19.93 & 3.5 $\pm$ 0.4 & 25.9 & 2.20 & 1.16 & 73.59 & 0.6 & 11.3 & 42.60 & 13.15 & 0.44 & 12.21\\
    & H & 18:47:35.85 &  -01:55:15.44 & 2.9 $\pm$ 0.3 & 16.5 & 1.57 & 1.16 & 128.28 & 0.7 & 7.2 & 36.03 & 12.80 & 0.43 & 10.33\\
    & I & 18:47:35.34 &  -01:55:21.99 & 2.2 $\pm$ 0.0 & 7.7 & 1.31 & 0.67 & 140.76 & 1.0 & 3.4 & 25.02 & 10.85 & 0.37 & 7.17\\
    & J & 18:47:35.41 &  -01:55:22.82 & 2.2 $\pm$ 0.1 & 8.4 & 1.19 & 0.93 & 151.89 & 0.7 & 3.7 & 28.06 & 12.35 & 0.42 & 8.04\\
    & K & 18:47:36.18 &  -01:55:20.52 & 1.6 $\pm$ 0.2 & 4.4 & 1.15 & 0.59 & 81.71 & 0.8 & 1.9 & 22.02 & 11.89 & 0.40 & 6.31\\
    & L & 18:47:35.78 &  -01:55:13.88 & 6.5 $\pm$ 0.1 & 62.2 & 2.43 & 1.39 & 58.72 & 1.0 & 27.1 & 48.91 & 10.44 & 0.35 & 14.02\\
    & M & 18:47:35.66 &  -01:55:15.74 & 3.1 $\pm$ 0.0 & 19.5 & 1.79 & 1.18 & 120.81 & 0.6 & 8.5 & 38.77 & 13.16 & 0.44 & 11.11\\
\\
\hline
\\
    & A & 18:47:39.65 &  -01:58:32.47 & 15.4 $\pm$ 0.8 & 26.9 & 0.57 & 0.49 & 131.27 & 15.6 & 11.7 & 14.05 & 2.23 & 0.06 & 3.91\\
    & B & 18:47:39.11 &  -01:58:39.98 & 12.3 $\pm$ 0.8 & 27.1 & 0.79 & 0.53 & 158.24 & 8.5 & 11.8 & 17.27 & 3.03 & 0.08 & 4.81\\
    & C & 18:47:39.61 &  -01:58:33.67 & 7.7 $\pm$ 1.2 & 22.9 & 1.14 & 0.67 & 114.18 & 3.0 & 10.0 & 23.18 & 5.12 & 0.14 & 6.46\\
    & D & 18:47:39.65 &  -01:58:35.04 & 5.8 $\pm$ 0.3 & 11.7 & 0.68 & 0.52 & 3.36 & 4.6 & 5.1 & 15.94 & 4.09 & 0.11 & 4.44\\
    & E & 18:47:39.80 &  -01:58:23.00 & 5.8 $\pm$ 0.1 & 7.2 & 0.33 & 0.27 & 39.47 & 23.1 & 3.1 & 7.95 & 1.83 & 0.05 & 2.22\\
    & F & 18:47:40.18 &  -01:58:38.78 & 5.7 $\pm$ 0.3 & 11.9 & 0.68 & 0.59 & 127.28 & 3.9 & 5.2 & 17.00 & 4.47 & 0.13 & 4.74\\
    & G & 18:47:40.19 &  -01:58:30.95 & 3.3 $\pm$ 0.2 & 4.0 & 0.39 & 0.20 & 80.71 & 15.2 & 1.7 & 7.50 & 2.26 & 0.06 & 2.09\\
    & H & 18:47:39.60 &  -01:58:28.16 & 3.8 $\pm$ 0.3 & 17.0 & 1.35 & 1.00 & 125.13 & 0.9 & 7.4 & 31.07 & 9.21 & 0.26 & 8.65\\
    & I & 18:47:38.50 &  -01:58:21.17 & 10.7 $\pm$ 0.9 & 21.7 & 0.75 & 0.54 & 87.22 & 7.3 & 9.5 & 16.85 & 3.26 & 0.09 & 4.69\\
    & J & 18:47:39.47 &  -01:58:30.78 & 2.6 $\pm$ 0.2 & 8.3 & 1.15 & 0.65 & 15.99 & 1.1 & 3.6 & 22.95 & 8.39 & 0.24 & 6.39\\
MM7 & K & 18:47:40.20 &  -01:58:40.30 & 5.6 $\pm$ 0.6 & 27.7 & 1.55 & 0.90 & 2.00 & 1.4 & 12.1 & 31.53 & 7.39 & 0.21 & 8.78\\
    & L & 18:47:39.63 &  -01:58:29.85 & 2.9 $\pm$ 0.2 & 13.1 & 1.56 & 0.80 & 157.22 & 0.8 & 5.7 & 29.75 & 9.86 & 0.28 & 8.29\\
    & M & 18:47:40.23 &  -01:58:36.58 & 2.7 $\pm$ 0.2 & 6.5 & 0.95 & 0.57 & 93.20 & 1.4 & 2.9 & 19.53 & 7.40 & 0.21 & 5.44\\
    & N & 18:47:39.89 &  -01:58:33.39 & 1.9 $\pm$ 0.1 & 4.9 & 1.29 & 0.36 & 59.49 & 1.3 & 2.1 & 18.10 & 7.64 & 0.21 & 5.04\\
    & O & 18:47:40.17 &  -01:58:31.81 & 2.1 $\pm$ 0.4 & 7.2 & 1.06 & 0.81 & 165.71 & 0.8 & 3.1 & 24.76 & 10.10 & 0.28 & 6.90\\
    & P & 18:47:39.44 &  -01:58:31.73 & 2.4 $\pm$ 0.1 & 11.7 & 1.76 & 0.83 & 63.42 & 0.6 & 5.1 & 32.11 & 11.66 & 0.33 & 8.94\\
    & Q & 18:47:39.09 &  -01:58:42.12 & 2.6 $\pm$ 0.3 & 10.8 & 1.65 & 0.56 & 6.57 & 1.0 & 4.7 & 25.53 & 8.61 & 0.24 & 7.11\\
    & R & 18:47:38.89 &  -01:58:20.15 & 3.3 $\pm$ 0.2 & 7.8 & 0.84 & 0.60 & 141.21 & 1.9 & 3.4 & 18.95 & 6.47 & 0.18 & 5.28\\
    & S & 18:47:38.90 &  -01:58:20.84 & 2.3 $\pm$ 0.1 & 8.5 & 1.20 & 0.86 & 132.71 & 0.7 & 3.7 & 27.02 & 10.61 & 0.30 & 7.53\\
    & T & 18:47:39.37 &  -01:58:23.73 & 2.4 $\pm$ 0.1 & 8.3 & 1.13 & 0.86 & 110.48 & 0.7 & 3.6 & 26.29 & 10.29 & 0.29 & 7.32\\
    & U & 18:47:39.43 &  -01:58:24.36 & 2.5 $\pm$ 0.0 & 15.7 & 2.06 & 1.01 & 91.72 & 0.4 & 6.8 & 38.33 & 13.16 & 0.37 & 10.67\\
    & V & 18:47:39.57 &  -01:58:25.41 & 2.5 $\pm$ 0.1 & 9.1 & 1.38 & 0.62 & 161.16 & 1.0 & 4.0 & 24.62 & 8.88 & 0.25 & 6.86\\
\\
\hline
\\
    & A & 18:47:36.52 &  -01:55:19.64 & 11.8 $\pm$ 0.2 & 13.3 & 0.22 & 0.20 & 77.40 & 149.3 & 5.8 & 5.57 & 0.87 & 0.03 & 1.60\\
    & B & 18:47:36.58 &  -01:55:33.17 & 14.7 $\pm$ 0.6 & 27.4 & 0.57 & 0.50 & 14.39 & 18.6 & 11.9 & 14.19 & 2.46 & 0.08 & 4.07\\
    & C & 18:47:36.49 &  -01:55:33.76 & 8.7 $\pm$ 0.8 & 18.3 & 0.86 & 0.42 & 68.01 & 8.6 & 8.0 & 16.03 & 3.61 & 0.12 & 4.60\\
    & D & 18:47:36.39 &  -01:55:35.19 & 14.9 $\pm$ 0.9 & 54.6 & 1.28 & 0.61 & 11.19 & 8.1 & 23.8 & 23.54 & 3.72 & 0.13 & 6.75\\
    & E & 18:47:36.60 &  -01:55:32.39 & 8.4 $\pm$ 0.4 & 34.8 & 1.54 & 0.55 & 6.58 & 4.6 & 15.2 & 24.51 & 4.95 & 0.17 & 7.02\\
    & F & 18:47:36.64 &  -01:55:35.14 & 4.5 $\pm$ 0.4 & 11.4 & 0.87 & 0.55 & 170.32 & 3.6 & 5.0 & 18.37 & 5.61 & 0.19 & 5.27\\
    & G & 18:47:36.44 &  -01:55:28.27 & 2.7 $\pm$ 0.3 & 4.6 & 0.57 & 0.40 & 50.33 & 4.2 & 2.0 & 12.78 & 5.15 & 0.17 & 3.66\\
    & H & 18:47:37.19 &  -01:55:26.31 & 1.4 $\pm$ 0.2 & 2.0 & 0.52 & 0.19 & 40.05 & 6.5 & 0.9 & 8.41 & 4.15 & 0.14 & 2.41\\
    & I & 18:47:37.25 &  -01:55:28.58 & 1.6 $\pm$ 0.1 & 4.9 & 0.98 & 0.64 & 12.62 & 1.0 & 2.1 & 21.15 & 10.59 & 0.36 & 6.06\\
MM8 & J & 18:47:36.74 &  -01:55:32.74 & 2.1 $\pm$ 0.2 & 8.6 & 1.56 & 0.67 & 81.53 & 0.8 & 3.7 & 27.37 & 11.77 & 0.40 & 7.85\\
    & K & 18:47:37.70 &  -01:55:24.23 & 1.3 $\pm$ 0.3 & 4.0 & 1.52 & 0.39 & 71.72 & 0.9 & 1.7 & 20.59 & 11.28 & 0.38 & 5.90\\
    & L & 18:47:35.90 &  -01:55:35.12 & 3.7 $\pm$ 0.3 & 8.8 & 0.71 & 0.67 & 49.76 & 2.8 & 3.8 & 18.31 & 6.36 & 0.21 & 5.25\\
    & M & 18:47:37.25 &  -01:55:27.56 & 1.5 $\pm$ 0.0 & 7.1 & 1.53 & 0.78 & 24.31 & 0.6 & 3.1 & 29.19 & 14.26 & 0.48 & 8.37\\
    & N & 18:47:37.42 &  -01:55:22.25 & 1.3 $\pm$ 0.1 & 8.0 & 1.52 & 1.17 & 20.39 & 0.3 & 3.5 & 35.68 & 18.15 & 0.61 & 10.23\\
    & O & 18:47:37.36 &  -01:55:24.55 & 1.2 $\pm$ 0.1 & 4.9 & 1.28 & 0.91 & 98.82 & 0.4 & 2.1 & 28.71 & 16.72 & 0.56 & 8.23\\
    & P & 18:47:36.17 &  -01:55:20.38 & 1.6 $\pm$ 0.2 & 6.4 & 1.36 & 0.77 & 101.12 & 0.6 & 2.8 & 27.23 & 13.53 & 0.46 & 7.80\\
    & Q & 18:47:36.01 &  -01:55:20.44 & 1.9 $\pm$ 0.3 & 5.9 & 1.24 & 0.62 & 87.22 & 0.9 & 2.6 & 23.34 & 11.12 & 0.38 & 6.69\\
    & R & 18:47:36.17 &  -01:55:30.99 & 1.1 $\pm$ 0.2 & 1.8 & 0.63 & 0.40 & 126.53 & 1.5 & 0.8 & 13.32 & 8.60 & 0.29 & 3.82\\
    & S & 18:47:37.26 &  -01:55:23.46 & 1.0 $\pm$ 0.1 & 6.5 & 1.61 & 1.16 & 153.02 & 0.3 & 2.8 & 36.43 & 20.76 & 0.70 & 10.44\\
    & T & 18:47:36.93 &  -01:55:28.86 & 0.9 $\pm$ 0.0 & 4.3 & 1.29 & 1.07 & 117.06 & 0.3 & 1.9 & 31.30 & 20.24 & 0.68 & 8.97\\
\\
\hline
\hline
\enddata
\tablenotetext{a}{Deconvolved major and minor axes are obtained from the Gaussian fitting procedure.}
\tablenotetext{b}{Volume number densities were calculated assuming spherical geometry.}
\tablenotetext{c}{The fragment size is calculated as d = 5500*$\sqrt{b_{maj} \times b_{min}}$ [pc].}
\end{deluxetable}
\end{longrotatetable}

\newpage

\startlongtable
\begin{deluxetable*}{c c c c c c c c c c c c c c c}        
\tablecolumns{15}
\tablewidth{0pt}
\tabletypesize{\scriptsize}
\tablecaption{The Table show the polarization parameters and the magnetic field estimations onto the plane of the sky using three versions of the DCF method.
The polarization parameters are calculated from the 3$\sigma$ data. We also show the mass to magnetic flux ratio ($\lambda_{B}$) 
for each of the cores where we have estimations of field strength. For the MM2 and MM3 clumps, some of our estimates for B$_{\mathrm{pos}}$ are lower limits which we indicate by using the $>$ sign. In the same way, this lower limits give upper limits for $\lambda_{B}$, which we indicate using the $<$ sign. \label{fieldValues}}
\tablehead{
    \colhead{Clump} &
    \colhead{Core} &
    \colhead{Region} &
    \colhead{N\tablenotemark{a}}&
    \colhead{$<\phi>$\tablenotemark{b}} &
    \colhead{$\delta \phi$\tablenotemark{b}} &
    \colhead{F$_{min}$\tablenotemark{b}} & 
    \colhead{F$_{max}$\tablenotemark{b}} & 
    \colhead{$<\mathrm{F}>$\tablenotemark{b}} & 
    \colhead{B$_{1}$\tablenotemark{c}} &  
    \colhead{B$_{2}$\tablenotemark{d}} &  
    \colhead{B$_{3}$\tablenotemark{e}} &  
    \colhead{$\lambda_{\mathrm{B_{1}}}$\tablenotemark{f}} & 
    \colhead{$\lambda_{\mathrm{B_{2}}}$\tablenotemark{g}} &
    \colhead{$\lambda_{\mathrm{B_{3}}}$\tablenotemark{h}}   \\ 
    \colhead{ } & 
    \colhead{ } & 
    \colhead{ } & 
    \colhead{[$10^{24}$ cm$^{-2}$ ]} & 
    \colhead{[$^{\circ}$]} & 
    \colhead{[$^{\circ}$]} & 
    \colhead{[\%]} & 
    \colhead{[\%]} & 
    \colhead{[\%]} & 
    \colhead{[mG]} & 
    \colhead{[mG]} & 
    \colhead{[mG]} &
    \colhead{ } & 
    \colhead{ } & 
    \colhead{ } 
}
\startdata 
      & A & 1 & 47.5 & -23.9 & 39.6 & 0.02 & 7.8 & 2.5 & $>$2.2 & $>$3.2 & $>$0.6 & $<$53.5 & $<$37.8 & $<$208.1\\
      & B & 3 & 3.6 & 7.9 & 43.6 & 0.07 & 26.3 & 5.1 & 1.0 & 1.4 & 0.5 & 8.6 & 6.4 & 18.9\\
MM2   & C & 4 & 3.3 & 26.1 & 40.2 & 0.45 & 7.0 & 3.5 & 0.8 & 1.1 & 0.2 & 10.8 & 7.7 & 47.4\\
      & D & 1 & 2.2 & -23.9 & 39.6 & 0.02 & 7.8 & 2.5 & $>$2.2 & $>$3.2 & $>$0.6 & $<$2.4 & $<$1.7 & $<$9.4\\
      & F & 2 & 1.5 & 2.2 & 66.1 & 0.19 & 14.2 & 5.1 & 0.5 & 0.5 & 0.1 & 7.4 & 8.6 & 36.4\\
      & J & 3 & 0.7 & 7.9 & 43.6 & 0.07 & 26.3 & 5.1 & 1.0 & 1.4 & 0.5 & 1.8 & 1.3 & 3.9\\
\\
\hline
\\
      & A & 3 & 9.5 & -8.5 & 42.8 & 0.01 & 8.3 & 2.4 & $>$1.0 & $>$1.4 & $>$0.1 & $<$23.3 & $<$17.0 & $<$292.0 \\
      & B & 3 & 4.8 & -8.5 & 42.8 & 0.01 & 8.3 & 2.4 & $>$1.0 & $>$1.4 & $>$0.1 & $<$12.0 & $<$8.7 & $<$149.7 \\
MM3   & C & 4 & 2.2 & 50.1 & 33.2 & 0.06 & 14.3 & 5.0 & 1.3 & 2.0 & 0.4 & 4.2 & 2.8 & 12.4 \\
      & D & 3 & 1.2 & -8.5 & 42.8 & 0.01 & 8.3 & 2.4 & $>$1.0 & $>$1.4 & $>$0.1 & $<$3.0 & $<$2.2 & $<$37.4 \\
      & E & 2 & 1.4 & -24.9 & 15.6 & 0.09 & 5.2 & 2.8 & 4.0 & 6.5 & 5.9 & 0.9 & 0.5 & 0.6 \\
      & F & 1 & 1.1 & -38.2 & 10.3 & 0.24 & 51.6 & 23.4 & 3.3 & 5.6 & 3.7 & 0.9 & 0.5 & 0.8 \\
\\
\hline
\\
MM4   & A & 1 & 1.3 & -56.7 & 22.1 & 0.11 & 6.1 & 2.2 & 1.4 & 2.3 & 0.1 & 2.4 & 1.5 & 32.6\\
      & F & 1 & 0.8 & -56.7 & 22.1 & 0.11 & 6.1 & 2.2 & 1.4 & 2.3 & 0.1 & 1.3 & 0.8 & 18.7\\
\\
\hline
\\
MM7   & H & 1 & 0.3 & 67.8 & 10.8 & 0.46 & 19.2 & 11.0 & 1.7 & 2.9 & 0.6 & 0.4 & 0.3 & 1.4 \\
      & K & 1 & 0.5 & 67.8 & 10.8 & 0.46 & 19.2 & 11.0 & 1.7 & 2.9 & 0.6 & 0.7 & 0.4 & 2.2 \\
\\
\hline
\hline
\enddata

\tablenotetext{a}{The column density displayed here corresponds to the core column density indicated by its letter, not the region
used to extract $\delta \phi$.}
\tablenotetext{b}{Here $<\phi>$ is the average EVPA, $\delta \phi$ is the EVPA dispersion 
(calculated using circular statistics), F$_{min}$ is the minumum fractional polarization,
F$_{max}$ is the maximun fractional polarization, and $<\mathrm{F}>$ is average fractional polarization value. 
All values are computed for the pixels in the region indicated in column 2.}
\tablenotetext{c}{Estimations of the magnetic field, in the plane of the sky, done with the original CF method (see equation \ref{cf} in the text)}
\tablenotetext{d}{Estimations of the magnetic field, in the plane of the sky, done using the corrections implemented by \citet{Falceta2008} equation 9}
\tablenotetext{e}{Estimations of the magnetic fieldin the plane of the sky,  done using the corrections implemented by \citet{Heitsch2001} equation 12}
\tablenotetext{f}{Mass to magnetic flux estimate using field strength estimate B$_{1}$}
\tablenotetext{g}{Mass to magnetic flux estimate using field strength estimate B$_{2}$}
\tablenotetext{h}{Mass to magnetic flux estimate using field strength estimate B$_{3}$}
\end{deluxetable*}

\newpage

\begin{deluxetable*}{c c c c c c c c c c c}
\tablecolumns{11}
\tablewidth{0pt}
\tabletypesize{\scriptsize}
\tablecaption{Results from the infall model fit\label{infParam}}              
\tablehead{
\colhead{Clump} &
\colhead{Line} &
\colhead{$\phi$} &
\colhead{$\tau_{0}$} &
\colhead{$J(T_{\mathrm{c}})$} &
\colhead{$J(T_{\mathrm{f}})$} &
\colhead{$J(T_{\mathrm{r}})$} &
\colhead{$v_{\mathrm{lsr}}$} &
\colhead{$\sigma$} &
\colhead{$v_{\mathrm{f}}$} &
\colhead{$v_{\mathrm{r}}$}\\
\colhead{} &
\colhead{} &
\colhead{} &
\colhead{} &
\colhead{[K]} &
\colhead{[K]} &
\colhead{[K]} &
\colhead{[\kms]} &
\colhead{[\kms]} &
\colhead{[\kms]} &
\colhead{[\kms]}
}
\startdata
MM2 & HCN       & 0.080 & 0.0 & 12.7 & 70.2  & 104.4 & 90.8 & 1.4 & 2.9 & 1.3\\
MM2 & HCN       & 0.080 & 2.2 & 12.7 & 1.9   & 1.8   & 90.8 & 1.7 & 2.8 & 0.1\\
MM2 & HCO$^{+}$ & 0.080 & 1.0 & 12.6 & 0.0   & 4.4   & 90.8 & 2.2 & 0.4 & -0.1\\
MM2 & HCO$^{+}$ & 0.080 & 1.0 & 12.6 & 0.0   & 4.4   & 90.8 & 2.2 & 0.4 & -0.0\\
MM3 & HCN       & 0.080 & 0.8 & 12.7 & 2.5   & 2.3   & 93.5 & 1.5 & 3.3 & 1.9\\
MM3 & HCN       & 0.080 & 0.0 & 12.7 & 110.0 & 193.0 & 93.5 & 2.0 & 3.5 & 1.8\\
MM3 & HCO$^{+}$ & 0.080 & 1.4 & 12.6 & 0.0   & 7.6   & 93.5 & 1.8 & 0.8 & -0.5\\
MM3 & HCO$^{+}$ & 0.080 & 1.3 & 12.6 & 0.0   & 7.5   & 93.5 & 1.9 & 0.9 & -0.6\\
\enddata
\end{deluxetable*}

\begin{deluxetable*}{c c c c c}
\tablecolumns{5}
\tablewidth{0pt}
\tablecaption{\scriptsize The velocity dispersion for each clump are presented here. 
	The values for $\sigma$ are derived assuming the turbulent power spectrum $\sigma^{2} = bL^{n}$ as
	indicate in the text. Here $\sigma_{\mathrm{CSO}}$ gives the velocity dispersion calculated from the
	H$^{13}$CO$^{+}(3 \rightarrow 2)$ taken from \citet[Table 2 in ][]{Motte2003}\footnote{See text for 
	a discussion on MM4, MM6, and MM8}, $\sigma_{\mathrm{B}}$ gives the values for the
	length-scales traced by the magnetic field region, and $\sigma_{\mathrm{core}}$ is velocity 
	dispersion for the core scales, or $1^{\prime \prime}$ \label{vdTab}}
\tablehead{
\colhead{Clump} &
\colhead{b} &
\colhead{$\sigma_{\mathrm{CSO}}$} &
\colhead{$\sigma_{\mathrm{B}}$} &
\colhead{$\sigma_{\mathrm{core}}$} \\
\colhead{}     &
\colhead{[km$^{2}$ s$^{-2}$ arcsec$^{-n}$]}     &
\colhead{[km s$^{-1}$]}   &
\colhead{[km s$^{-1}$]}   &
\colhead{[km s$^{-1}$]}
}
\startdata
MM2 & 0.54 & 1.78 & 0.99 & 0.74\\
MM3 & 0.57 & 1.87 & 1.01 & 0.76\\
MM4 & 0.35 & 1.14 & 0.79 & 0.59 \\
MM6 & 0.49 & 1.61 & - & 0.7\\
MM7 & 0.48 & 1.57 & 0.93 & 0.69 \\
MM8 & 0.49 & 1.61 & - & 0.7\\
\enddata
\end{deluxetable*}

\newpage
\clearpage

\bibliographystyle{aasjournal}
\bibliography{biblio}

\end{document}